%% file: main_arxiv.tex
\documentclass{article}
 
\usepackage[a4paper,bindingoffset=0.2in,
             left=0.9in,right=0.9in,top=1in,bottom=1in,
             footskip=.25in]{geometry}
\usepackage[hang,flushmargin, multiple]{footmisc} 
\usepackage[utf8]{inputenc}
\usepackage{pgfplots}
\DeclareUnicodeCharacter{2212}{−}
\usepgfplotslibrary{groupplots,dateplot}
\usetikzlibrary{patterns,shapes.arrows}
\pgfplotsset{compat=newest}
\usetikzlibrary{arrows,matrix,positioning}
\usepackage[normalem]{ulem}
\usepgfplotslibrary{fillbetween}
\usepackage{hyperref}
\usepackage{algorithm}
\usepackage{algpseudocode}
\usepackage{pifont}
\usepackage[normalem]{ulem}
\usepackage{caption}
\usepackage{amssymb}
\usepackage{amsfonts} 
\usepackage{amsmath} 
\usepackage{url}
\usepackage{cite}
\usepackage{amsthm}
\theoremstyle{plain}

\theoremstyle{definition}

\theoremstyle{remark}

\usepackage{multirow, booktabs}
\usepackage{color}
\usepackage{tcolorbox}
\usepackage{todonotes}
\newcommand{\myfont}{\fontsize{27}{60}\selectfont}

\newcommand{\mbf}{\mathbf}

\newcommand{\mcl}{\mathcal}
 
\usepackage{xstring}
\def\alphabet{abcdefghijklmnopqrstuvwxyzABCDEFGHIJKLMNOPQRST123456789}
\renewcommand{\vec}[1]{
\IfSubStr{\alphabet}{#1}{
\ensuremath{\mathbf{\MakeLowercase{#1}}}
}{
\ensuremath{\boldsymbol{\MakeLowercase{#1}}}
}
}
\newcommand{\mat}[1]{
\IfSubStr{\alphabet}{#1}{
\ensuremath{\mathbf{\MakeUppercase{#1}}}
}{
\ensuremath{\boldsymbol{\MakeUppercase{#1}}}
}
}

\def\R{\mathbb R}
\def\C{\mathbb C}

\newcommand*{\norm}[1]{\left\|#1\right\|}

\newcommand*{\card}[1]{\left|#1\right|}
\newcommand*{\diag}[1]{\text{diag}\left(#1\right)}
\newcommand*{\vct}[1]{\text{vec}\left(#1\right)}

\newcommand{\AB}[1]{\textcolor{red}{#1}}

\def\defeq{:=}

\usepackage{acronym}
\acrodef{PACBED}{position-averaged convergent beam electron diffraction pattern}
\acrodef{PIE}{Ptychographic Iterative Engine}
\acrodef{ADMM}{Alternating Direction Method of Multipliers}
\acrodef{GS}{Gerchberg Saxton}
\acrodef{HIO}{Hybrid Input-Output}
\acrodef{TEM}{transmission electron microscopy}
\acrodef{STEM}{scanning transmission electron microscopy}
\acrodef{MoS$_2$}{Molybdenum disulfide}
\acrodef{GaAs}{Gallium arsenide}
\acrodef{SrTiO$_3$}{Strontium titanate}
\title{Inverse Multislice Ptychography by Layer-wise Optimisation and Sparse Matrix Decomposition }
\author{Arya Bangun\thanks{Ernst Ruska-Centre for Microscopy and Spectroscopy with Electrons, Forschungszentrum J{\"u}lich, Wilhelm-Johnen-Strasse, 52425 J{\"u}lich, Germany.}, Oleh Melnyk\thanks{Mathematical Imaging and Data Analysis, Helmholtz Center Munich.}  \thanks{Department of Mathematics, Technical University of Munich.}, Benjamin März\thanks{Department of Chemistry and Center for NanoScience, Ludwig-Maximilians-University of Munich (LMU), Butenandtstr.~11, 81377 M\"unich, Germany.}, Benedikt Diederichs\footnotemark[1] \footnotemark[2],\\ Alexander Clausen\footnotemark[1], Dieter Weber\footnotemark[1], Frank Filbir\footnotemark[2], Knut M{\"u}ller-Caspary\footnotemark[1] \footnotemark[4].}

\begin{document}
 
\maketitle
\begin{abstract}
We propose algorithms based on an optimisation method for inverse multislice ptychography in, e.g. electron microscopy. The multislice method is widely used to model the  interaction between relativistic electrons and thick specimens. Since only the intensity of diffraction patterns can be recorded, the challenge in applying inverse multislice ptychography is to uniquely reconstruct the electrostatic potential in each slice up to some ambiguities. In this conceptual study, we show that a unique separation of atomic layers for simulated data is possible when considering a low acceleration voltage. We also introduce an adaptation for estimating the illuminating probe. For the sake of practical application, we finally present slice reconstructions using experimental 4D scanning transmission electron microscopy (STEM) data. 
\end{abstract} 
 

 

\section{Introduction} \label{Sec1:Intro}

One of the fundamental challenges in electron microscopy is dealing with phase retrieval from the intensity of diffraction patterns. The reason for this problem is that current detectors used in electron microscopy are unable to record phase information, which
is necessary for example to improve the image resolution, to understand the interaction of electrons and atoms within a material, and in particular to recover the electrostatic potential of the specimen. 

Several methodologies and approaches have been developed for solving the phase problem. One of the most prevalent techniques is ptychography. Instead of only exploiting the intensity of a single diffraction pattern, ptychography takes advantage of a large set of subsequently recorded diffraction patterns stemming from multiple, partly overlapping illuminations of the object. {Here, the illumination, given by the electron wave incident, is sometimes also called as the probe. }
In essence, the acquisition of diffracted intensities from adjacent scanning positions provides additional information enabling to solve the phase problem \cite{Hoppe1969,Hoppe1969a,Hoppe1969b}. 
Following on from these original approaches, various contributions to the phase retrieval from a single diffraction pattern have led to the introduction of new algorithms, i.e. enhanced methods adapted for ptychographic reconstructions.
For instance, adoptions from alternating projection-based algorithms like classical \ac{GS} \cite{gerchberg1972practical} and Fienup \ac{HIO} \cite{fienup1982phase} are referred to under the term of \ac{PIE} algorithms \cite{rodenburg2004phase,maiden2009improved,YANG2017173}. Another approach for solving the phase problem by direct inversion has been proposed in refs. \cite{rodenburg1992theory,li2014ptychographic}. It utilises the property of the ambiguity function, sometimes also called Wigner function, which naturally appears by reformulating the equation for deriving the intensity in terms of the probe and object transfer functions.

In addition to the aforementioned methods, further approaches for modelling ptychography as an optimisation problem have been developed over the last few years. 
As the phase retrieval problem is generally non-convex, there is no certainty that the global optimum can be attained. However, several contributions \cite{wang2017solving,candes2015phase,tan2019phase} manage to achieve the convergence to a local optimum. 
The crucial fundamental assumption for most studies is the single multiplicative approximation used for modelling the interaction between the electron beam and a thin specimen. However, this assumption does not necessarily hold when investigating thick specimens due to strong dynamical electron scattering effects \cite{cowley1957scattering,kirkland1998advanced}. 
For this purpose, one should take into account the theory of multiple scattering and propagation when solving the phase problem for thick specimens, e.g. 
via the multislice approach \cite{cowley1957scattering,goodman1974numerical}, Bloch waves \cite{bethe1928theorie, howie1961diffraction}, i.e. scattering matrix-based formulations \cite{donatelli2020inversion}.

Several attempts have been made to adapt the phase retrieval model for thick specimens by incorporating the scattering matrix, as discussed in \cite{donatelli2020inversion,brown2020three,pelz2021phase}. In \cite{donatelli2020inversion}, the authors developed an iterative projection algorithm called $N-phaser$ for estimating the scattering matrix. The key idea stems from the specific eigenvalue structure of the scattering matrix. It can therefore be used for estimating the object transfer function from a thick specimen while eliminating the unwanted scattering artefacts in the recorded diffraction patterns.
Another approach has been proposed in \cite{pelz2021phase}, where the authors used optimisation methods, e.g. \ac{ADMM} and block coordinate descent, in order to estimate both the scattering matrix as well as the probe. {A similar approach to estimate object and probe is presented in \cite{zhong2016nonlinear}, where the reconstruction is done iteratively by a modified Gauss–Newton method.}
\vspace{-0.2cm}

\subsection{Related work}
Implementations of inverse multislice ptychography have been applied for instance in \cite{maiden2012ptychographic,li2018multi, kahnt2021multi}. The key idea in these studies bears a strong resemblance to extending the established algorithms, such as the extended \ac{PIE}. As three-dimensional specimens were investigated, these algorithms were named 3\ac{PIE}. 
The forward model deals with the propagation of the specimen entrance wave (probe) and calculates the complex wave function for the observed specimen at a specific thickness. 
 The backward model constructs an estimation of the entrance wave by applying an inverse Fourier transform to the product of the estimated phase and the intensity of diffraction patterns acquired by the measurement. {However, mentioned works focus on the reconstruction of visible light and x-ray datasets.} {The same algorithm has been applied to reconstruct images from the LED microscope data in \cite{tian20153d}}. 
 {Another approach are gradient-based methods, where the gradients are calculated over the whole multislice model all at once.
Examples can be found in \cite{van2012method, van2013general, schloz2020overcoming, chen2021electron}.}

In order to address the inverse multislice ptychography problem for electron microscopy data sets we present two different approaches, an adaptation from the Amplitude Flow method and a matrix decomposition, respectively.
Amplitude Flow is a gradient-based method, which has been analysed for a randomised one-dimensional phase retrieval \cite{wang2017solving}. This analysis was later enhanced for arbitrary measurements and in particular for ptychography \cite{Xu.2018}.
In the second approach, a matrix factorisation technique adopted from the field of optimisation and dictionary learning \cite{le2016flexible} is incorporated into the estimation of the matrix from intensity measurements. 
Apart from proposing different techniques to solve inverse multislice ptychography we have also reformulated the forward multislice model. This adjustment enables separation of the effect of the illuminating probe from interaction with the specimen, which in turn yields only the construction of a thick object transmission function in respect of a single matrix.
Additionally,  we outline a methodology to reconstruct each atomic plane by applying the proposed algorithm to synthetic data simulated for a low acceleration voltage. However, it should be noted that increasing the number of slices affects the decomposition performance.
\vspace{-0.2cm}
\subsection{Summary of Contributions}
\begin{itemize}
    \item The forward multislice model is reformulated to disentangle the effect of the probe from the object transfer function at any thickness. This allows modelling a thick object as a matrix consisting only of the product of phase gratings and Fresnel propagators.
    \item Two approaches for estimating the slices of a thick object are proposed, namely layer-wise optimisation and sparse matrix decomposition. In the first approach we cycle over slices applying the Amplitude Flow algorithm, only optimising with respect to a single slice. The matrix which represents the phase gratings and the Fresnel propagations were recovered by applying a second algorithm. Further factorisation of this matrix was then carried out in order to extract the slices of the object.
    \item Simulations of diffraction data of specimens (MoS$_2$, SrTiO$_3$ and GaAs) with different crystal structures have been carried out. These simulations were performed for different energies of the incoming electrons, i.e. different wavelengths. They serve as the ground truth for determining the error of the reconstructions, which were carried out using the proposed algorithms. 
    {Since the depth resolution in an electron microscope is limited, we investigated the impact of the slice thickness and under what conditions a unique reconstruction is possible.} {We also adapted the algorithm to estimate the illuminating probe and present the probe reconstruction}.
    \item To highlight the practical use of our proposed method we provide results for the first applications of the algorithms in respect of experimental data, notably in our reconstructions of the object transfer function of a MoS$_2$ specimen using a four-dimensional data set acquired by scanning transmission electron microscopy (STEM).
\end{itemize}
\vspace{-0.3cm}
\subsection{Notations}
Vectors are written in bold small-cap letters $\mathbf{x} \in \C^L$ and matrices are written as a bold big-cap letter $\mathbf{A} \in \C^{K \times L}$ for a complex field $\C$ and for a real field $\R$.  A matrix can also be written by indexing its elements 
\[\mathbf{A} = \left(a_{k\ell}\right), \quad \text{where} \quad k \in [K], \ell \in [L].
\]
The set of integers is written as $[N] := \{1,2,\hdots,N\}$ and calligraphic letters are used to define functions $\mathcal{A} : \C \rightarrow \C$. Specifically, we denote the discrete two-dimensional Fourier transform by $\mathcal{F}$.
For both matrices and vectors, the notation $\circ$ is used to represent element-wise or Hadamard product. The $\mbf{A}^H$ is used to represent conjugate transpose. For a vector $\mbf x \in \C^L$, the $\ell_p$-norm is given by $\norm{\mbf x}_p = \left(\sum_{\ell = 1}^L \card{x_\ell}^p\right)^{1/p},$ $1 \le p < \infty$ and for $p = \infty$ we have $\underset{\ell \in [L]}{\text{max}} \card{x_\ell}$. 
For a matrix $\mathbf{X} \in \C^{K \times L}$, the Frobenius norm is denoted by $\norm{\mbf X}_F \defeq \sqrt{\sum_{k = 1}^K \sum_{\ell = 1}^L \card{X_{k\ell}}^2} $ and the spectral norm is given by $\norm{\mbf X} = \underset{\norm{\mbf v}_2 = 1}{\text{max}}   \norm{\mbf X \mbf v}_2$. The trace operator $\text{Tr}\left(.\right)$ is the operator to sum all elements in the diagonal of square matrices. {Indices are wrapped around, so that $\mbf x_{-p}=\mbf x_{N-p}$.}

\section{Problem Statement}\label{Sec2:ProbState}


\subsection{Forward Multislice Model}

\begin{figure}[htb!]
\centering
\includegraphics[scale=0.4]{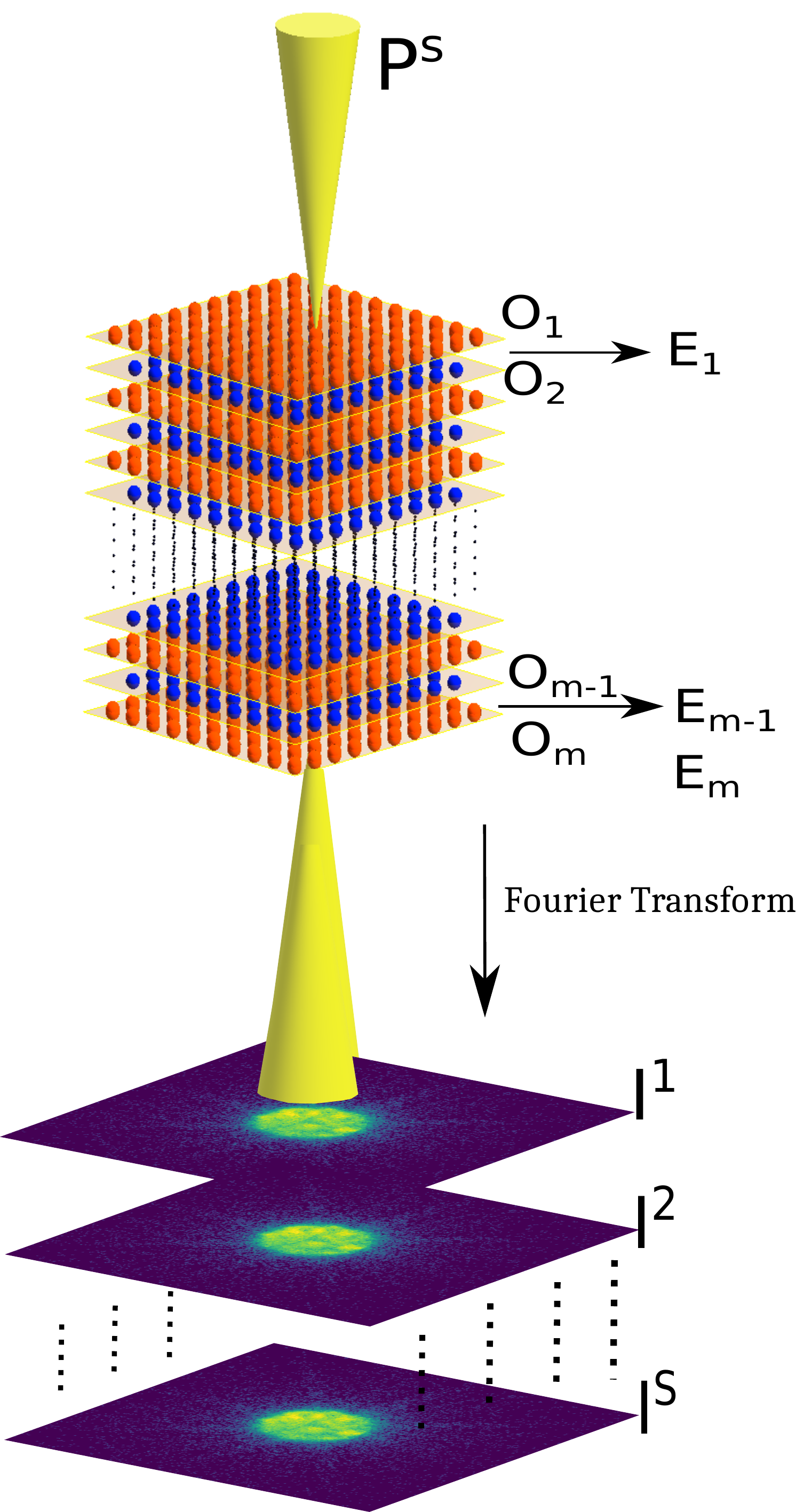}
\vspace{0.2cm}
\caption{Illustration of the multislice method showing a probe focused on a specimen and the resulting diffraction intensity $\mbf{I}^s$ at scanning position $s$ recorded in the far field. After each slice $\mbf X_M$ of the specimen an exit wave $\mbf E^s_M$ is produced.}
\label{Fig:Multislice}
\end{figure}

The forward multislice model is based on the idea that a thick object can be approximated by multiple thin slices stacked on top of each other. For each slice of the specimen the interaction between the electron wave incident on this slice and the potential of the slice can be modelled by a multiplicative approximation in analogy to the standard model in ptychography. Furthermore, as the illumination progresses through the object, the exit wave of the previous slice propagates through potential-free space to the subsequent slice, where it acts as the new illumination for this slice, as schematically shown in Figure \ref{Fig:Multislice}.  
%

Considering an aberration-free probe, the two-dimensional probe $\mbf{P} \in \C^{N \times N}$ can be described 
for different aperture sizes $q_{\text{max}}$ with entries given by 
\begin{equation*}
\begin{aligned}
   \small
p_{y,x} &= \pi q_{\text{max}}^2 \left(\frac{2J_1\left(2 \pi q_{\text{max}} \card{\mbf{r}} \right)}{2\pi q_{\text{max}}\card{\mbf{r}}} \right),\text{where} \\ & \quad y,x \in [N], \ \card{\mbf{r}} = \sqrt{ x^2 + y^2}, 
\end{aligned}
\end{equation*}
where $J_1$ is a Bessel function of the first kind of order 1. {The intensity of this function is called Airy disk. In general, this function can be derived analytically by applying a two-dimensional inverse Fourier transform to a circular aperture.}
The probe shifted to the scanning position $(x_s, y_s), s \in [S]$ is denoted by a matrix $\mbf P^s \in \C^{N \times N}$ with entries $p^s_{x,y} = p_{x - x_s, y - y_s}$. {In general aberrations exist and affect the probe formation. In this study the focus is on the aberration-free condition when generating the simulated data. For a more general treatment of this subject please refer to \cite{kirkland1998advanced}.}

The interaction between the probe $\mbf P^s \in \C^{N \times N}$ at 
scanning point $s \in [S]$ and the first slice $\mbf X_1 \in \C^{N \times N}$ is given by the element-wise product and in turn produces an exit wave {of slice~1}
\begin{equation*}
\mbf E^s_1 = \mbf P^s \circ \mbf X_1.
\end{equation*}
After passing through the first slice the propagation of the exit wave between the slices is modelled by the Fresnel transform $\mathcal{V}_z$ which is given by
\[
\mathcal{V}_z\left(\mathbf{E}\right) \defeq \mcl{F}^{-1}\left(\mcl{F} \left( \mathbf{E}\right) \circ \mbf{H}_m \right),
\]
where $\mcl{F}$ is the Fourier operator and $\mbf{H}_m \in \C^{N \times N}$ is the Fresnel propagator matrix with entries
\begin{equation}
h_{y,x} \defeq e^{-\pi i\Delta_m \lambda \left( \left( q_y^2 + q_x^2 \right) + 2\left( q_x \frac{\sin \theta_x}{\lambda} + q_y \frac{\sin \theta_y}{\lambda} \right)\right)}, \quad y,x\in [N].
\label{eq:fresnel_element}
\end{equation}
The parameters $q_y,q_x$ denote the discrete grid in the reciprocal space and hence represent spatial frequencies, $\Delta_m$ is the distance of the wave propagation, and $\theta_x, \theta_y$ are the two-dimensional tilt angles. In this article, the illumination was set to be perpendicular to the object surface along a major crystallographic axis, i.e. tilt angles are zero.

As the beam reaches the second slice, it is described by $\mathcal{V}_z\left(\mbf E^s_1\right)$ and the next exit wave is given by
\begin{equation*}
\mbf E^s_2 = \mathcal{V}_z\left(\mbf E^s_1\right) \circ \mbf X_2.
\end{equation*}
%
Consequently, the general representation of the $m$-th observed exit wave is written as
$$
\mathbf{E}^s_m = \mathcal{V}_z\left( \mathbf{E}^s_{m-1}\right) \circ \mbf{X}_m \quad \text{for} \quad  m \in [M], m \neq 1,
$$
Finally, the intensity of the Fraunhofer diffraction pattern that is recorded by a detector in the far field is given by
\begin{equation}\label{eq: intensity measurements}
\mbf{I}^s = \left|{\mcl{F}\left(\mbf{E}^s_M \right)}\right|^2.
\end{equation}
In \ac{STEM} the illuminating probe is rastered across the specimen. Therefore, the illumination is varied to yield a set of $S$ diffraction pattern intensities collected throughout an experiment. This four-dimensional data set is then subjected to phase retrieval by multislice ptychography.

\vspace{-0.2cm}
\subsection{Reformulation of Multislice Ptychography}
The measurement model in \eqref{eq: intensity measurements} can be further reformulated in order to separate the probe- and the object-related terms. This reformulation is based on the following property of the Hadamard product.   
For matrices $\mbf A, \mbf B \in \C^{N \times N}$ the Hadamard product $\mbf A \circ \mbf B$ can be written according to
\begin{equation}
\begin{aligned}
\vct{\mbf A \circ \mbf B} &= \diag{\vct{\mbf A}} \vct{\mbf B}\\
&= \diag{\vct{\mbf B}}  \vct{\mbf A}.
\end{aligned}
\label{eq:haddamard}
\end{equation}
The notation $\text{vec}: \C^{N \times N} \rightarrow \C^{N^2}$ is an operator
that vectorises the matrix and $\text{diag}: \C^{N^2} \rightarrow \C^{N^2 \times N^2}$ constructs a diagonal matrix by placing the elements of the given vector on the main diagonal. Additionally, the second equality in \eqref{eq:haddamard} is valid from the commutative property of the Hadamard product. By using \eqref{eq:haddamard} the first exit wave for $s$-th position of the probe can be rewritten as 
\[
\vct{\mbf E^s_1} = \vct{\mbf P^s \circ \mbf X_1} = \diag{\vct{\mbf X_1}} \vct{\mbf P^s}.
\]
For convenience, the following notations are introduced
\begin{equation*}
\begin{aligned}
 \mbf{o}_m &\defeq \vct{\mbf{X}_m}\in \C^{N^2}, \\
\mbf{O}_m &\defeq \diag{\mbf{o}_m}  \in \C^{N^2 \times N^2}\,\text{for}\, m \in [M],
\end{aligned}
\end{equation*}
 
and
\[
\mbf{p}^s \defeq \vct{\mbf{P}^s} \in \C^{N^2}\, \text{for} \, s \in [S].
\]
As a result, the first exit wave can be simplified as
\[
\vct{\mbf E^s_1} = \mbf{O}_1 \mbf{p}^s.
\]
For the second exit wave the action of the Fresnel propagator is required. Again, by facilitating \eqref{eq:haddamard}, it is given by
\begin{align*}
\vct{ \mathcal{V}_z\left(\mathbf{E^s_1}\right) } 
& = \vct{ \mcl{F}^{-1}\left(\mcl{F} \left( \mathbf{E^s_1}\right) \circ \mbf{H}_1 \right) } \\
&= \mathbf{F}^{-1}_{2D} \vct{ \left(\mcl{F} \left( \mathbf{E^s_1}\right) \circ \mbf{H}_1 \right) } \\
& = \mathbf{F}^{-1}_{2D} \diag{ \vct{\mbf{H}_1} }  \vct{ \mcl{F} \left( \mathbf{E^s_1}\right) } \\
&= \mathbf{F}^{-1}_{2D} \diag{ \vct{\mbf{H}_1} } \mathbf{F}_{2D} \vct{ \mathbf{E^s_1} },
\end{align*}
where matrices $\mathbf{F}_{2D},\mathbf{F}^{-1}_{2D} \in \C^{N^2 \times N^2}$ are two-dimensional Fourier and inverse Fourier matrices, respectively. The term one-dimensional Fourier matrix represents the discrete implementation of the Fourier basis, i.e. when the Fourier basis is sampled and stored as a matrix. Along the same line, the two-dimensional Fourier matrix can be constructed by using the Kronecker product between two one-dimensional Fourier matrices.

Hence, the Fresnel propagator is a multiplication of the vectorised exit wave $\mathbf E^s_m$ with a matrix
\[
\mathbf{G}_m \defeq \mathbf{F}^{-1}_{2D} \text{diag}\left(\vct{\mathbf{H}_m}\right)\mathbf{F}_{2D} \in \C^{N^2 \times N^2}, \ m \in [M-1].
\]
Substituting the obtained representation to the second exit wave results in
\begin{equation*}
\begin{aligned}
\vct{ \mathbf{E^s_2} } &= \vct{\mathcal{V}_z\left(\mbf E^s_1\right) \circ \mbf X_2} \\
&= \diag{ \vct{\mbf X_2} } \vct{\mathcal{V}_z\left(\mbf E^s_1\right)}
= \mbf{O}_2 \mathbf{G}_1 \mbf{O}_1 \mbf{p}^s.
\end{aligned}
\end{equation*}
Consequently, the $M$-th exit wave is given by
\begin{equation}
\begin{aligned}
\vct{ \mathbf{E}_M^s} &=  \left(\mbf{O}_M\prod_{m = 1}^{M-1}\mathbf{G}_{m}\mbf{O}_m\right)\mbf{p}^s
&=: \mathbf{A}_M \mbf{p}^s  
\label{eq.exit_wave_mslices}
\end{aligned}
\end{equation}
and the resulting far-field vectorised intensity is
\begin{align*}
    \mbf{i}^s 
&\defeq \vct{\mbf{I}^s} 
= \vct{ \left|{\mcl{F}\left(\mbf{E}^s_M \right)}\right|^2 }
= \left|\vct{\mcl{F}\left(\mbf{E}^s_M \right)}\right|^2\\
&= \left|\mathbf{F}_{2D} \vct{ \mathbf{E}_M^s} \right|^2
= \left|\mathbf{F}_{2D} \mathbf{A}_M \mbf{p}^s \right|^2.
\end{align*}
By combining all vectorised intensities $\mbf{i}^s$ and probes $\mbf{p}^s$ as columns of the matrices
\begin{align*}
\mbf I &\defeq \left({\mbf{i}^1},{\mbf{i}^2},\dots,{\mbf{i}^S} \right) \in \C^{N^2 \times S}
\text{ and }\\
\mbf P &\defeq \left(\mbf{p}^1,\mbf{p}^2,\dots,\mbf{p}^S \right) \in \C^{N^2 \times S},
\end{align*}
respectively, can be simplified to
\begin{equation}\label{eq: measurement FAP}
\mathbf{I} = \card{\mathbf{F}_{2D}\mathbf{A}_M\mathbf{P}}^2.
\end{equation}
There are at least two benefits of this reformulation. Firstly, the object transfer function for an arbitrary thickness $M$ is now represented by the matrix $\mbf A_M \in \C^{N^2 \times N^2}$. {Note that} $\mbf A_M$ purely represents the properties of the thick specimen without being affected by the probe, in opposition to the model \eqref{eq: intensity measurements}, where the probe's illumination is entangled with the slices. Secondly, the matrix $\mbf A_M$ decomposes into the product according to \eqref{eq.exit_wave_mslices} and each slice $\mbf{O}_m$ can therefore be separated from other multipliers, {i.e. Fresnel propagator $\mbf{G}_m$},  which will be convenient in the next section where the recovery of a thick object is discussed.


With this reformulation the matrices now have an ambient dimension of $N^2\times N^2$ which increases the computational complexity for processing data of this form. However, because most of these matrices are diagonal matrices they may therefore allow a more efficient treatment and storage in comparison to, e.g. the Bloch wave method, in which an eigenvalue decomposition is directly performed on a scattering matrix {of dimension} $N^2\times N^2$.
\section{Methods and Algorithms}\label{Sec4:Algo}
This section considers the recovery of a thick object and the probe from intensity measurements \eqref{eq: measurement FAP} in diffraction space. Firstly it is posed as a constrained optimisation problem, then two algorithms are proposed for solving the optimisation problem under the assumption that the probes are known. Finally concepts for incorporating the probe estimation into the suggested methods are provided.
\vspace{-0.2cm}
\subsection{Inverse Multislice Ptychography as an optimisation problem}
One of the standard approaches for the recovery of the object from intensity measurements is the optimisation of the data fidelity, which is represented by the least squares problem
\begin{equation}
\begin{aligned}
&\underset{\mbf{P},\mbf{O}_m , m\in[M]}{\text{minimize}}
& & \norm{\sqrt{\mbf I} - \card{\mathbf{F}_{2D}\mathbf{A}_M\mathbf{P} } }^2_F\, \\
& \text{subject to}
& & \mathbf{A}_M =  \mbf{O}_M\prod_{m= 1}^{M-1}\mathbf{G}_{m}\mbf{O}_m.\\
\label{Eq:Opt_Prob}
\end{aligned}
\vspace{-0.5cm}
\end{equation}
The challenge in obtaining the minimiser of \eqref{Eq:Opt_Prob} is the non-convexity of the objective function, which results from the absolute value and product representation of the matrix $\mathbf{A}_M$ as well as the multiplication with the probes $\mbf{P}$. In general, non-convex functions are known to require non-polynomial time to find the global optima \cite{Pardalos1991}. 

One common method for tackling a non-convex minimisation is the alternating projections method \cite{Beck.2015,Lu.2019}. It is based on minimising the objective {function} with respect to a single selected unknown at a time, while other unknowns remain fixed. This usually results in simpler intermediate subproblems which can be solved efficiently. Afterwards, the next unknown is chosen for optimisation and this process is continued until the minimisation with respect to any of the variables does not improve further. Whereas the alternating minimisation is well-understood for convex functions \cite{Beck.2015}, it often acts as a heuristic for non-convex ones. Nevertheless, in applications such as ptychography \cite{chang2019blind}, the estimates obtained by alternating minimisation are quite accurate, which motivates applying this technique to inverse multislice ptychography.
\vspace{-0.2cm}
\subsection{Phase retrieval via Amplitude Flow} \label{Sec: gradient}
As observed throughout this section, the minimisation of the objective {function} in \eqref{Eq:Opt_Prob} with respect to a single unknown, either $\mbf{O}_m, \mathbf{A}_M$ or $\mathbf{P}$, leads to the phase retrieval problem.
It concerns the recovery of an unknown vector $\mbf z \in \C^{L}$ from the measurements of the form
\[
\mbf y = \card{\mbf Q \mbf z}^2 \in \mathbb R^{K},
\]
with the measurement matrix $\mbf Q \in \mathbb \C^{K \times L}$.
One popular approach for the reconstruction of $\mbf z$ is the Amplitude Flow algorithm \cite{wang2017solving,Xu.2018}. It applies the gradient descent in order to minimise the least squares objective 
\begin{equation}\label{eq: phase retrieval}
\mcl{A}\left(\mbf z\right) = \norm{\sqrt{\mbf y} - |\mbf Q \mbf z|}^2_2.
\end{equation}
The generalised Wirtinger gradient of the function $\mcl A$ is given by
\[
\nabla \mcl{A}\left(\mbf z \right) = \mbf{Q}^H ( \mbf Q \mbf z - \frac{\mbf Q \mbf z}{\card{\mbf Q \mbf z}} \circ \sqrt{\mbf y} ),
\]
where each element $k \in [K]$ in the fraction $\left(\frac{\mbf Q \mbf z}{\card{\mbf Q \mbf z}}\right)_k$ is set to $0$ whenever $(\mbf Q \mbf z)_k=0$.
{Then, starting from a position} $\mbf{z}^0$, the $t$-th iteration is obtained via the gradient step 
\begin{equation} \label{Eq:Opt_Prob_Amplitude_Flow_Step}
\mbf{z}^{t} = \mbf{z}^{t-1} - \mu \nabla \mcl{A}\left(\mbf{z}^{t-1}\right),
\end{equation}
with learning rate $\mu = \norm{\mbf Q}^{-2}$ given by the squared inverse of the spectral norm of the matrix $\mbf Q$. 
This approach with the chosen learning rate $\mu$ was adopted such that the convergence of the algorithm to the critical point of the objective  {function} \eqref{eq: phase retrieval} can be guaranteed \cite{Xu.2018}. 
\subsection{Layer-wise Optimisation}
For the layer-wise optimisation, the alternating minimisation was adopted so as to optimise \eqref{Eq:Opt_Prob} with respect to a single slice $\mbf{O}_m$ at a time, while keeping all other slices unchanged. Note that the objective in the optimisation problem \eqref{Eq:Opt_Prob} can be understood as a sum of errors for each scanning point $s$,
\begin{equation}
\begin{aligned}
& \underset{\mbf{O}_m, m \in [M]}{\text{minimize}}
&  \sum_{s = 1}^S\norm{\sqrt{\mbf{i}^s} - \card{\mbf{F}_{2D}  \mbf{O}_M\prod_{m = 1}^{M - 1}  \mathbf{G}_{m}\mbf{O}_m \mathbf{p}^s}}^2_2
\label{Eq:Opt_Prob_Layerwise}
\end{aligned}
\end{equation}
Using initial guesses of the object transfer functions $\mathbf{O}_1^0,\ldots, \mathbf{O}_M^0$, the alternating minimisation technique is employed in order to optimise with respect to a single slice $\mathbf{O}_\ell$, $\ell \in [M]$ by solving
{\begin{equation}
\begin{aligned}
\small
\mbf{O}_\ell^{t+1} &= 
 \underset{\mbf{O}_\ell}{\text{arg min}}
  \sum_{s \in [S] } \norm{\sqrt{\mbf{i}^s} - \card{\mathbf{F}_{2D} \mathbf{R}^t_\ell \mathbf{O}_\ell
\mathbf{S}^t_{\ell} \mathbf{p}^s }}^2_2,
\label{Eq:Opt_Prob_Single_Layer}
\end{aligned}
\end{equation}
where the supporting prefix and suffix matrices are given by
\begin{align}
\mathbf{R}^t_\ell 
& := \prod_{m=\ell+1}^{M} \mathbf{O}_m^t \mathbf{G}_{m-1} \,
\,\\
\mathbf{S}^t_{\ell} & := \prod_{m=1}^{\ell-1}  \mathbf{G}_{m} \mathbf{O}_m^{t+1}.
\label{Eq:Layerwise_Prefix_Suffix}
\end{align}
Once an estimate for the $\ell$-th slice is produced, the algorithm continues with the $\ell+1$-th slice. After the $L$-th slice is estimated, the estimation process is repeated from the first slice until a desired stopping criterion is reached.
}
By applying \eqref{eq:haddamard}, the intensity measurement for a single probe $\mathbf{p}^s, s  \in [S]$ can be rearranged according to
\begin{align*}
 \card{\mathbf{F}_{2D}  \mathbf{R}_\ell^t \mathbf{O}_\ell \mathbf{S}_{\ell}^t \mathbf{p}^s }
&= \card{\mathbf{F}_{2D}  \mathbf{R}_\ell^t ~\diag{\mathbf{o}_\ell} 
\mathbf{S}_{\ell}^t \mathbf{p}^s}\\
&= \card{\mathbf{F}_{2D}  \mathbf{R}_\ell^t ~\diag{ \mathbf{S}_{\ell}^t \mathbf{p}^s }
\mathbf{o}_\ell }.
\end{align*}

The optimisation problem \eqref{Eq:Opt_Prob_Single_Layer} is equivalent to the phase retrieval problem
\begin{align}
\mbf{o}_\ell^{t+1} = 
&\, \underset{\mbf{o}_\ell}{\text{argmin}} \norm{\sqrt{\mbf y} - \card{ \mathbf{Q}_\ell^t
\mathbf{o}_\ell }}^2_2, \label{Eq:Opt_Prob_Single_Layer_Simplified}
\end{align}
with the measurement matrix and the measurements given by
\begin{equation}\label{Eq:Opt_Prob_Single_Layer_Matrix}
\mathbf{Q}_\ell^t 
:=
\begin{bmatrix}
\mathbf{F}_{2D}  \mathbf{R}_\ell^t ~\diag{ \mathbf{S}_{\ell}^t \mathbf{p}^1 } \\
\vdots\\
\mathbf{F}_{2D}  \mathbf{R}_\ell^t ~\diag{ \mathbf{S}_{\ell}^t \mathbf{p}^S }
\end{bmatrix}
\text{ and }
\mbf y = 
\begin{bmatrix}
\mbf{i}^1 \\
\vdots\\
\mbf{i}^S
\end{bmatrix}
,
\end{equation}
respectively.
The problem in \eqref{Eq:Opt_Prob_Single_Layer_Simplified} can be solved by running the gradient descent method as discussed in Section \ref{Sec: gradient} for a fixed number of iterations.

{Overall, it grants us the Algorithm \ref{Algo:Layerwise} summarised below.}

\begin{algorithm}[H]
\caption{Layer-wise Estimation}
\label{Algo:Layerwise}
\begin{algorithmic}[1]
\State \textbf{Initialisation:} 
\begin{itemize}
    \item Initial object transfer functions $\mathbf{O}_1^0,\ldots, \mathbf{O}_M^0$.
    \item Intensity measurement $\mbf I \in \R^{N^2 \times S}$.
    \item Number of iterations $T$ and number of gradient steps $K$.
\end{itemize}
\For{each iteration $t \in [T]$}
\For{each layer $\ell \in [M]$}
\State Compute prefix and suffix matrices $\mathbf{\hat{R}}_\ell^t$ and $\mathbf{\hat{S}}_{\ell}^t$ via \eqref{Eq:Layerwise_Prefix_Suffix}.
\State Construct the measurement matrix $\mathbf{Q}_\ell^t$ and measurement $\mbf y$ as in \eqref{Eq:Opt_Prob_Single_Layer_Matrix}.
\State Produce estimate $\mathbf{O}_\ell^{t+1} = \diag{\mathbf{o}_\ell^{t+1}}$ as in \eqref{Eq:Opt_Prob_Single_Layer_Simplified} by performing $K$ gradient steps \eqref{Eq:Opt_Prob_Amplitude_Flow_Step} with starting point $\mathbf{o}_\ell^t$ corresponding to the diagonal elements of $\mathbf{O}_\ell^t$. 
\EndFor
\State If convergence criteria is reached $\to$ Stop
\EndFor
\end{algorithmic}
\end{algorithm}



\subsection{Sparse Matrix Decomposition}
Another approach for solving the optimisation problem \eqref{Eq:Opt_Prob} is separating it into two subproblems. At first, the matrix $\mbf A_M \in \C^{N^2 \times S}$, which represents the complex object transfer function of a thick specimen is estimated from the measurements.
In the second step, this estimated matrix $\mbf A_M$ is to be decomposed in order to determine the slices $\mbf{O}_m^0$, $m \in [M]$, {which is sparse in the sense that only diagonal elements are non-zero}. Both steps are then reiterated. The detailed procedure is described below.

In the first step, the object transfer function is estimated by solving the optimisation problem
\begin{equation}
\begin{aligned}
\mbf{\hat  A} = 
& \underset{\mbf{A}}{\text{arg min}}
&  \frac{1}{2}\norm{\sqrt{\mbf{I}} - \card{\mathbf{F}_{2D} \mbf{A} \mbf{P}}}_F^2.
\label{Eq:Opt_Prob_Matrix}
\end{aligned}
\end{equation}
In case the matrix $\mbf{A}$ is vectorised, \eqref{Eq:Opt_Prob_Matrix} can be treated as a phase retrieval problem of the form \eqref{eq: phase retrieval}, which gives the gradient step
$$ 
\mbf{A}^{t+1}  = \mbf{A}^t - \mu\mbf{F}_{2D}^H \left(\left(\card{\mbf{F}_{2D} \mbf{A}^t \mbf{P}} - \sqrt{\mbf{I}} \right) \circ \frac{\mbf{F}_{2D} \mbf{A}^t \mbf{P}}{\card{\mbf{F}_{2D} \mbf{A}^t \mbf{P}}}\right) \mbf{P}^H,
$$
with a learning rate $\mu = \frac{1}{\norm{\mbf{P}}^2\norm{\mbf{F}_{2D}}^2}    = \frac{1}{N^2 \norm{\mbf{P}}^2} $. The initial guess of the matrix $\mbf{A}$ is  
\begin{equation}\label{Eq:Initial}
\mbf A^0 = \mbf{O}_M^0 \prod_{m = 1}^{M-1} \mbf{G}_m \mbf{O}_m^0,
\end{equation}
with $\mbf{O}_m^0, m \in [M]$ being the intialisations for each slice.
Once matrix $\mbf A$ is estimated using $\mbf{\hat A}$, its sparse decomposition \cite{le2016flexible} is achieved by solving the following problem
\begin{equation*}
\begin{aligned}
\underset{\substack{\lambda, ~ \mbf{O}_m, m \in [M]  \\ \mbf{O}_m \text{-- diagonal} \\ \norm{\mbf{O}_m}_F = 1 }}{\text{minimize}}
& \ \frac{1}{2}\norm{\mbf{\hat A} - \lambda \mbf{O}_M\prod_{m = 1}^{M - 1}  \mathbf{G}_{m}\mbf{O}_m}_F^2. 
\end{aligned}
\end{equation*}
{In a view of fact that for any set of multipliers $\{ \alpha_m \mbf{O}_m, m \in [M] \}$ such that $\prod_{m=1}^M \alpha_m = 1$ the slices $\alpha_m \mbf{O}_m$ will generate the same $\mbf{A}_M$, this ambiguity is taken care of} by normalising $\mbf{O}_m$ during the optimisation and by introducing the data fidelity parameter $\lambda$.
Simultaneous minimisation with respect to all unknowns is cumbersome. Instead the alternating minimisation technique is employed. 

Starting with initial guesses $\mbf{O}_m^0, m \in [M]$ as used in \eqref{Eq:Initial} and $\lambda^0 =  1$ for the $\ell$-th slice, $\ell \in [M]$, the new estimate is obtained by minimising
\begin{align*}
\mbf{O}_\ell^{t+1} 
& = \underset{ \substack{ \mbf{O}_\ell \text{ -- diagonal} \\ \norm{\mbf{O}_\ell}_F = 1 } }{\text{arg min}}
\ \frac{1}{2}\norm{\mbf{\hat A} - \lambda^t \mathbf{R}_\ell^t \mbf{O}_{\ell} \mathbf{S}_\ell^t }_F^2,
\end{align*}
where the objective is reformulated in terms of the prefix and suffix \eqref{Eq:Layerwise_Prefix_Suffix} matrices. Thereby, proximal gradient descent methods \cite{parikh2014proximal} can be applied, which grants an update of the form
\begin{equation}
\mbf{O}_{\ell}^{t + 1} =  \mcl{P}\left(\mbf{O}_{\ell}^{t} -  \mu \lambda^t \left(\mathbf{R}^t_\ell \right)^H    \left( \lambda^t \mathbf{R}^t_{\ell} \mbf{O}_{\ell} \mbf{S}^{t}_{\ell} - \mbf{\hat  A}  \right) \left( \mbf{S}^{t}_{\ell} \right)^H\right),
\label{Eq:Prox_Min}
\end{equation}
with the projection operator $\mcl{P}$ acting onto the space of diagonally normalised matrices given by
$$
(\mcl{P}(\mathbf{X}))_{k,j}
=
\begin{cases} \frac{\mathbf{X}_{k,k}}{\sqrt{\sum_{n = 1}^{N^2} \card{\mathbf{X}_{n,n}}^2}}
  , & k = j,\\
0, & k \neq j,
\end{cases}
\quad
k,j \in [N^2],
$$
and the learning rate $\mu = \frac{1}{c}$ where $c \geq \left(\lambda^t \norm{\mathbf{R}^t_\ell} \norm{\mbf{S}^{t}_{\ell}}\right)^2$, as discussed in \cite{le2016flexible}.

For the minimisation with respect to $\lambda$, estimates of the $M$-th slice are combined according to $\mbf{\tilde A} = \mbf{O}_M^{t+1}\prod_{m = 1}^{M-1} \mbf{G}_m \mbf{O}_m^{t+1} $ and $\lambda$ is updated by minimising it according to the one-parameter least squares problem
\begin{equation*}
\begin{aligned}
 \lambda^{t + 1} &= \underset{ \lambda}{\text{arg min}} \frac{1}{2}\norm{\mbf{\hat  A} - \lambda \mbf{\tilde A} }_F^2 \\
\end{aligned}.
\end{equation*}
Therefore, the update for $\lambda^{t + 1}$ is given by
\begin{equation}
\lambda^{t+1} = \frac{ \text{Tr}\left(\mbf{\hat  A}^H \mbf{\tilde  A}\right)}{\text{Tr}\left(\mbf{\tilde  A}^H \mbf{\tilde  A}\right)},
\label{Eq:Update Lambda}
\end{equation}
which concludes the second step of our method. These two steps are reiterated by using the new initialisation for the first step $\mbf{A}^{t+1 } = \lambda^{t+1} \mbf{O}_M^{t+1}\prod_{m = 1}^{M-1} \mbf{G}_m \mbf{O}_m^{t+1}$. The summary of the procedure is given in Algorithm \ref{Algo:Sparse Decom}.

\begin{algorithm}
\caption{Sparse Matrix Decomposition}
\label{Algo:Sparse Decom}
\begin{algorithmic}[1]
\State \textbf{Initialisation:} 
\begin{itemize}
    \item Initial matrix $\mbf A^0 = \mbf{O}_M^0 \prod_{m = 1}^{M-1} \mbf{G}_m \mbf{O}_m^0$
    \item Intensity measurement $\mbf I \in \R^{N^2 \times S}$
    \item Number of iterations $T$ and regularization $\lambda^0$
\end{itemize}
\For{each iteration $t \in [T]$ }
\State Estimate matrix $\mbf{\hat A}$ with initial $\mbf{A}^t$ by solving \eqref{Eq:Opt_Prob_Matrix}.
\For{each layer $\ell \in [M]$} 
\State Compute prefix matrix : $\mathbf{R}^t_\ell $ 
\State Compute suffix matrix : $\mathbf{S}^t_\ell $ 
\State Estimate : $\mbf{O}^{t+1}_\ell$ by solving \eqref{Eq:Prox_Min}
\EndFor
\State Set $\mbf{\tilde A} = \mbf{O}_M^{t+1}\prod_{m = 1}^{M-1} \mbf{G}_m \mbf{O}_m^{t+1} $
\State Update $\lambda^{t+1}$ via \eqref{Eq:Update Lambda}
\State Update $\mbf{A}^{t+1} =  \lambda^{t+1}\mbf{O}_M^{t+1}\prod_{m = 1}^{M-1} \mbf{G}_m \mbf{O}_m^{t+1}$
\State If convergence criteria is reached $\to$ Stop
\EndFor
\State \Return $(\lambda^{T})^{\frac 1 M} \mbf{O}_1^{T}, \ldots,\ (\lambda^{T})^{\frac 1 M} \mbf{O}_M^{T}$
\end{algorithmic}
\end{algorithm}

\subsection{Probe Reconstruction}
After estimating the object i.e. the phase gratings of a specimen the optimisation method can be adapted related to Amplitude Flow in \eqref{eq: phase retrieval} in order to estimate the centered probe $\mbf{p}^c \in \C^{N^2}$, by utilising the intensity of diffraction patterns at the same position, i.e. $\mbf{i}^c \in \R^{N^2}$,

\begin{equation}
\begin{aligned}
& \underset{\mbf{p}^c}{\text{minimize}}
&  \mcl{A}\left(\mbf{p}^c\right) \defeq \frac{1}{2}\norm{\sqrt{\mbf{i}^c} - \card{\mbf{F}_{2D}  \mbf{\hat A} \mbf{p}^c}}^2_2.
\label{Eq:Opt_Est_Probe}
\end{aligned}
\end{equation}
Additionally, the gradient update for the $t$-th iteration is similar to \eqref{Eq:Opt_Prob_Amplitude_Flow_Step}, where there is
\begin{equation} 
\mbf{p}^{t} = \mbf{p}^{t-1} - \mu \nabla \mcl{A}\left(\mbf{p}^{t-1}\right).
\end{equation}
In this case, the learning rate $\mu$ is calculated by using the spectral norm of the estimated matrix $ \mbf{F}_{2D}  \mbf{\hat A}$, i.e. $\norm{ \mbf{F}_{2D}  \mbf{\hat A}}$.
\section{Simulation and experimental details}\label{Sec3:Datasets}
In this section, information regarding the simulated dataset of a specimen used as the ground truth, including its type and crystal structure, is given. Furthermore, the microscope and a description of the experimental conditions used for obtaining actual experimental diffraction data are provided.
\begin{table*}[t]
  \caption{Parameters for generating simulated datasets taken from \textcolor{pink}{Ga}\textcolor{violet}{As} \cite{wyckoff1963interscience}, \textcolor{teal}{Mo}\textcolor{yellow}{S$_2$} \cite{schonfeld1983anisotropic}, \textcolor{green}{Sr}\textcolor{gray}{Ti}\textcolor{red}{O$_3$} \cite{tsuda1995refinement}. In the simulated data, the hexagonal cell was transformed into an orthogonal cell.}
\label{tab:params_multislice}
  \small\centering
  \begin{tabular}{llll}

    \toprule
     \bfseries Parameters & \bfseries \bfseries GaAs & \bfseries SrTiO$_3$ & \bfseries MoS$_2$ \\
     Unitcell (a,b,c) (nm) & $(0.56533, 0.56533, 0.56533)$ & $(0.3905,0.3905,0.3905)$ & $(0.3161,0.54750,1.2295)$ \\
      Supercell (Na,Nb)  & $(2,2)$ & $(2,2)$ & $(2,2)$ \\
      Semiconv. angle (mrad) & $32$ & $32$ & $32$ \\
      Accel. voltage (keV) & $80, 200$ & $80, 200$ & $80, 200$ \\
      Scan and detector size & $(40,40,40,40)$& $(40,40,40,40)$& $(40,40,40,40)$\\
      Fresnel distance monolayer/$3$ slices (nm) & $(0.1413, 0.1413, 0.1413)$ & $(0.1952, 0.1952, 0.1952)$ & $(0.1561, 0.1561, 0.1561)$\\
    \bottomrule
  \end{tabular}
\vspace{-0.4cm}
\end{table*}
\vspace{-0.25cm}
\subsection{Simulated data sets}
\begin{figure}[!htb]
\centering 
     \scalebox{0.4}{\input{Figure/Structure_Specimen/specimen.tex}}  

\caption{Structure of materials, (a)  2D projection of \textcolor{pink}{Ga}\textcolor{violet}{As}  (b) 3D structure of \textcolor{pink}{Ga}\textcolor{violet}{As}, (c) 2D projection of \textcolor{green}{Sr}\textcolor{gray}{Ti}\textcolor{red}{O$_3$}  (d) 3D structure of \textcolor{green}{Sr}\textcolor{gray}{Ti}\textcolor{red}{O$_3$} , (e) 2D projection of \textcolor{teal}{Mo}\textcolor{yellow}{S$_2$} (f) 3D structure of \textcolor{teal}{Mo}\textcolor{yellow}{S$_2$}. Structural parameters were taken from \textcolor{pink}{Ga}\textcolor{violet}{As} \cite{wyckoff1963interscience}, \textcolor{teal}{Mo}\textcolor{yellow}{S$_2$} \cite{schonfeld1983anisotropic}, \textcolor{green}{Sr}\textcolor{gray}{Ti}\textcolor{red}{O$_3$} \cite{tsuda1995refinement}. In the simulated data, the hexagonal cell was transformed into an orthogonal cell.}
\label{Fig3. Materials}
\end{figure}
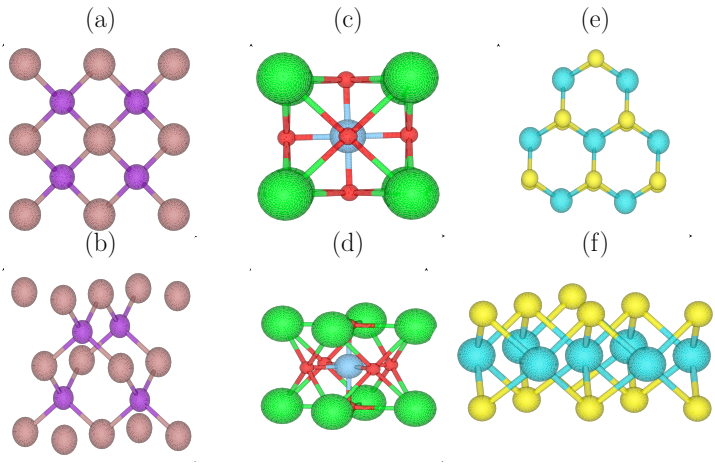

Intensities of simulated diffraction patterns from \ac{MoS$_2$}, \ac{SrTiO$_3$}, and \ac{GaAs} specimens with elevated thicknesses were generated by using a forward multislice algorithm. 
In Figure \ref{Fig3. Materials}, their $3$D structural representations as well as $2$D projections along [0 0 1] of the unit cells are shown. The structural as well as simulation parameters 
are given in Table \ref{tab:params_multislice}.
\vspace{-0.1cm}
In addition, the parameter unit cell presents the most simple repeated lattice point in the crystal. The collection of several unit cells is called a supercell. At last, the semi-convergence angle represents the semi-angle that appears in a cone shape when a convergent electron beam illuminates a specimen.
\subsection{Experimental dataset} 
\begin{figure}[!htb]
\centering
     \scalebox{1.5}{\input{Figure/PACBED/idp_probe.tex}}  

\vspace{0.1cm}
\caption{(a) \ac{PACBED} of an experimental data set of \ac{MoS$_2$} acquired using a Hitachi HF5000 microscope in \ac{STEM} mode with a Medipix3 Merlin camera, (b) Amplitude of the probe initialisation, or the so-called Airy disk, which is generated by taking the absolute value of the two-dimensional inverse Fourier transform of the circular aperture generated from the \ac{PACBED}.}
\label{fig:pacbed_mos2}
\end{figure}
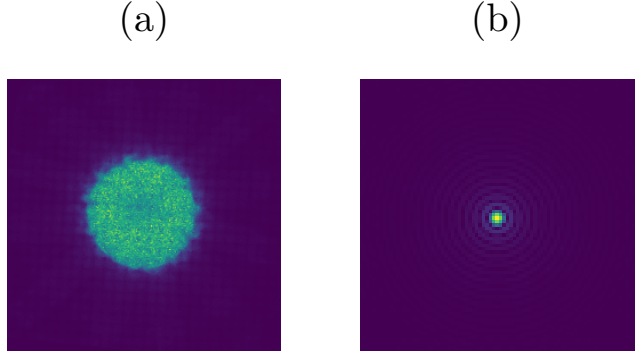

Besides simulated datasets, numerical evaluations of experimental datasets were also performed. From a bulk crystal of 2H-MoS$_2$, sheets were exfoliated by using a poly-dimethylsiloxane elastomeric film supported on a glass slide and transferred onto a holey silicon nitride membrane for the use in \ac{TEM}. 
Experimental data of MoS$_2$ was acquired using a probe corrected Hitachi HF5000 field emission microscope in \ac{STEM} mode and with an acceleration voltage of $200$~keV as well as a beam current of about $7.4$~pA. Intensities of diffraction patterns were recorded by using a Medipix3 Merlin4EM camera with $256 \times 256$ pixels. The distance between neighbouring scan points was 26.5\,pm in $x$, i.e. horizontal, and $y$, i.e. vertical, scanning directions. In addition, the acquisition time per diffraction pattern was $0.5$~ms and data was acquired using a dynamic range of $6$~bit. The \ac{PACBED} is depicted in Figure \ref{fig:pacbed_mos2}, where the intensity of all diffraction patterns from $128 \times 128$ scanning points is averaged.
\vspace{-0.45cm}
\vspace{-0.5cm}

\section{Numerical Results}\label{Sec5:Numeric}
\vspace{-0.1cm}
Several numerical evaluations that measure performance of the proposed algorithms are presented in this section. The object $\mbf{O}_m^0$ and the probe are initialised by using an identity matrix and an Airy disk previously outlined in Section \ref{Sec2:ProbState}. Initially the error metric used for measuring the quality of a reconstruction will be defined.
\vspace{-0.2cm}
\subsection{Error metrics for Evaluation of the Algorithm}
The error metric used for evaluating the reconstruction of each slice is calculated as the mean square error
\begin{equation}
 \frac{1}{M}\sum_{m = 1}^M\frac{\norm{\mbf{O}_m - \mbf{\hat O}_m}_F}{\norm{\mbf{O}_m }_F},
\label{eq:error_metric_object}
\end{equation}
with $\mbf{O}_m$ and $\mbf{\hat O}_m$ being the ground truth and the estimated object at the $m$-th slice, respectively.

The objective of the presented optimisation problem is to minimise the error between measured and estimated intensities of diffraction patterns. Accordingly, it is necessary to introduce an additional error metric
\begin{equation}
\frac{\norm{\sqrt{\mbf{I}} - \card{\mathbf{F}_{2D} \mbf{\hat A}_{M} \mbf{P}}}_F}{\norm{\sqrt{\mbf{I}}}_F},
\label{eq:error_metric_intensity}
\end{equation}
with $\mbf{\hat A}_M$ being the total estimated object transfer function at slice $M$. In \eqref{eq:error_metric_object} the error metric is referred to as the relative reconstruction error whereas, the error metric in \eqref{eq:error_metric_intensity} is referred to as the relative measurement error.

Two settings, the reconstruction of an arbitrary thickness and the reconstruction for decomposing into the different atomic planes, are evaluated. Both differ by the conducted Fresnel propagation distance. Using the latter a reconstruction of the phase grating for each slice with the actual Fresnel propagation distance was attempted, i.e. the thickness resulting from the crystal structure of the specimen.
\subsection{Reconstruction of Arbitrary Slice Thickness}
The intensities of diffraction patterns of GaAs, MoS$_2$ and SrTiO$_3$ specimens were generated by using the forward multislice method for a thickness of $20$~nm. The simulation parameters are given in Table \ref{tab:params_multislice}. 
\begin{figure}[!htb] 
\centering
     \scalebox{0.4}{\input{Figure/Result/Adapt_Thick/Adaptive_Thick_4nm.tex}}  

\vspace{0.1cm}
\caption{Slice phase reconstruction, {in radian}, of $4$~nm Fresnel propagation distance, specimen thickness $20$~nm, using an acceleration voltage of $200$ keV, observed at $100$ iterations, for the following specimens: (a) GaAs, (b) MoS$_2$ and (c) SrTiO$_3$. GT is the ground truth, L is the layer-wise optimisation, and S is the sparse matrix decomposition. }
\label{fig:adapt_thick_4nm}
\vspace{-0.3cm}
\end{figure}
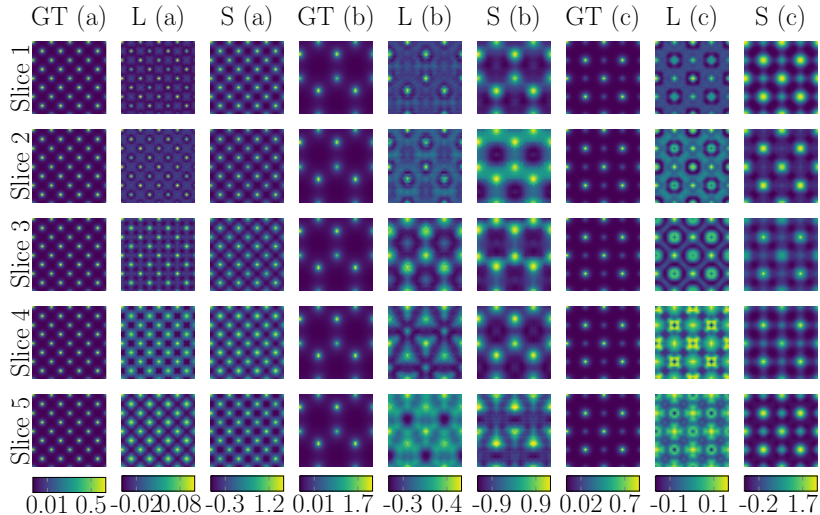
\begin{figure}[!ht] 
\centering
     \scalebox{0.35}{\input{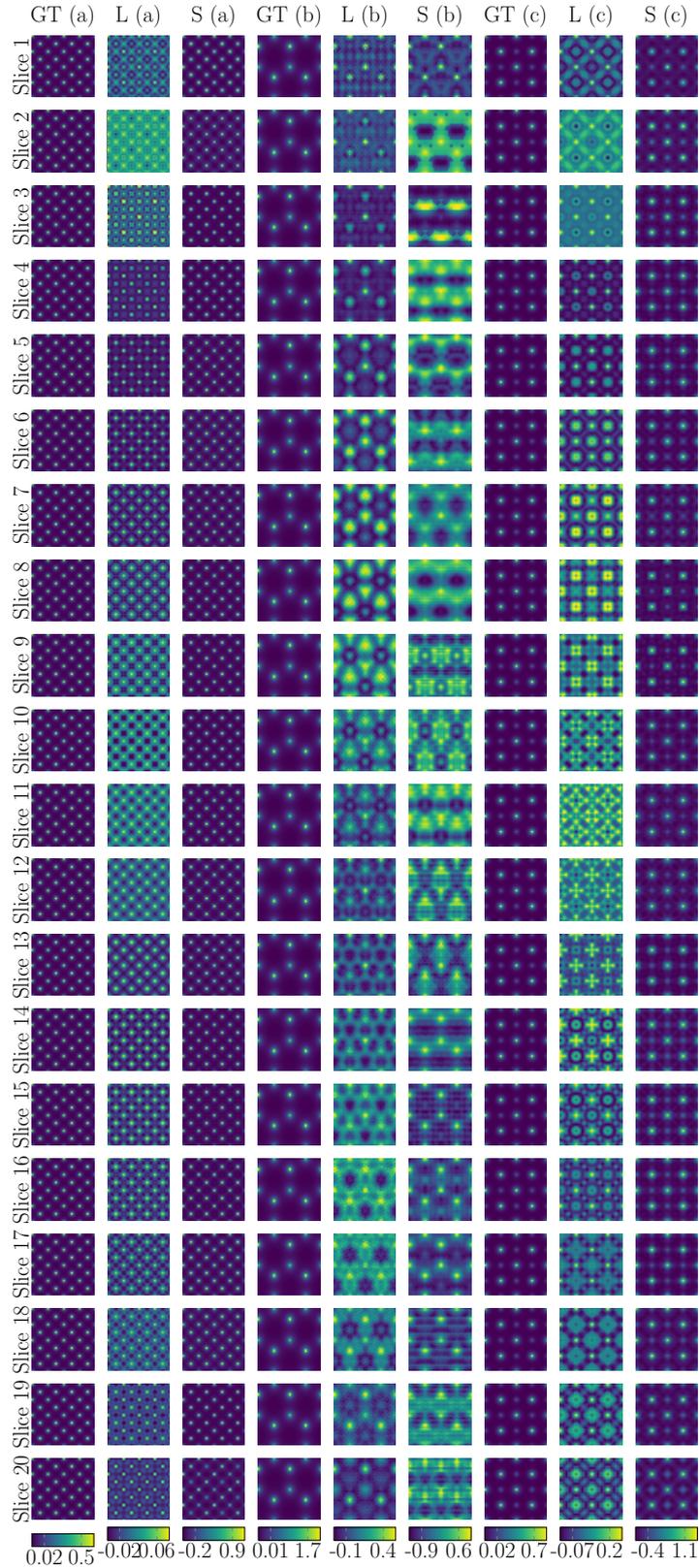}}  
 
\caption{Slice phase reconstruction, {in radian}, of $1$~nm Fresnel propagation distance, specimen thickness $20$~nm, using an acceleration voltage of $200$~keV, observed at $100$ iterations, for the following specimens: (a) GaAs, (b) MoS$_2$, and (c) SrTiO$_3$. GT is the ground truth, L is the layer-wise optimisation, and S is the sparse matrix decomposition.}
\label{fig:adapt_thick_1nm}
\vspace{-0.5cm}
\end{figure}
In this case, the reconstructions of $1$~nm and $4$~nm Fresnel propagation distance in the inversion process are evaluated. It should be noted that, depending on the thickness, the reconstructions will accumulate all the atom positions of each phase grating into one slice. Thereby, all atom positions are projected onto one image. 
This setting is necessary to evaluate experimental data as the Fresnel propagation distance of the specimen cannot be exactly determined at the atomic scale. Thus, one can heuristically approximate the correct reconstruction by using an initial, comparably large slice thickness and evaluating the atomic positions.

In Figure \ref{fig:adapt_thick_4nm} the phase reconstructions of five slices with a Fresnel propagation distance of $4$~nm are shown for the layer-wise optimisation as well as the sparse matrix decomposition. The sparse matrix decomposition yielded a higher range of the phase reconstruction and in comparison outperformed the layer-wise optimisation in terms of the accumulated atom positions of the MoS$_2$ and SrTiO$_3$ specimens. Due to the large slice thickness the reconstructed phases differed significantly from the ground truth GT, in particular when using the sparse matrix method, although the average structure of each slice was correctly reconstructed. It was observed that a 20\,nm thick specimen leads to very strong multiple scattering effects.

Numerical evaluations with $1$\,nm slice thickness, i.e. Fresnel propagation distance, have been performed as depicted in Figure \ref{fig:adapt_thick_1nm}. Similar to the $4$~nm case, except for MoS$_2$ where the phase reconstruction appears unstable after the second slice, the sparse matrix decomposition generally performed better than the layer-wise optimisation. This confirmed that adding another constraint to impose the solution as a diagonal matrix can significantly improve the reconstruction. The reconstruction performed best for the GaAs and SrTiO$_3$ cases, this could be explained due to their slice thicknesses being closer to an integer multiple of the lattice parameter in comparison to MoS$_2$. This caused a slight beating effect in dependence of the thickness. In these evaluations knowledge of the correct lattice parameter from Table~\ref{tab:params_multislice} was intentionally not used as a prior to verify the outcome for the realistic case where the structure of the investigated material is also unknown.

\subsection{Atomic Plane Decomposition}
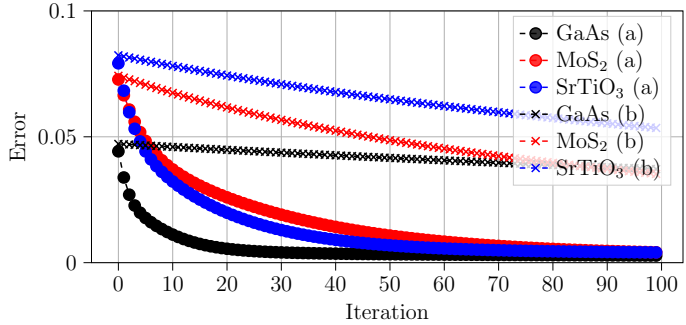
\begin{figure}[h!] 
\centering
     \scalebox{0.7}{\input{Figure/Result/Figure_6/error_idp.tex}}  

\caption{Relative measurement error as in \eqref{eq:error_metric_intensity} with (a) sparse matrix decomposition, (b) layer-wise optimisation for three slices}
 \label{fig:error_idp}
\end{figure}

The reconstruction of each atomic plane, i.e. direct recontruction of each slice in the inverse process at the Fresnel propagation distance similar to the forward multislice model described in Table \ref{tab:params_multislice}, was also examined. This approach proved quite challenging due to ambiguities concerning the atom positions along the direction of the electron beam which occurred in the reconstruction. The relative measurement error is depicted in Figure \ref{fig:error_idp}.

\begin{figure}[!htb] 
\centering
     \scalebox{0.4}{\input{Figure/Result/Figure_7/Fig_7_Ambiguity_Atomic_Decomposition_new.tex}}  

\caption{Slice phase reconstruction, {in radian}, of atomic slice decomposition, using layer-wise optimisation and sparse matrix decomposition as well as an acceleration voltage of $200$ keV for the following specimens: (a) GaAs, (b) MoS$_2$ and (c) SrTiO$_3$. {GT is the ground truth, L is the layer-wise optimisation, and S is the sparse matrix decomposition.}}
 \label{fig:recon_atomic_decom}
 \vspace{-0.2cm}
\end{figure}
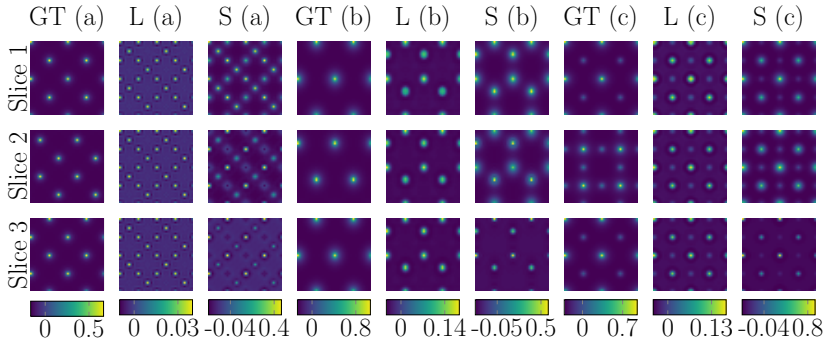

In terms of the error between the ground truth and the reconstructed intensity of the diffraction patterns it was observed that the sparse matrix decomposition converged faster than seen in the layer-wise optimisation. The reconstructions of each slice for both algorithms are presented in Figure~\ref{fig:recon_atomic_decom}. While attempting to uniquely decompose each slice in the atomic plane, at the original Fresnel propagation distance given in Table \ref{tab:params_multislice}, ambiguities could be observed. This can be evidenced by comparing the layer-wise optimisation and the GT columns in Figure~\ref{fig:recon_atomic_decom}. Atom positions appear to have been combined and decomposition of the layers was not successful. This phenomenon stems from the fact that the structure of the Fresnel matrix $\mbf G$ becomes similar to the identity matrix for very small Fresnel propagation distances, i.e. close to zero.
The total matrix at the $M$-th slice $\mbf{A}_M$, generated in the forward multislice model, therefore is simply the product of each atomic plane. Consequently it was difficult to uniquely decompose each slice. 
Since the intensity of the diffraction patterns and the depth resolution depend highly on the convergence angle of the electron beam, they also affect the reconstruction. Conceptual approaches for improving the phase reconstruction in order to resolve each atomic plane are provided below.
\begin{figure}[!ht]
\centering
     \scalebox{0.7}{\input{Figure/Result/Figure_8/recon_error_w_woprobe_200.tex}}  

\caption{Relative reconstruction error as in \eqref{eq:error_metric_object} of a probe initialised with an Airy disk, using an acceleration voltage of $200$ keV, by using (a) sparse matrix decomposition and (b) layer-wise optimisation.}
\label{fig:error_blind_recon}
\end{figure}
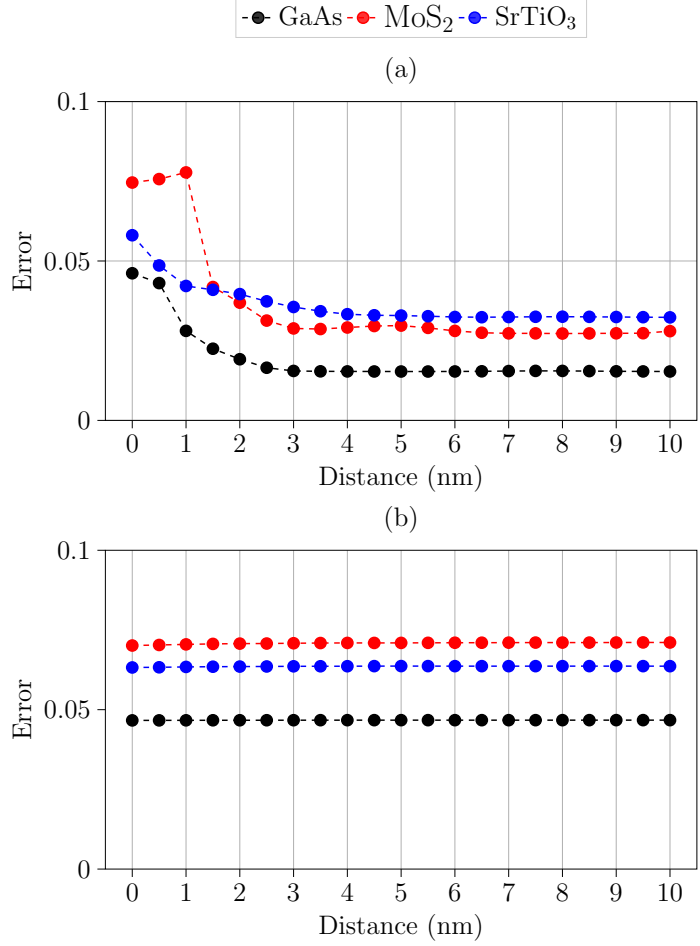
\subsubsection{Increasing the Fresnel propagation distance}
A huge advantage of conducting a conceptual study is the ability to set the Fresnel propagation distance in the forward multislice model to any desired value within the simulation in order to improve the convergence of the sparse matrix decomposition, as presented in Figure \ref{fig:error_blind_recon}. 
\begin{figure}[!htb] 
\centering
     \scalebox{0.35}{\input{Figure/Result/Figure_Both_Algo_Dist2nm/Both_Algo_Dist_2nm.tex}}  

\caption{Slice phase reconstruction, {in radian}, of Fresnel propagation distance $2$~nm, using an acceleration voltage of $200$ keV, by layer-wise optimisation and sparse matrix decomposition, observed at $100$ iterations, for the following specimens: (a) GaAs, (b) MoS$_2$ and (c) SrTiO$_3$. GT is the ground truth, L is the layer-wise optimisation, and S is the sparse matrix decomposition. }
\vspace{-0.3cm}
 \label{fig:slice_recon_none_2}
\end{figure}
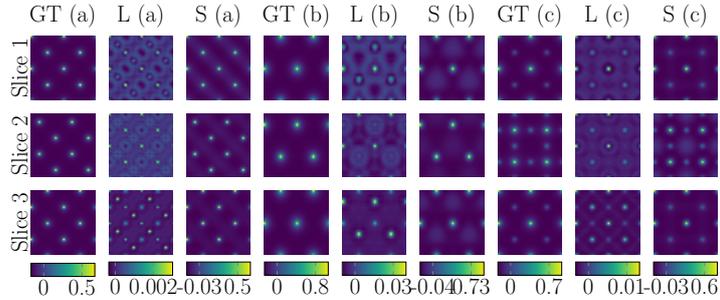
The relative reconstruction error, as in \eqref{eq:error_metric_object}, was evaluated for each slice. Compared to an arbitrary thickness that also determines the Fresnel propagation distance in the inversion process, the Fresnel propagation distance for both the forward and the inverse processes were deliberately increased. 
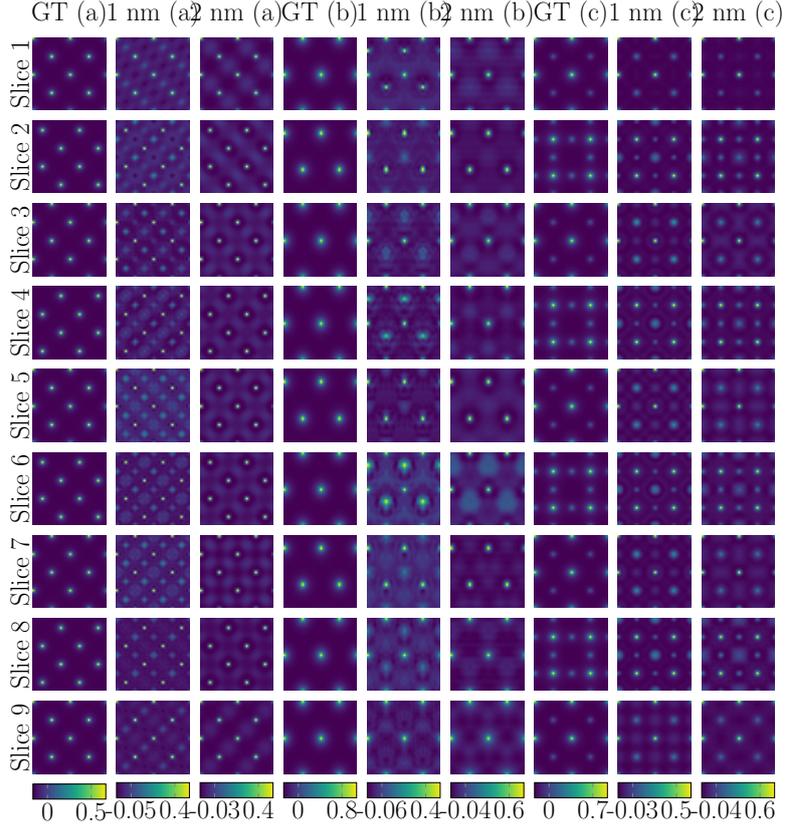
\begin{figure}[!htb] 
\centering
     \scalebox{0.4}{\input{ Figure/Result/Figure_10/Fig_10_Increase_Distance_Sparse_Dec_1nm_2nm_new.tex}}  

\vspace{0.2cm}
\caption{Slice phase reconstruction, {in radian},  of several Fresnel propagation distances, using an acceleration voltage of $200$ keV, by sparse matrix decomposition, observed at $100$ iterations, for the following specimens: (a) GaAs, (b) MoS$_2$ and (c) SrTiO$_3$. GT is the ground truth and S is the sparse matrix decomposition.}
 \label{fig:slice_recon_1_2nm_sparse_slice9}
\vspace{-0.4cm}
\end{figure}

Reconstructions with both layer-wise optimisation and sparse matrix decomposition for the first three atomic slices of GaAs, MoS$_2$ and SrTiO$_3$ specimens are presented in Figure \ref{fig:slice_recon_none_2}. The reconstructions resulting from the layer-wise optimisation still suffer from ambiguities even after the Fresnel propagation distance was increased. In contrast to the layer-wise optimisation, sparse matrix decomposition could uniquely reconstruct each slice after increasing the Fresnel propagation distance to $2$~nm.

The investigation of the capability of slice-wise reconstructions for crystals with larger thicknesses as a function of the Fresnel propagation distance was also conducted. This distance was artificially increased to 1\,nm and 2\,nm, respectively. From the results in Figure \ref{fig:slice_recon_1_2nm_sparse_slice9} it is apparent that increasing the specimen thickness also affects the performance of the sparse matrix decomposition, since some ambiguities appear despite a Fresnel propagation distance of $2$~nm. Apart from ambiguities that occurred in the phase retrieval problem, the performance of both algorithms generally depends on a trade-off between the Fresnel propagation distance and the number of slices to be reconstructed.

Increasing the Fresnel propagation distance would correspond to artificially increasing the lattice parameter in electron beam direction. This is only possible in a simulation study. However, the conceptual insight is that the phase fronts of electron waves with 200\,keV energy would only change significantly after propagating 2\,nm for the algorithms to separate the Fresnel propagation from the interaction with the Coulomb potential of the slices.

\subsubsection{Low electron energy}
The Fresnel propagator on eq.~(\ref{eq:fresnel_element}) contains the product of the wavelength and the propagation distance as the governing parameters. Therefore, a realistic  reconstruction with atomic layer sensitivity needs to use larger wavelengths if the Fresnel distance is reduced to atomic spacings in the range of 0.1\,nm. Furthermore, the same potential in a given specimen leads to a larger phase change of low-energy electrons as compared to high energies due to the effect of the so-called interaction constant on the phase grating.
The relation between electron acceleration voltage $U$ and wavelength $\lambda$ is given by
$\lambda = \frac{h c}{\sqrt{e^2 U^2+2e U m c^2}}
\label{eq:lambda},$
with $c$ being the speed of light, $h$ the Planck constant, $m$ the electron mass, and $e$ the elementary charge. For 200\,keV electrons, the wavelength is approximately 2.5\,pm, whereas it increases to 4.18\,pm for electrons with 80\,keV energy. The interaction constant increases by about 40\%. Note that both acceleration voltages, 200 and 80\,kV, are common settings in STEM such that atomic resolution can be obtained readily in aberration corrected machines. 
\begin{figure}[!htb] 
\centering
     \scalebox{0.5}{\input{Figure/Result/Figure_Low_Acc_Iter_5000_Slice3/Low_Acc_Iter_5000_Slice3.tex}}  

\vspace{0.2cm}
\caption{Slice phase reconstruction, {in radian}, using an acceleration voltage of $80$ keV, by sparse matrix decomposition, observed at $5000$ iterations, for the following specimens: (a) GaAs, (b) MoS$_2$ and (c) SrTiO$_3$. GT is the ground truth and S is the sparse matrix decomposition.}
 \label{fig:slice_recon_low_acc_slice3}
\end{figure}
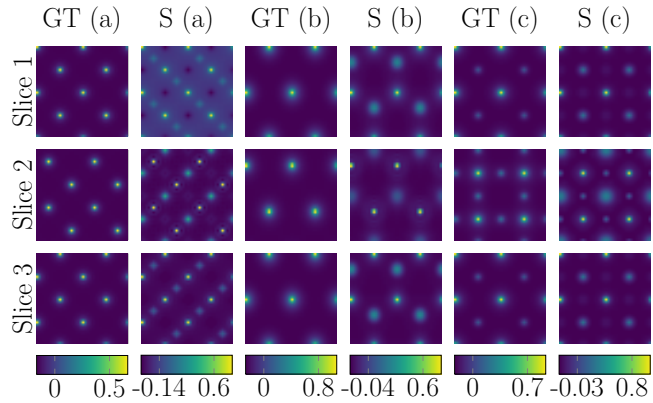
Figure~\ref{fig:slice_recon_low_acc_slice3} shows a reconstruction of the atomic planes using an acceleration voltage of $80$ keV. Despite the low-signal artefacts related to the location of the atoms in the different slices, the exact location of the atoms in each slice can now be correctly determined. As discussed earlier, it should be noted that a unique decomposition highly depends on the number of slices to be reconstructed, as shown in Figure~\ref{fig:slice_recon_low_acc_slice6}. It can be seen that having more slices affects the reconstruction since the ambiguities related to the atom positions were still present even after observing the reconstruction at $10000$ iterations.
\begin{figure}[!htb] 
\centering
     \scalebox{0.5}{\input{Figure/Result/Figure_Low_Acc_Iter_10000_Slice6/Low_Acc_Iter_10000_Slice6.tex}}  

\vspace{0.2cm}
\caption{Slice phase reconstructions, {in radian}, using an acceleration voltage of $80$ keV, by sparse matrix decomposition, observed at $10000$ iterations, for the following specimens: (a) GaAs, (b) MoS$_2$ and (c) SrTiO$_3$. GT is the ground truth and S is the sparse matrix decomposition.}
\label{fig:slice_recon_low_acc_slice6}
\end{figure}
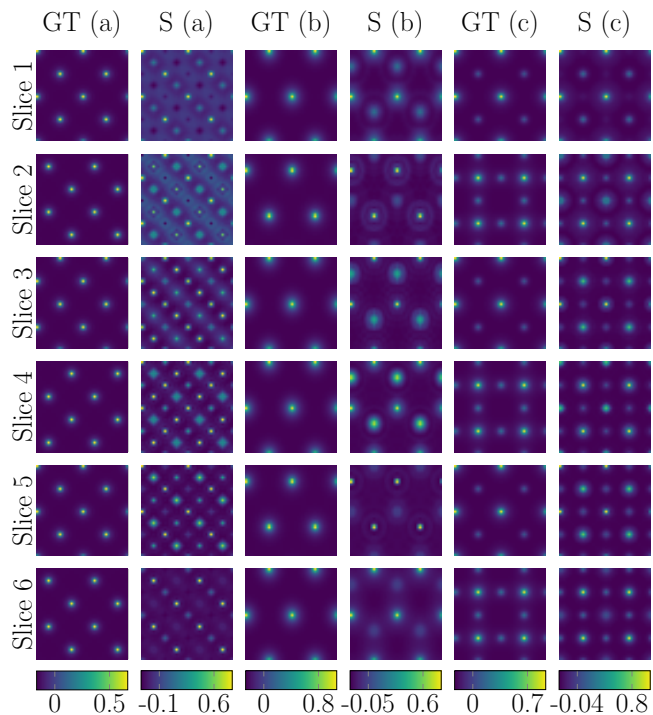 
\subsection{Probe Reconstruction}
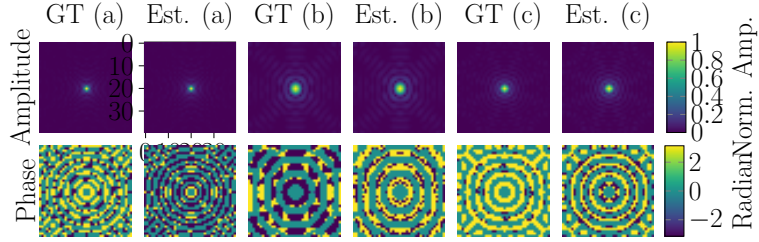
\begin{figure}[!htb] 
\centering
     \scalebox{0.5}{\input{Figure/Result/Figure_13/Fig_13_Probe_Est_OTF_Low_Acc_Sparse_3_Slice.tex}}  

\caption{Probe reconstructions from the object transfer function estimated by sparse matrix decomposition in Figure \ref{fig:slice_recon_low_acc_slice3}, for the following specimens: (a) GaAs, (b) MoS$_2$ and (c) SrTiO$_3$. GT is the ground truth, Est. is the reconstruction.}
\vspace*{-0.2cm}
 \label{fig:probe_reconstruction_slice3}  
\end{figure}

The reconstruction of the probe after estimating the matrix at the $M$-th slice, i.e. $\mbf{A}_M$ by using the optimisation problem in \eqref{Eq:Opt_Est_Probe}, is presented below. As discussed in the forward multislice model, the estimated matrix $\mbf{A}_M$ can be generated by calculating the product of each slice of the object $\mbf{O}_m$ and the Fresnel matrix $\mathbf{G}_{m}$,
$
 \mathbf{A}_M =  \mbf{O}_M\prod_{m= 1}^{M-1}\mathbf{G}_{m}\mbf{O}_m.
$
In Figure \ref{fig:probe_reconstruction_slice3}, the reconstructed probes observed at $100$ iterations are shown after estimating the slices in Figure~\ref{fig:slice_recon_low_acc_slice3}. The estimated probe matches the ground truth up to a global phase factor, which is in general an undefined quantity.
\subsection{Experimental Data}
Numerical evaluations on diffraction intensity measurements acquired experimentally from a MoS$_2$ specimen are presented below. The thickness of the experimental specimen was determined by comparison with simulated position-averaged diffraction patterns to be approximately $35$~nm. Further experimental details are described in Section \ref{Sec3:Datasets}.
Applying the same algorithms to experimental instead of simulated data is a crucial aspect to demonstrate the practical usefulness of the methods. However, reconstructing based on real data is also a critical point, because experiments are affected by additional parameters that are difficult or even impossible to include in the algorithmic setups above. For example, the recording is inherently containing Poissonian counting noise, the camera has a modulation transfer function which leads to a blurring of diffraction space features, and the projection system of the microscope can cause geometrical distortions of the diffraction patterns. Furthermore, the scan positions of the STEM probe usually deviate slightly from the ideal regular raster due to instabilities of the scan engine.

Due to this a detailed analysis of the performance of layer-wise optimisation and sparse matrix decomposition algorithms in dependence of the experimental conditions are set aside for a future task. Instead this preliminary evaluation focuses on demonstrating the principal applicability by targeting the qualitative reconstruction of the MoS$_2$ structure using a relatively small number of five slices, similar to the example in Figure~\ref{fig:adapt_thick_4nm}. In particular, this was necessary due to computational efficiency and the much higher dimensionality of the experimental data as compared to the simulations, i.e. the large number of probe positions and camera pixels.
\subsubsection{Slice reconstruction}
\begin{figure}[!htb] 
\centering
     \scalebox{0.8}{\input{Figure/Result/Experimental/Slice_5_MoS2/Slice_5_MoS2.tex}}  

 \caption{Slice phase reconstruction of an experimental data set of MoS$_2$, in radian, by (a) sparse matrix decomposition and (b) layer-wise optimisation with Fresnel propagation distance of around $7.377$~nm. The reconstruction is observed after $50$ iterations.}
\label{fig:sparse_dec_layer_exp_slice}
\vspace*{-0.4cm}
\end{figure}
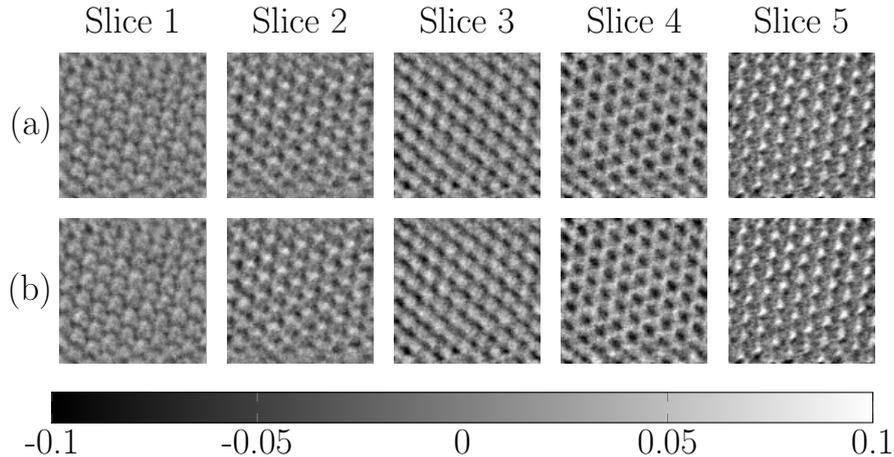

Consequently, the phases of the individual slice reconstructions in Figure~\ref{fig:sparse_dec_layer_exp_slice} are not expected to quantitatively represent the actual phase gratings on the one hand. On the other hand, they are supposed to resemble the atomic structure of the specimen in the respective slices, taking a large portion of the dynamical scattering into account. Indeed, the atomic structure is consistently visible in all slices, opposite to single-slice models for which evaluations at thicknesses of tens of nanometers are by far out of range. The dynamic range of the phase is comparably low, most probably because more slices would be needed to disentangle the slice potentials and Fresnel propagation between the slices completely.

Figure \ref{fig:sparse_dec_layer_exp_slice} shows the reconstructions for both sparse matrix decomposition and layer-wise optimisation, in which both algorithms are able to reconstruct the atom positions of MoS$_2$. In addition the object transfer function matrix $\mbf A$ was generated by considering the products of the Fresnel propagation matrices and all reconstructed slices.  
\begin{figure}[!htb] 
\centering
     \scalebox{1}{\input{Figure/Result/Experimental/Projection_3D_2D/Proj3Dto2D.tex}}  

\caption{Projection of the phase reconstruction, in radian, of \textcolor{teal}{Mo}\textcolor{yellow}{S$_2$}, in radian, by (a) sparse matrix decomposition and (b) layer-wise optimisation, from Figure \ref{fig:sparse_dec_layer_exp_slice}.}
\label{fig:projection_3D_2D}
\end{figure}
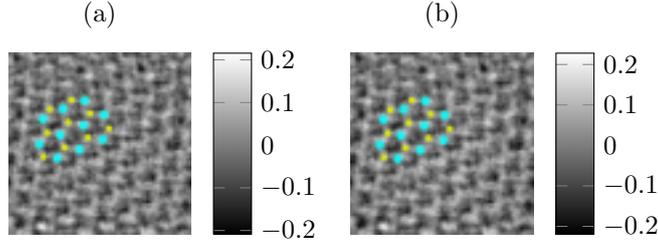
Furthermore, the two-dimensional projection of all atoms can also be easily produced, as presented in Figure \ref{fig:projection_3D_2D}, where all atom positions of MoS$_2$ are present in the projection. A coloured overlay of the Molybdenum (Mo) and Sulfur (S) atoms were added to the figure in order to better visualise the reconstructed atomic arrangement.

\subsubsection{Probe Reconstruction}
A direct implementation of the Amplitude Flow as in \eqref{eq: phase retrieval} was adopted in order to reconstruct the illuminating probe after estimating the object transfer function matrix $\mbf A$. At this setting the focus was solely on the intensity of the diffraction patterns acquired at the center position of the illuminated area on the specimen.
\vspace*{-0.2cm}
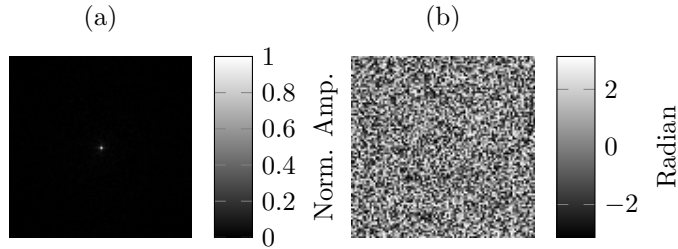
\begin{figure}[!htb] 
\centering
     \scalebox{1}{\input{Figure/Result/Experimental/Probe_Sparse/Probe_Recon.tex}}  

 \caption{Probe reconstruction after estimating using an experimental data set of \ac{MoS$_2$}: (a) normalised amplitude and (b) phase observed at $50$ iterations}
 \label{fig:probe_est_experiment}
 \vspace*{-0.2cm}
\end{figure}
The resulting reconstructions of amplitude and phase of the probe are presented in Figure \ref{fig:probe_est_experiment}. This shows that the data was taken with a well-focused probe as indicated by the sharp peak in the amplitude and a flat phase except for the noise. Note that the reconstruction of the probe is in general a robust check whether the algorithm and the parameters used for the reconstruction are suitable to separate illumination and specimen. In the present case, one can, therefore, conclude that the large slice thickness did not affect this, because no specimen details are visible in the reconstructed probe.
\vspace*{-0.25cm}
\section{Discussion}
Two algorithms based on optimisation methods for inverse multislice ptychography have been presented, namely sparse matrix decomposition and layer-wise optimisation, which are derived from the reformulation of the forward multislice model. The connections between reformulation of the multislice and other models to represent thick specimens are discussed and the possible direction of future research is outlined.

The numerical observations showed that the type of specimens and the number of slices impinges on the reconstruction performance of the algorithms. Theoretically it would be interesting to examine the fundamental limit of the algorithms with respect to the number of slices needed to disentangle interaction and Fresnel propagation sufficiently well. Moreover, a systematic study addressing the impact of the probe semi-convergence angle, the electron energy, aberrations of the electron-optical system and coherence effects could shed light on the robustness of the presented methodology in respect to the multitude of experimental parameters in real measurements. {If one wants to work with low dose data, where Poisson noise is dominant, one has to modifiy the objective function.
Poisson maximum likelihood has been used with success in case of single slice ptychography \cite{bian2016fourier} and can be adapted to our layer-wise estimation.
For the sparse matrix decomposition, one can even use a Poisson phase retrieval method \cite{li2021algorithms} in the first step, estimating $\mbf A$, without changing the decomposition method at all.}

Since the reformulation of the forward multislice model in this article yielded purely a matrix representing the transfer function of a thick specimen, it would be possible to relate such a matrix to a scattering matrix constructed from the Bloch wave method and observe the differences between both approaches. The latter requires an intensive computational effort of eigenvalue decomposition for huge scattering matrices. 

Apart from the comparison between the proposed algorithms to the eigenvalue decomposition with the Bloch wave method, it should be possible to estimate the specimen thickness directly from the algorithms. One possibility could be to incorporate information from the high-angle intensity of diffraction patterns, or to start from a coarse slicing first with large slice thicknesses, and then increase the number of slices subsequently. In case a sufficiently high total thickness is assumed, empty slices should emerge, indicating that the specimen is actually compact along the electron beam direction. In general, a suitable regularisation should be developed and applied in future works. {An interesting approach is a suitable sparsity model, as applied in \cite{jagatap2019sample} for the case of single slice ptychography.}

Finally, the reformulation of the multislice scheme as a simple, though large, one-step matrix multiplication circumvents the successive forward and backward Fourier transform, which characterises the conventional multislice implementations that compute both the Fresnel propagation and its interaction with slice potentials in real space. In that respect, studying the capabilities and performance of the reformulation is not only relevant for solving inverse problems, but also interesting with respect to conventional forward simulations.
\vspace*{-0.2cm}

\section{Conclusion and Summary}\label{Sec6:Conclusion}
We proposed reformulation of the forward multislice method such that the transfer function of a thick specimen can be directly determined. In combination with the ptychographic approach we presented two optimisation models for solving the inverse multislice ptychography problem for both arbitrary thickness and atomic plane decomposition. In the first case, given the intensity of diffraction patterns, several atomic planes were jointly processed into a single reconstruction in order to show the total potential. In the atomic plane decomposition each slice was reconstructed at its atomic plane and given only the intensity of the diffraction patterns the results showed the unique atomic positions in each slice.

Although the resulting phase reconstructions by layer-wise optimisation still contained ambiguities, both algorithms could recover the locations of the atoms in the inversion process.
 Furthermore, this showed that for simulated data the sparse matrix decomposition could reconstruct the atom locations unambiguously for each atomic layer given the intensity of diffraction patterns with low acceleration voltage $80$~keV. However, it could be observed that the reconstruction using both algorithms were dependent highly on the number of estimated slices.  After the thickness of the specimen was increased in terms of the number of slices it resulted in ambiguities of the locations of the atoms. This indicated that the algorithm failed to accurately reconstruct each atomic layer in respect of the different slices.  

We also supported our numerical observations with the reconstruction of the object given the intensity of diffraction patterns acquired from the experimental data set of MoS$_2$. It was shown that both algorithms can reconstruct the phase of the specimen. Additionally, by using the reformulation of the forward multislice method, a matrix was constructed that represented the thick object transfer function, i.e. scattering matrix. This reformulation can be used to directly generate the two-dimensional projection of the atom arrangement.
\vspace{-0.25cm}

\section*{Acknowledgements}
K.\,M.-C. and B.M. acknowledge support from the Helmholtz Association under contract No.~VH-NG\,1317 (\textit{more}STEM), and from the Deutsche Forschungsgemeinschaft under DFG grant EXC 2089/1-390776260. K.\,M.-C., D.W., A.B., A.C., B.M., O.M., B.D. and F.F. acknowledge support from Helmholtz under contract No.~ZT-I-0025 (Ptychography 4.0). Helmholtz support under grant No.~ZT-I-PF-5-28 (EDARTI) for B.D., F.F, and K.\,M.-C. is gratefully acknowledged.
\vspace{-0.2cm}
\bibliographystyle{IEEEbib}
\bibliography{bibliography}

\end{document}

%% file: Figure/Structure_Specimen/specimen.tex
\begin{tikzpicture}
\begin{groupplot}[group style={group size=3 by 2,horizontal sep=5em, vertical sep=3em},width = 0.5*\textwidth,
         height=(0.5)*\textwidth]
\nextgroupplot[
tick align=outside,
tick pos=left,
title={\myfont(a)},
axis lines=left, 
xtick=\empty, 
ytick=\empty,
y axis line style={draw opacity=0}, 
x axis line style={draw opacity=0},
]
\addplot graphics [includegraphics cmd=\pgfimage,xmin=-0.5, xmax=8.5, ymin=20.5, ymax=-0.5] {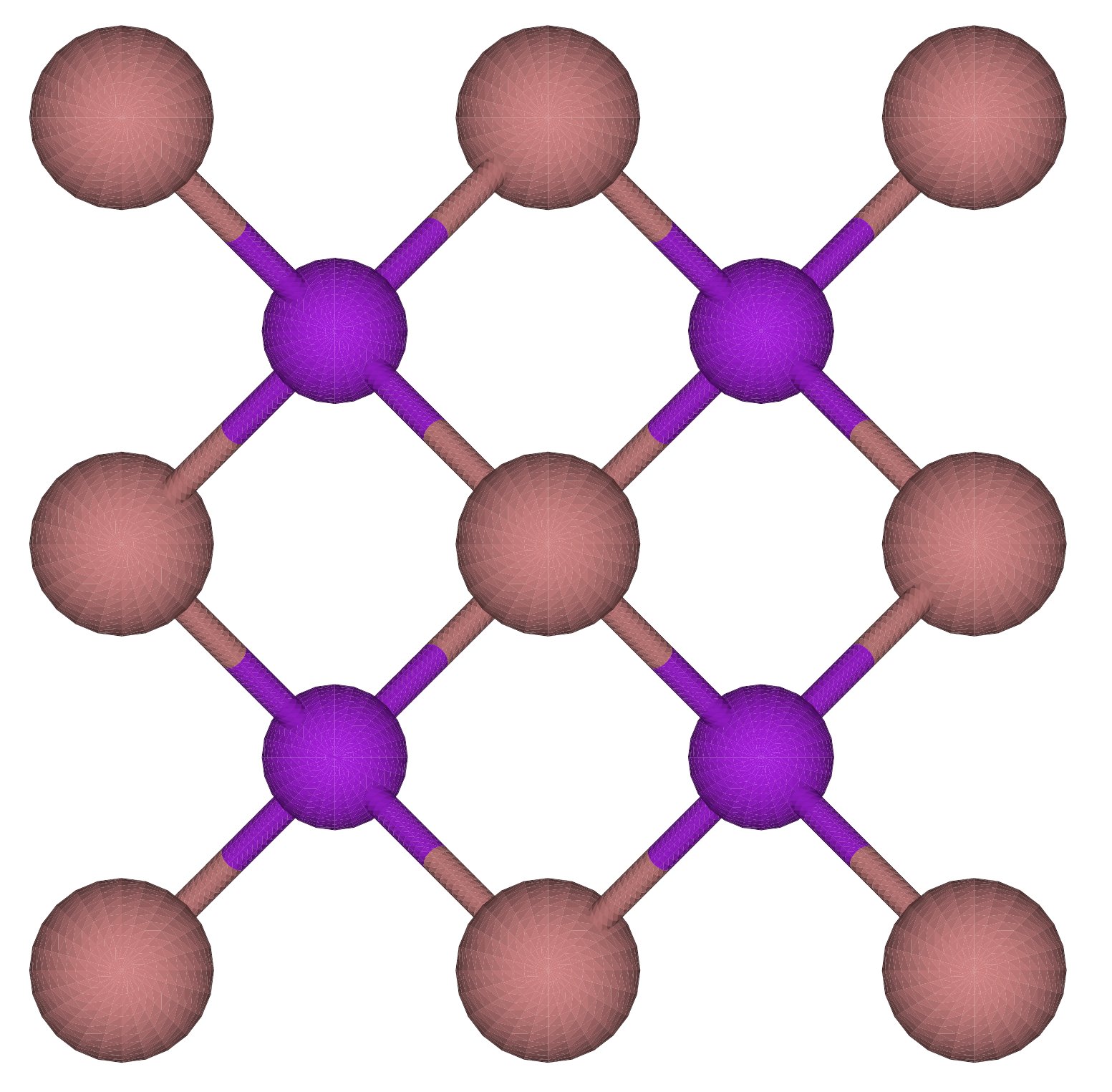};

\nextgroupplot[
tick align=outside,
tick pos=left,
title={\myfont (c)},
axis lines=left, 
xtick=\empty, 
ytick=\empty,
y axis line style={draw opacity=0}, 
x axis line style={draw opacity=0},
]
\addplot graphics [includegraphics cmd=\pgfimage,xmin=-0.5, xmax=8.5, ymin=20.5, ymax=-0.5] {Figure/Structure_Specimen/SrTiO3_inverse.pdf};

\nextgroupplot[
tick align=outside,
tick pos=left,
title={\myfont (e)},
axis lines=left, 
xtick=\empty, 
ytick=\empty,
y axis line style={draw opacity=0}, 
x axis line style={draw opacity=0},
]
\addplot graphics [includegraphics cmd=\pgfimage,xmin=2.5, xmax=5.5, ymin=3.5, ymax=-0.5] {Figure/Structure_Specimen/MoS2_inverse.pdf};

\nextgroupplot[
tick align=outside,
tick pos=left,
title={\myfont (b)},
axis lines=left, 
xtick=\empty, 
ytick=\empty,
y axis line style={draw opacity=0}, 
x axis line style={draw opacity=0},
]
\addplot graphics [includegraphics cmd=\pgfimage,xmin=-0.5, xmax=8.5, ymin=20.5, ymax=-0.5] {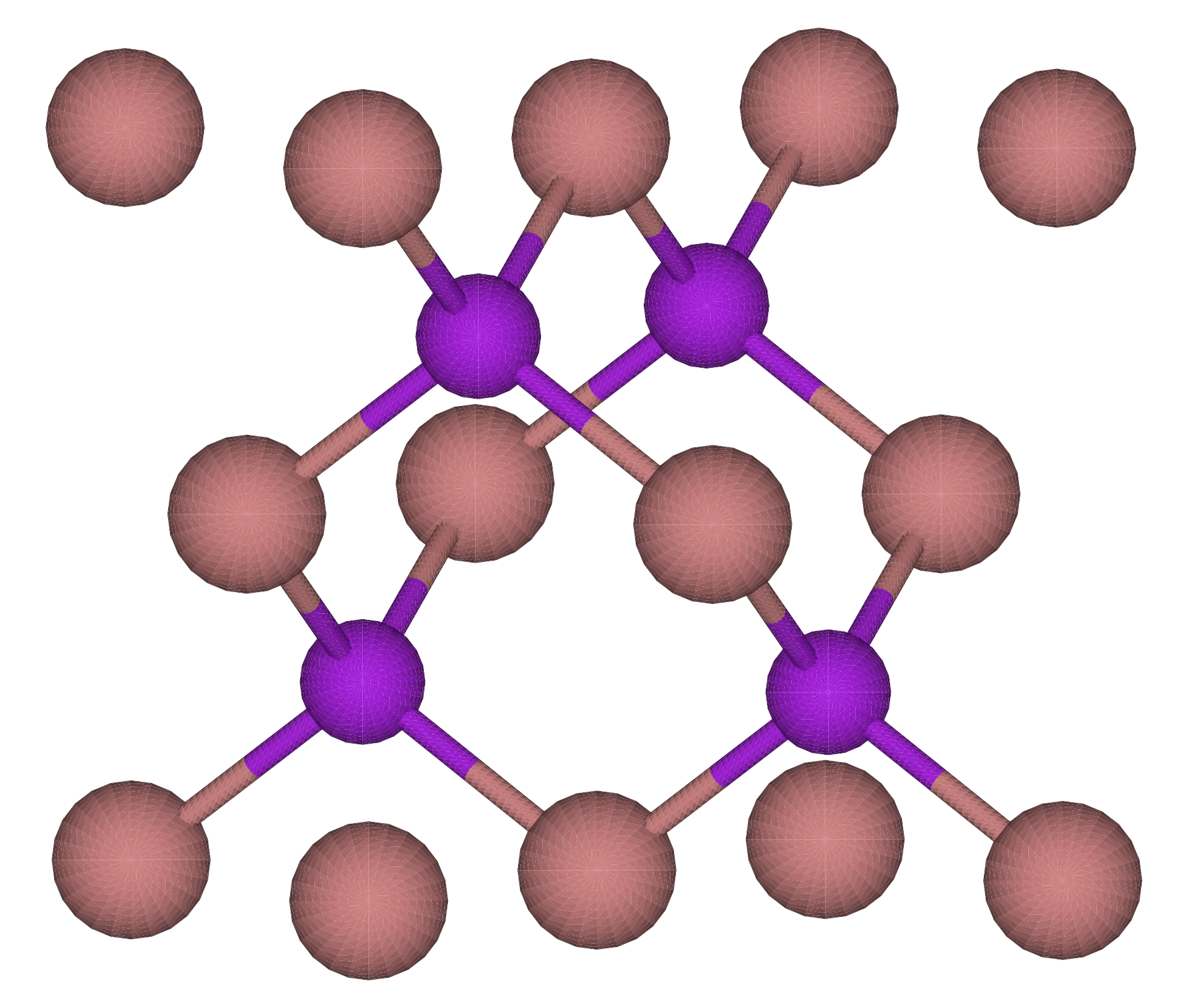};

\nextgroupplot[
tick align=outside,
tick pos=left,
title={\myfont(d)},
axis lines=left, 
xtick=\empty, 
ytick=\empty,
y axis line style={draw opacity=0}, 
x axis line style={draw opacity=0},
]
\addplot graphics [includegraphics cmd=\pgfimage,xmin=-0.5, xmax=8.5, ymin=20.5, ymax=-0.5] {Figure/Structure_Specimen/SrTiO3_inverse_.pdf};

\nextgroupplot[
width=5in,
height=3in,
tick align=outside,
tick pos=left,
title={\myfont (f)},
axis lines=left, 
xtick=\empty, 
ytick=\empty,
y axis line style={draw opacity=0}, 
x axis line style={draw opacity=0},
]
\addplot graphics [includegraphics cmd=\pgfimage,xmin=1.5, xmax=8.5, ymin=20.5, ymax=-0.5] {Figure/Structure_Specimen/MoS2_inverse_.pdf};
\end{groupplot}
 
\end{tikzpicture}

%% file: Figure/PACBED/idp_probe.tex
\begin{tikzpicture}
\begin{groupplot}[group style={group size=2 by 1,vertical sep=1em,horizontal sep=2em},width = 0.25*\textwidth, height=(0.25)*\textwidth]
\nextgroupplot[
title = (a),
hide x axis,
hide y axis,
tick align=outside,
tick pos=left,
x grid style={white!69.0196078431373!black},
xmin=-0.5, xmax=239.5,
xtick style={color=black},
y dir=reverse,
y grid style={white!69.0196078431373!black},
ymin=-0.5, ymax=239.5,
ytick style={color=black}
]
\addplot graphics [includegraphics cmd=\pgfimage,xmin=-0.5, xmax=239.5, ymin=239.5, ymax=-0.5] {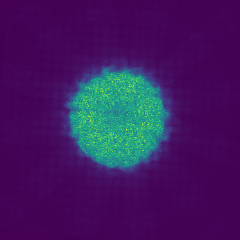};

\nextgroupplot[
title = (b),
hide x axis,
hide y axis,
tick align=outside,
tick pos=left,
x grid style={white!69.0196078431373!black},
xmin=80, xmax=160,
xtick style={color=black},
y grid style={white!69.0196078431373!black},
ymin=80, ymax=160,
ytick style={color=black}
]
\addplot graphics [includegraphics cmd=\pgfimage,xmin=-0.5, xmax=239.5, ymin=239.5, ymax=-0.5] {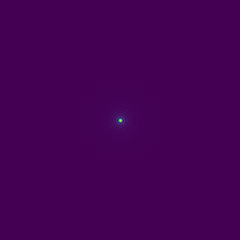};

\end{groupplot}

\end{tikzpicture}

%% file: Figure/Result/Adapt_Thick/Adaptive_Thick_4nm.tex
\pgfplotsset{every tick label/.append style={font=\myfont}}
\begin{tikzpicture}

\begin{groupplot}[group style={group size=9 by 5,vertical sep = 1.5em,horizontal sep=1.5em},width = 0.25*\textwidth, height=(0.25)*\textwidth]
\nextgroupplot[
title = \myfont GT (a),
hide x axis,
ticks=none,
ylabel = \myfont Slice 1,
tick align=outside,
tick pos=left,
x grid style={white!69.0196078431373!black},
xmin=-0.5, xmax=39.5,
xtick style={color=black},
y dir=reverse,
y grid style={white!69.0196078431373!black},
ymin=-0.5, ymax=39.5,
ytick style={color=black}
]
\addplot graphics [includegraphics cmd=\pgfimage,xmin=-0.5, xmax=39.5, ymin=39.5, ymax=-0.5] {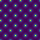};

\nextgroupplot[
title = \myfont L (a),
hide x axis,
hide y axis,
tick align=outside,
tick pos=left,
x grid style={white!69.0196078431373!black},
xmin=-0.5, xmax=39.5,
xtick style={color=black},
y dir=reverse,
y grid style={white!69.0196078431373!black},
ymin=-0.5, ymax=39.5,
ytick style={color=black}
]
\addplot graphics [includegraphics cmd=\pgfimage,xmin=-0.5, xmax=39.5, ymin=39.5, ymax=-0.5] {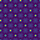};

\nextgroupplot[
hide x axis,
hide y axis,
title = \myfont S (a),
tick align=outside,
tick pos=left,
x grid style={white!69.0196078431373!black},
xmin=-0.5, xmax=39.5,
xtick style={color=black},
y dir=reverse,
y grid style={white!69.0196078431373!black},
ymin=-0.5, ymax=39.5,
ytick style={color=black}
]
\addplot graphics [includegraphics cmd=\pgfimage,xmin=-0.5, xmax=39.5, ymin=39.5, ymax=-0.5] {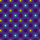};

\nextgroupplot[
title = \myfont GT (b),
hide x axis,
hide y axis,
tick align=outside,
tick pos=left,
x grid style={white!69.0196078431373!black},
xmin=-0.5, xmax=39.5,
xtick style={color=black},
y dir=reverse,
y grid style={white!69.0196078431373!black},
ymin=-0.5, ymax=39.5,
ytick style={color=black}
]
\addplot graphics [includegraphics cmd=\pgfimage,xmin=-0.5, xmax=39.5, ymin=39.5, ymax=-0.5] {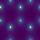};

\nextgroupplot[
title = \myfont L (b),
hide x axis,
hide y axis,
tick align=outside,
tick pos=left,
x grid style={white!69.0196078431373!black},
xmin=-0.5, xmax=39.5,
xtick style={color=black},
y dir=reverse,
y grid style={white!69.0196078431373!black},
ymin=-0.5, ymax=39.5,
ytick style={color=black}
]
\addplot graphics [includegraphics cmd=\pgfimage,xmin=-0.5, xmax=39.5, ymin=39.5, ymax=-0.5] {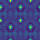};

\nextgroupplot[
title = \myfont S (b),
hide x axis,
hide y axis,
tick align=outside,
tick pos=left,
x grid style={white!69.0196078431373!black},
xmin=-0.5, xmax=39.5,
xtick style={color=black},
y dir=reverse,
y grid style={white!69.0196078431373!black},
ymin=-0.5, ymax=39.5,
ytick style={color=black}
]
\addplot graphics [includegraphics cmd=\pgfimage,xmin=-0.5, xmax=39.5, ymin=39.5, ymax=-0.5] {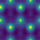};

\nextgroupplot[
title = \myfont GT (c),
hide x axis,
hide y axis,
tick align=outside,
tick pos=left,
x grid style={white!69.0196078431373!black},
xmin=-0.5, xmax=39.5,
xtick style={color=black},
y dir=reverse,
y grid style={white!69.0196078431373!black},
ymin=-0.5, ymax=39.5,
ytick style={color=black}
]
\addplot graphics [includegraphics cmd=\pgfimage,xmin=-0.5, xmax=39.5, ymin=39.5, ymax=-0.5] {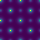};

\nextgroupplot[
title = \myfont L (c),
hide x axis,
hide y axis,
tick align=outside,
tick pos=left,
x grid style={white!69.0196078431373!black},
xmin=-0.5, xmax=39.5,
xtick style={color=black},
y dir=reverse,
y grid style={white!69.0196078431373!black},
ymin=-0.5, ymax=39.5,
ytick style={color=black}
]
\addplot graphics [includegraphics cmd=\pgfimage,xmin=-0.5, xmax=39.5, ymin=39.5, ymax=-0.5] {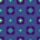};

\nextgroupplot[
title = \myfont S (c),
hide x axis,
hide y axis,
tick align=outside,
tick pos=left,
x grid style={white!69.0196078431373!black},
xmin=-0.5, xmax=39.5,
xtick style={color=black},
y dir=reverse,
y grid style={white!69.0196078431373!black},
ymin=-0.5, ymax=39.5,
ytick style={color=black}
]
\addplot graphics [includegraphics cmd=\pgfimage,xmin=-0.5, xmax=39.5, ymin=39.5, ymax=-0.5] {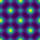};

\nextgroupplot[
hide x axis,
ticks=none,
ylabel = \myfont Slice 2,
tick align=outside,
tick pos=left,
x grid style={white!69.0196078431373!black},
xmin=-0.5, xmax=39.5,
xtick style={color=black},
y dir=reverse,
y grid style={white!69.0196078431373!black},
ymin=-0.5, ymax=39.5,
ytick style={color=black}
]
\addplot graphics [includegraphics cmd=\pgfimage,xmin=-0.5, xmax=39.5, ymin=39.5, ymax=-0.5] {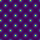};

\nextgroupplot[
hide x axis,
hide y axis,
tick align=outside,
tick pos=left,
x grid style={white!69.0196078431373!black},
xmin=-0.5, xmax=39.5,
xtick style={color=black},
y dir=reverse,
y grid style={white!69.0196078431373!black},
ymin=-0.5, ymax=39.5,
ytick style={color=black}
]
\addplot graphics [includegraphics cmd=\pgfimage,xmin=-0.5, xmax=39.5, ymin=39.5, ymax=-0.5] {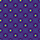};

\nextgroupplot[
hide x axis,
hide y axis,
tick align=outside,
tick pos=left,
x grid style={white!69.0196078431373!black},
xmin=-0.5, xmax=39.5,
xtick style={color=black},
y dir=reverse,
y grid style={white!69.0196078431373!black},
ymin=-0.5, ymax=39.5,
ytick style={color=black}
]
\addplot graphics [includegraphics cmd=\pgfimage,xmin=-0.5, xmax=39.5, ymin=39.5, ymax=-0.5] {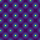};

\nextgroupplot[
hide x axis,
hide y axis,
tick align=outside,
tick pos=left,
x grid style={white!69.0196078431373!black},
xmin=-0.5, xmax=39.5,
xtick style={color=black},
y dir=reverse,
y grid style={white!69.0196078431373!black},
ymin=-0.5, ymax=39.5,
ytick style={color=black}
]
\addplot graphics [includegraphics cmd=\pgfimage,xmin=-0.5, xmax=39.5, ymin=39.5, ymax=-0.5] {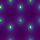};

\nextgroupplot[
hide x axis,
hide y axis,
tick align=outside,
tick pos=left,
x grid style={white!69.0196078431373!black},
xmin=-0.5, xmax=39.5,
xtick style={color=black},
y dir=reverse,
y grid style={white!69.0196078431373!black},
ymin=-0.5, ymax=39.5,
ytick style={color=black}
]
\addplot graphics [includegraphics cmd=\pgfimage,xmin=-0.5, xmax=39.5, ymin=39.5, ymax=-0.5] {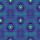};

\nextgroupplot[
hide x axis,
hide y axis,
tick align=outside,
tick pos=left,
x grid style={white!69.0196078431373!black},
xmin=-0.5, xmax=39.5,
xtick style={color=black},
y dir=reverse,
y grid style={white!69.0196078431373!black},
ymin=-0.5, ymax=39.5,
ytick style={color=black}
]
\addplot graphics [includegraphics cmd=\pgfimage,xmin=-0.5, xmax=39.5, ymin=39.5, ymax=-0.5] {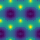};

\nextgroupplot[
hide x axis,
hide y axis,
tick align=outside,
tick pos=left,
x grid style={white!69.0196078431373!black},
xmin=-0.5, xmax=39.5,
xtick style={color=black},
y dir=reverse,
y grid style={white!69.0196078431373!black},
ymin=-0.5, ymax=39.5,
ytick style={color=black}
]
\addplot graphics [includegraphics cmd=\pgfimage,xmin=-0.5, xmax=39.5, ymin=39.5, ymax=-0.5] {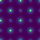};

\nextgroupplot[
hide x axis,
hide y axis,
tick align=outside,
tick pos=left,
x grid style={white!69.0196078431373!black},
xmin=-0.5, xmax=39.5,
xtick style={color=black},
y dir=reverse,
y grid style={white!69.0196078431373!black},
ymin=-0.5, ymax=39.5,
ytick style={color=black}
]
\addplot graphics [includegraphics cmd=\pgfimage,xmin=-0.5, xmax=39.5, ymin=39.5, ymax=-0.5] {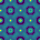};

\nextgroupplot[
hide x axis,
hide y axis,
tick align=outside,
tick pos=left,
x grid style={white!69.0196078431373!black},
xmin=-0.5, xmax=39.5,
xtick style={color=black},
y dir=reverse,
y grid style={white!69.0196078431373!black},
ymin=-0.5, ymax=39.5,
ytick style={color=black}
]
\addplot graphics [includegraphics cmd=\pgfimage,xmin=-0.5, xmax=39.5, ymin=39.5, ymax=-0.5] {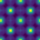};

\nextgroupplot[
hide x axis,
ticks=none,
ylabel = \myfont Slice 3,
tick align=outside,
tick pos=left,
x grid style={white!69.0196078431373!black},
xmin=-0.5, xmax=39.5,
xtick style={color=black},
y dir=reverse,
y grid style={white!69.0196078431373!black},
ymin=-0.5, ymax=39.5,
ytick style={color=black}
]
\addplot graphics [includegraphics cmd=\pgfimage,xmin=-0.5, xmax=39.5, ymin=39.5, ymax=-0.5] {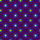};

\nextgroupplot[
hide x axis,
hide y axis,
tick align=outside,
tick pos=left,
x grid style={white!69.0196078431373!black},
xmin=-0.5, xmax=39.5,
xtick style={color=black},
y dir=reverse,
y grid style={white!69.0196078431373!black},
ymin=-0.5, ymax=39.5,
ytick style={color=black}
]
\addplot graphics [includegraphics cmd=\pgfimage,xmin=-0.5, xmax=39.5, ymin=39.5, ymax=-0.5] {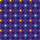};

\nextgroupplot[
hide x axis,
hide y axis,
tick align=outside,
tick pos=left,
x grid style={white!69.0196078431373!black},
xmin=-0.5, xmax=39.5,
xtick style={color=black},
y dir=reverse,
y grid style={white!69.0196078431373!black},
ymin=-0.5, ymax=39.5,
ytick style={color=black}
]
\addplot graphics [includegraphics cmd=\pgfimage,xmin=-0.5, xmax=39.5, ymin=39.5, ymax=-0.5] {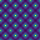};

\nextgroupplot[
hide x axis,
hide y axis,
tick align=outside,
tick pos=left,
x grid style={white!69.0196078431373!black},
xmin=-0.5, xmax=39.5,
xtick style={color=black},
y dir=reverse,
y grid style={white!69.0196078431373!black},
ymin=-0.5, ymax=39.5,
ytick style={color=black}
]
\addplot graphics [includegraphics cmd=\pgfimage,xmin=-0.5, xmax=39.5, ymin=39.5, ymax=-0.5] {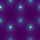};

\nextgroupplot[
hide x axis,
hide y axis,
tick align=outside,
tick pos=left,
x grid style={white!69.0196078431373!black},
xmin=-0.5, xmax=39.5,
xtick style={color=black},
y dir=reverse,
y grid style={white!69.0196078431373!black},
ymin=-0.5, ymax=39.5,
ytick style={color=black}
]
\addplot graphics [includegraphics cmd=\pgfimage,xmin=-0.5, xmax=39.5, ymin=39.5, ymax=-0.5] {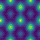};

\nextgroupplot[
hide x axis,
hide y axis,
tick align=outside,
tick pos=left,
x grid style={white!69.0196078431373!black},
xmin=-0.5, xmax=39.5,
xtick style={color=black},
y dir=reverse,
y grid style={white!69.0196078431373!black},
ymin=-0.5, ymax=39.5,
ytick style={color=black}
]
\addplot graphics [includegraphics cmd=\pgfimage,xmin=-0.5, xmax=39.5, ymin=39.5, ymax=-0.5] {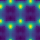};

\nextgroupplot[
hide x axis,
hide y axis,
tick align=outside,
tick pos=left,
x grid style={white!69.0196078431373!black},
xmin=-0.5, xmax=39.5,
xtick style={color=black},
y dir=reverse,
y grid style={white!69.0196078431373!black},
ymin=-0.5, ymax=39.5,
ytick style={color=black}
]
\addplot graphics [includegraphics cmd=\pgfimage,xmin=-0.5, xmax=39.5, ymin=39.5, ymax=-0.5] {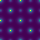};

\nextgroupplot[
hide x axis,
hide y axis,
tick align=outside,
tick pos=left,
x grid style={white!69.0196078431373!black},
xmin=-0.5, xmax=39.5,
xtick style={color=black},
y dir=reverse,
y grid style={white!69.0196078431373!black},
ymin=-0.5, ymax=39.5,
ytick style={color=black}
]
\addplot graphics [includegraphics cmd=\pgfimage,xmin=-0.5, xmax=39.5, ymin=39.5, ymax=-0.5] {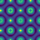};

\nextgroupplot[
hide x axis,
hide y axis,
tick align=outside,
tick pos=left,
x grid style={white!69.0196078431373!black},
xmin=-0.5, xmax=39.5,
xtick style={color=black},
y dir=reverse,
y grid style={white!69.0196078431373!black},
ymin=-0.5, ymax=39.5,
ytick style={color=black}
]
\addplot graphics [includegraphics cmd=\pgfimage,xmin=-0.5, xmax=39.5, ymin=39.5, ymax=-0.5] {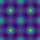};

\nextgroupplot[
hide x axis,
ticks=none,
ylabel = \myfont Slice 4,
tick align=outside,
tick pos=left,
x grid style={white!69.0196078431373!black},
xmin=-0.5, xmax=39.5,
xtick style={color=black},
y dir=reverse,
y grid style={white!69.0196078431373!black},
ymin=-0.5, ymax=39.5,
ytick style={color=black}
]
\addplot graphics [includegraphics cmd=\pgfimage,xmin=-0.5, xmax=39.5, ymin=39.5, ymax=-0.5] {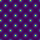};

\nextgroupplot[
hide x axis,
hide y axis,
tick align=outside,
tick pos=left,
x grid style={white!69.0196078431373!black},
xmin=-0.5, xmax=39.5,
xtick style={color=black},
y dir=reverse,
y grid style={white!69.0196078431373!black},
ymin=-0.5, ymax=39.5,
ytick style={color=black}
]
\addplot graphics [includegraphics cmd=\pgfimage,xmin=-0.5, xmax=39.5, ymin=39.5, ymax=-0.5] {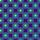};

\nextgroupplot[
hide x axis,
hide y axis,
tick align=outside,
tick pos=left,
x grid style={white!69.0196078431373!black},
xmin=-0.5, xmax=39.5,
xtick style={color=black},
y dir=reverse,
y grid style={white!69.0196078431373!black},
ymin=-0.5, ymax=39.5,
ytick style={color=black}
]
\addplot graphics [includegraphics cmd=\pgfimage,xmin=-0.5, xmax=39.5, ymin=39.5, ymax=-0.5] {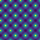};

\nextgroupplot[
hide x axis,
hide y axis,
tick align=outside,
tick pos=left,
x grid style={white!69.0196078431373!black},
xmin=-0.5, xmax=39.5,
xtick style={color=black},
y dir=reverse,
y grid style={white!69.0196078431373!black},
ymin=-0.5, ymax=39.5,
ytick style={color=black}
]
\addplot graphics [includegraphics cmd=\pgfimage,xmin=-0.5, xmax=39.5, ymin=39.5, ymax=-0.5] {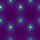};

\nextgroupplot[
hide x axis,
hide y axis,
tick align=outside,
tick pos=left,
x grid style={white!69.0196078431373!black},
xmin=-0.5, xmax=39.5,
xtick style={color=black},
y dir=reverse,
y grid style={white!69.0196078431373!black},
ymin=-0.5, ymax=39.5,
ytick style={color=black}
]
\addplot graphics [includegraphics cmd=\pgfimage,xmin=-0.5, xmax=39.5, ymin=39.5, ymax=-0.5] {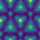};

\nextgroupplot[
hide x axis,
hide y axis,
tick align=outside,
tick pos=left,
x grid style={white!69.0196078431373!black},
xmin=-0.5, xmax=39.5,
xtick style={color=black},
y dir=reverse,
y grid style={white!69.0196078431373!black},
ymin=-0.5, ymax=39.5,
ytick style={color=black}
]
\addplot graphics [includegraphics cmd=\pgfimage,xmin=-0.5, xmax=39.5, ymin=39.5, ymax=-0.5] {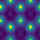};

\nextgroupplot[
hide x axis,
hide y axis,
tick align=outside,
tick pos=left,
x grid style={white!69.0196078431373!black},
xmin=-0.5, xmax=39.5,
xtick style={color=black},
y dir=reverse,
y grid style={white!69.0196078431373!black},
ymin=-0.5, ymax=39.5,
ytick style={color=black}
]
\addplot graphics [includegraphics cmd=\pgfimage,xmin=-0.5, xmax=39.5, ymin=39.5, ymax=-0.5] {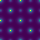};

\nextgroupplot[
hide x axis,
hide y axis,
tick align=outside,
tick pos=left,
x grid style={white!69.0196078431373!black},
xmin=-0.5, xmax=39.5,
xtick style={color=black},
y dir=reverse,
y grid style={white!69.0196078431373!black},
ymin=-0.5, ymax=39.5,
ytick style={color=black}
]
\addplot graphics [includegraphics cmd=\pgfimage,xmin=-0.5, xmax=39.5, ymin=39.5, ymax=-0.5] {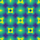};

\nextgroupplot[
hide x axis,
hide y axis,
tick align=outside,
tick pos=left,
x grid style={white!69.0196078431373!black},
xmin=-0.5, xmax=39.5,
xtick style={color=black},
y dir=reverse,
y grid style={white!69.0196078431373!black},
ymin=-0.5, ymax=39.5,
ytick style={color=black}
]
\addplot graphics [includegraphics cmd=\pgfimage,xmin=-0.5, xmax=39.5, ymin=39.5, ymax=-0.5] {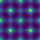};

\nextgroupplot[
colorbar horizontal,
colormap/viridis,
hide x axis,
ticks=none,
ylabel = \myfont Slice 5,
colorbar style={xtick={0.2, 0.8},xticklabels={0.01, 0.5} },
tick align=outside,
tick pos=left,
x grid style={white!69.0196078431373!black},
xmin=-0.5, xmax=39.5,
xtick style={color=black},
y dir=reverse,
y grid style={white!69.0196078431373!black},
ymin=-0.5, ymax=39.5,
ytick style={color=black}
]
\addplot graphics [includegraphics cmd=\pgfimage,xmin=-0.5, xmax=39.5, ymin=39.5, ymax=-0.5] {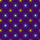};

\nextgroupplot[
colorbar horizontal,
colormap/viridis,
hide x axis,
hide y axis,
colorbar style={xtick={0.2, 0.8},xticklabels={-0.02, 0.08} },
tick align=outside,
tick pos=left,
x grid style={white!69.0196078431373!black},
xmin=-0.5, xmax=39.5,
xtick style={color=black},
y dir=reverse,
y grid style={white!69.0196078431373!black},
ymin=-0.5, ymax=39.5,
ytick style={color=black}
]
\addplot graphics [includegraphics cmd=\pgfimage,xmin=-0.5, xmax=39.5, ymin=39.5, ymax=-0.5] {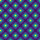};

\nextgroupplot[
colorbar horizontal,
colormap/viridis,
hide x axis,
hide y axis,
colorbar style={xtick={0.2, 0.8},xticklabels={-0.3, 1.2} },
tick align=outside,
tick pos=left,
x grid style={white!69.0196078431373!black},
xmin=-0.5, xmax=39.5,
xtick style={color=black},
y dir=reverse,
y grid style={white!69.0196078431373!black},
ymin=-0.5, ymax=39.5,
ytick style={color=black}
]
\addplot graphics [includegraphics cmd=\pgfimage,xmin=-0.5, xmax=39.5, ymin=39.5, ymax=-0.5] {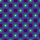};

\nextgroupplot[
colorbar horizontal,
colormap/viridis,
hide x axis,
hide y axis,
colorbar style={xtick={0.2, 0.8},xticklabels={0.01, 1.7} },
tick align=outside,
tick pos=left,
x grid style={white!69.0196078431373!black},
xmin=-0.5, xmax=39.5,
xtick style={color=black},
y dir=reverse,
y grid style={white!69.0196078431373!black},
ymin=-0.5, ymax=39.5,
ytick style={color=black}
]
\addplot graphics [includegraphics cmd=\pgfimage,xmin=-0.5, xmax=39.5, ymin=39.5, ymax=-0.5] {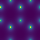};

\nextgroupplot[
colorbar horizontal,
colormap/viridis,
hide x axis,
hide y axis,
colorbar style={xtick={0.2, 0.8},xticklabels={-0.3, 0.4} },
tick align=outside,
tick pos=left,
x grid style={white!69.0196078431373!black},
xmin=-0.5, xmax=39.5,
xtick style={color=black},
y dir=reverse,
y grid style={white!69.0196078431373!black},
ymin=-0.5, ymax=39.5,
ytick style={color=black}
]
\addplot graphics [includegraphics cmd=\pgfimage,xmin=-0.5, xmax=39.5, ymin=39.5, ymax=-0.5] {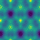};

\nextgroupplot[
colorbar horizontal,
colormap/viridis,
hide x axis,
hide y axis,
colorbar style={xtick={0.2, 0.8},xticklabels={-0.9, 0.9} },
tick align=outside,
tick pos=left,
x grid style={white!69.0196078431373!black},
xmin=-0.5, xmax=39.5,
xtick style={color=black},
y dir=reverse,
y grid style={white!69.0196078431373!black},
ymin=-0.5, ymax=39.5,
ytick style={color=black}
]
\addplot graphics [includegraphics cmd=\pgfimage,xmin=-0.5, xmax=39.5, ymin=39.5, ymax=-0.5] {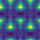};

\nextgroupplot[
colorbar horizontal,
colormap/viridis,
hide x axis,
hide y axis,
colorbar style={xtick={0.2, 0.8},xticklabels={0.02, 0.7} },
tick align=outside,
tick pos=left,
x grid style={white!69.0196078431373!black},
xmin=-0.5, xmax=39.5,
xtick style={color=black},
y dir=reverse,
y grid style={white!69.0196078431373!black},
ymin=-0.5, ymax=39.5,
ytick style={color=black}
]
\addplot graphics [includegraphics cmd=\pgfimage,xmin=-0.5, xmax=39.5, ymin=39.5, ymax=-0.5] {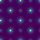};

\nextgroupplot[
colorbar horizontal,
colormap/viridis,
hide x axis,
hide y axis,
colorbar style={xtick={0.2, 0.8},xticklabels={-0.1, 0.1} },
tick align=outside,
tick pos=left,
x grid style={white!69.0196078431373!black},
xmin=-0.5, xmax=39.5,
xtick style={color=black},
y dir=reverse,
y grid style={white!69.0196078431373!black},
ymin=-0.5, ymax=39.5,
ytick style={color=black}
]
\addplot graphics [includegraphics cmd=\pgfimage,xmin=-0.5, xmax=39.5, ymin=39.5, ymax=-0.5] {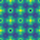};

\nextgroupplot[
colorbar horizontal,
colormap/viridis,
hide x axis,
hide y axis,
colorbar style={xtick={0.2, 0.8},xticklabels={-0.2, 1.7} },
tick align=outside,
tick pos=left,
x grid style={white!69.0196078431373!black},
xmin=-0.5, xmax=39.5,
xtick style={color=black},
y dir=reverse,
y grid style={white!69.0196078431373!black},
ymin=-0.5, ymax=39.5,
ytick style={color=black}
]
\addplot graphics [includegraphics cmd=\pgfimage,xmin=-0.5, xmax=39.5, ymin=39.5, ymax=-0.5] {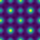};
\end{groupplot}

\end{tikzpicture}

%% file: Figure/Result/Figure_6/error_idp.tex
\pgfplotsset{every tick label/.append style={font=\large}}
\begin{tikzpicture}

\begin{axis}[
width=5in,
height=2.5in,
legend cell align={left},
legend style={fill opacity=0.8, draw opacity=1, text opacity=1, draw=white!80!black},
tick align=outside,
tick pos=left,
x grid style={white!69.0196078431373!black},
xlabel={\large Iteration},
xmajorgrids,
xmin=-4.95, xmax=103.95,
xtick style={color=black},
y grid style={white!69.0196078431373!black},
ylabel={\large Error},
ymajorgrids,
ymin=0, ymax=0.1,
ytick style={color=black},
ytick={0,0.05,0.1},
yticklabels={0,0.05,0.1},
]
\addplot [thick, black, dashed, mark=*, mark size=3, mark options={solid}]
table {%
0 0.044229853069541
1 0.0338360553251608
2 0.0270005606499141
3 0.0227065958216948
4 0.0198501421728552
5 0.0177547184512787
6 0.0160682858681714
7 0.0146229775960554
8 0.0133417881005626
9 0.0121919497679826
10 0.0111577091112367
11 0.0102299485809178
12 0.00940122119974355
13 0.00866457960838096
14 0.00801413527784574
15 0.00744372261463317
16 0.00694563159812247
17 0.00651202700113522
18 0.0061353631735936
19 0.00580887717579912
20 0.00552599751280423
21 0.00528078127240126
22 0.00506774588567791
23 0.00488204767672354
24 0.00471983570740483
25 0.00457760954747312
26 0.00445233061033835
27 0.00434145380023923
28 0.00424282108415434
29 0.00415476834439384
30 0.00407581846068054
31 0.00400471778248863
32 0.00394041447983458
33 0.00388201733303062
34 0.00382880572976988
35 0.00378015454551113
36 0.00373552652032932
37 0.00369445323822483
38 0.00365653671535509
39 0.00362142967643082
40 0.00358882285427389
41 0.00355845000860719
42 0.00353008000068249
43 0.00350350628983791
44 0.00347854464017249
45 0.0034550325982964
46 0.00343282635197937
47 0.0034117945125504
48 0.0033918249299293
49 0.00337281539101889
50 0.00335467418361611
51 0.00333731864692866
52 0.00332067321847968
53 0.00330467575075958
54 0.00328926809900241
55 0.00327439648523566
56 0.00326001415834094
57 0.00324607761440663
58 0.00323254943999769
59 0.00321939463924582
60 0.00320658364669032
61 0.00319408878813134
62 0.00318188427906492
63 0.00316994896720464
64 0.00315825985775109
65 0.003146801558835
66 0.00313555484326791
67 0.00312450706399985
68 0.00311364331522223
69 0.00310295259931562
70 0.00309242242330218
71 0.00308204334076447
72 0.00307180737128213
73 0.00306170470064726
74 0.00305172930910359
75 0.00304187305115887
76 0.003032131723663
77 0.00302249699002674
78 0.00301296607233724
79 0.00300353234252985
80 0.00299419448250924
81 0.00298494559251785
82 0.00297578459969423
83 0.00296670518923269
84 0.00295770622191009
85 0.0029487854749526
86 0.00293993889432711
87 0.00293116564932552
88 0.00292246216978056
89 0.00291382794559234
90 0.00290525931900504
91 0.00289675567407515
92 0.00288831521911136
93 0.00287993656820497
94 0.00287161749248143
95 0.00286335732168473
96 0.00285515484064986
97 0.00284700890671711
98 0.00283891906009449
99 0.00283088230033831
};
\addlegendentry{\large GaAs (a)}
\addplot [thick, red, dashed, mark=*, mark size=3, mark options={solid}]
table {%
0 0.0727590953875519
1 0.0664898679450686
2 0.0608206471987432
3 0.0559678505231069
4 0.0518557465476887
5 0.048371669227479
6 0.0454073705337803
7 0.0428669236776979
8 0.0406685584849985
9 0.0387447112404738
10 0.0370408333351726
11 0.0355136329991358
12 0.0341290064352868
13 0.032860162205948
14 0.0316863001487206
15 0.0305912809459384
16 0.0295625794512296
17 0.0285904645227158
18 0.0276672406638132
19 0.0267871001202713
20 0.0259452290016003
21 0.0251379289728723
22 0.0243622388343136
23 0.0236157166487841
24 0.0228965330165569
25 0.0222029099405282
26 0.0215334398716593
27 0.0208868912847096
28 0.0202622367769514
29 0.0196585528042071
30 0.0190751223234175
31 0.0185112430519214
32 0.0179662067597835
33 0.0174393735796608
34 0.0169301531575038
35 0.0164380339680513
36 0.0159624605509549
37 0.015502880565444
38 0.0150587824641021
39 0.0146297055400989
40 0.0142152114586803
41 0.0138147948053429
42 0.0134280478558091
43 0.0130545602307037
44 0.0126939055535745
45 0.0123456593475168
46 0.0120094121784413
47 0.0116847768270582
48 0.0113713167338222
49 0.0110686632807382
50 0.0107764624842545
51 0.0104943823698111
52 0.0102220814184009
53 0.00995921699455066
54 0.00970546820446452
55 0.00946052468689054
56 0.00922407203060462
57 0.0089958100251282
58 0.00877547108839701
59 0.00856279324132697
60 0.00835750946935575
61 0.00815937482785693
62 0.00796814025939183
63 0.00778356333921238
64 0.00760541445673428
65 0.00743345833067465
66 0.00726749230322845
67 0.00710732444918869
68 0.00695273951947214
69 0.00680355693210907
70 0.00665959291461694
71 0.00652067036845514
72 0.00638661305240745
73 0.00625726628260595
74 0.00613246391750969
75 0.00601206006160313
76 0.00589589897981698
77 0.00578384731684487
78 0.0056757478738536
79 0.00557148974967939
80 0.00547092816080512
81 0.00537395051714497
82 0.00528042311359633
83 0.00519024425884804
84 0.00510328426522951
85 0.0050194541499899
86 0.00493862877167001
87 0.00486072420606119
88 0.00478563050470112
89 0.00471325893319341
90 0.00464350737625596
91 0.00457629115386587
92 0.00451152378596857
93 0.0044491304512425
94 0.00438902081817435
95 0.00433112773896364
96 0.00427536498194459
97 0.00422166046892575
98 0.004169934524338
99 0.00412012295519041
};
\addlegendentry{\large MoS$_2$ (a)}
\addplot [thick, blue, dashed, mark=*, mark size=3, mark options={solid}]
table {%
0 0.0791895132153937
1 0.0683209847237362
2 0.0597405547896897
3 0.0531872318111767
4 0.0481542930233655
5 0.0442080686902558
6 0.0410208246462618
7 0.0383613790732352
8 0.0360749161804292
9 0.0340586768831196
10 0.0322450125907697
11 0.0305894711308235
12 0.029061981272179
13 0.027641858099014
14 0.0263145456684426
15 0.0250693605313911
16 0.0238980798447909
17 0.0227941907130859
18 0.0217529603142406
19 0.0207701543096797
20 0.0198419563356273
21 0.0189651045158883
22 0.0181362770610881
23 0.0173523882986153
24 0.0166116507478193
25 0.0159118295801347
26 0.0152508552743522
27 0.0146265014286226
28 0.0140364860191572
29 0.0134792065211464
30 0.0129529725866708
31 0.0124561541994475
32 0.0119870129195256
33 0.0115440455492621
34 0.011125861966269
35 0.0107311311084856
36 0.0103585389125018
37 0.010006892203569
38 0.00967501789915271
39 0.00936179370895695
40 0.00906618549291923
41 0.00878715868070241
42 0.00852378285784482
43 0.00827473859944216
44 0.00803957877798918
45 0.00781758569898101
46 0.00760799907799493
47 0.00741010860086488
48 0.00722324218784075
49 0.00704676130582675
50 0.0068800724957963
51 0.00672261219899352
52 0.00657384923123615
53 0.00643328952779726
54 0.00630047254080902
55 0.00617496056472087
56 0.00605632977086939
57 0.00594417408242129
58 0.00583812283779776
59 0.0057378224691771
60 0.0056429425312384
61 0.0055531607592284
62 0.00546819421673323
63 0.00538774937280052
64 0.0053115709183921
65 0.00523940785065884
66 0.00517102413255496
67 0.00510620140316698
68 0.00504472976276552
69 0.00498641025274372
70 0.0049310563800413
71 0.00487848794519715
72 0.00482854213649514
73 0.00478106301248202
74 0.00473590765753635
75 0.00469292531046172
76 0.00465198889578606
77 0.00461298305382305
78 0.00457578785871817
79 0.00454029837013537
80 0.00450641192469688
81 0.00447403061878751
82 0.00444306941079294
83 0.00441343771607511
84 0.00438506414046512
85 0.00435787146416053
86 0.0043317871290458
87 0.00430674807246919
88 0.00428269423127439
89 0.00425956200779555
90 0.00423730290272851
91 0.00421586509312756
92 0.00419520184566952
93 0.00417526927025658
94 0.00415602208391983
95 0.00413742400432802
96 0.00411943768034864
97 0.00410202608110955
98 0.00408516103464319
99 0.00406880295634912
};
\addlegendentry{\large SrTiO$_3$ (a)}
\addplot [thick, black, dashed, mark=x, mark size=3, mark options={solid}]
table {%
0 0.0471588788962028
1 0.0470463440822711
2 0.0469275192178002
3 0.0468076969331975
4 0.0466876994456679
5 0.0465678205334057
6 0.0464481955774744
7 0.0463288994694682
8 0.0462099760422904
9 0.0460914564173774
10 0.0459733575378451
11 0.0458556946462443
12 0.0457384753845417
13 0.0456217075263751
14 0.0455053965430314
15 0.0453895445162329
16 0.0452741537575041
17 0.0451592251976946
18 0.0450447617467066
19 0.0449307617138762
20 0.0448172258541628
21 0.0447041535794983
22 0.0445915443422649
23 0.0444793963721908
24 0.0443677077288677
25 0.0442564784264223
26 0.0441457075376702
27 0.0440353919213968
28 0.0439255301770674
29 0.0438161213866155
30 0.0437071631631842
31 0.0435986532840013
32 0.0434905902489546
33 0.0433829726112161
34 0.0432757967590331
35 0.0431690620813534
36 0.0430627647836452
37 0.042956905681868
38 0.0428514797533166
39 0.0427464867185369
40 0.0426419233340653
41 0.0425377896954845
42 0.0424340809598761
43 0.0423307960394464
44 0.0422279329109156
45 0.0421254894745608
46 0.0420234636095217
47 0.0419218530065406
48 0.0418206571217284
49 0.0417198712825325
50 0.0416194953696033
51 0.0415195273242083
52 0.041419963571344
53 0.0413208032683896
54 0.0412220436210194
55 0.0411236836410377
56 0.0410257196973508
57 0.0409281506966594
58 0.0408309745001056
59 0.040734189941836
60 0.0406377921376656
61 0.0405417821589134
62 0.0404461564874699
63 0.0403509130493927
64 0.0402560513013063
65 0.0401615676803425
66 0.0400674604297504
67 0.039973727838093
68 0.0398803675977726
69 0.0397873792668773
70 0.0396947593610694
71 0.03960250613214
72 0.0395106185274718
73 0.0394190937306979
74 0.0393279302556695
75 0.0392371260315565
76 0.0391466803942136
77 0.0390565895577594
78 0.0389668533072976
79 0.0388774685049725
80 0.0387884348273426
81 0.0386997490017707
82 0.0386114090216002
83 0.0385234148067911
84 0.0384357634695936
85 0.0383484536509889
86 0.0382614831232312
87 0.0381748505661722
88 0.0380885543018891
89 0.0380025927517601
90 0.0379169626782259
91 0.0378316645666333
92 0.0377466947443293
93 0.0376620532560967
94 0.0375777375770633
95 0.0374937471199497
96 0.0374100782509181
97 0.0373267302764868
98 0.0372437016293425
99 0.037160990712999
};
\addlegendentry{\large GaAs (b)}
\addplot [thick, red, dashed, mark=x, mark size=3, mark options={solid}]
table {%
0 0.0742249046591165
1 0.0735091923957812
2 0.0727882974074104
3 0.0720780471495763
4 0.0713795885050015
5 0.070693048090207
6 0.0700182767172915
7 0.0693550376775176
8 0.0687030597879071
9 0.0680620715923517
10 0.0674318002310879
11 0.0668119676804309
12 0.066202308371355
13 0.0656025628764075
14 0.0650124848723299
15 0.0644318324431641
16 0.0638603861261655
17 0.0632979203896556
18 0.0627442187532433
19 0.0621990649621238
20 0.0616622558210387
21 0.06113359053243
22 0.0606128770700279
23 0.0600999301781518
24 0.0595945676310401
25 0.0590966173347765
26 0.0586059081730998
27 0.0581222769701666
28 0.057645568008154
29 0.0571756230805379
30 0.0567122996888604
31 0.0562554450649005
32 0.0558049250725952
33 0.0553606032580631
34 0.0549223449384731
35 0.0544900253612885
36 0.0540635165523815
37 0.0536427001400314
38 0.053227457193971
39 0.0528176747944159
40 0.0524132410435859
41 0.0520140484883024
42 0.0516199937407961
43 0.0512309726943252
44 0.0508468860616079
45 0.0504676390620062
46 0.0500931373625998
47 0.0497232871840173
48 0.0493580020253797
49 0.0489971948972163
50 0.0486407832093491
51 0.048288680350364
52 0.0479408099528059
53 0.0475970920918889
54 0.047257450172781
55 0.0469218104174413
56 0.0465901026106135
57 0.0462622530322229
58 0.0459381947572398
59 0.0456178596101786
60 0.0453011814587373
61 0.0449880979977169
62 0.0446785445958748
63 0.0443724630897029
64 0.0440697930335169
65 0.0437704732104887
66 0.0434744518275973
67 0.0431816706589346
68 0.0428920739874134
69 0.042605613830081
70 0.0423222334644576
71 0.0420418839740886
72 0.0417645179928443
73 0.0414900834914094
74 0.0412185365988291
75 0.0409498272577319
76 0.0406839138169734
77 0.0404207503130351
78 0.0401602938921871
79 0.0399025039847399
80 0.0396473352850777
81 0.039394749055462
82 0.0391447059771391
83 0.0388971679578871
84 0.0386520949129531
85 0.0384094500544441
86 0.0381691989846892
87 0.0379313037274778
88 0.0376957296075983
89 0.03746244144072
90 0.0372314097527139
91 0.0370025993444086
92 0.0367759749510112
93 0.0365515108717597
94 0.0363291693310007
95 0.0361089265243077
96 0.0358907484936944
97 0.0356746070987938
98 0.0354604738453088
99 0.0352483210834299
};
\addlegendentry{\large MoS$_2$ (b)}
\addplot [thick, blue, dashed, mark=x, mark size=3, mark options={solid}]
table {%
0 0.0823631484953083
1 0.0819290700260686
2 0.0814906106748897
3 0.0810562187203595
4 0.0806264558174576
5 0.0802013974833904
6 0.0797810130090665
7 0.0793652386813045
8 0.0789539946572149
9 0.0785472015656549
10 0.0781447780564562
11 0.0777466361425226
12 0.0773527036513198
13 0.0769628970743645
14 0.0765771407052487
15 0.0761953610223817
16 0.0758174861630085
17 0.0754434398545809
18 0.0750731571553739
19 0.0747065715562414
20 0.0743436166719053
21 0.073984227960002
22 0.073628344466566
23 0.0732759051317362
24 0.072926851068436
25 0.0725811255995542
26 0.0722386732790808
27 0.071899437232295
28 0.0715633685431136
29 0.07123040886214
30 0.0709005125536416
31 0.0705736310848552
32 0.0702497110924174
33 0.0699287101445564
34 0.0696105797256097
35 0.0692952766545263
36 0.068982755320993
37 0.0686729744481068
38 0.0683658920617096
39 0.0680614651486144
40 0.0677596549290001
41 0.0674604231226331
42 0.0671637320445137
43 0.0668695438551993
44 0.0665778226296633
45 0.0662885288790402
46 0.0660016325585303
47 0.0657170964541055
48 0.0654348873540999
49 0.0651549728868024
50 0.0648773191888231
51 0.0646018981868861
52 0.0643286729615391
53 0.0640576198586322
54 0.0637887023992547
55 0.0635218974400649
56 0.0632571736442289
57 0.0629945019701762
58 0.0627338578607607
59 0.0624752100590975
60 0.0622185359153432
61 0.061963807920724
62 0.0617110016020098
63 0.0614600915168602
64 0.0612110529240651
65 0.0609638610064211
66 0.0607184937259748
67 0.0604749267072604
68 0.0602331390994144
69 0.0599931079066004
70 0.0597548084542674
71 0.0595182220321741
72 0.0592833289281881
73 0.0590501050884464
74 0.0588185327208324
75 0.058588590686244
76 0.0583602584060798
77 0.0581335184095167
78 0.0579083515032518
79 0.0576847356095458
80 0.0574626565053495
81 0.0572420957926653
82 0.0570230349538822
83 0.0568054539093394
84 0.0565893391750011
85 0.0563746717707621
86 0.0561614373336774
87 0.055949617290363
88 0.0557391963157222
89 0.0555301591254427
90 0.0553224895767265
91 0.0551161728393191
92 0.0549111928884995
93 0.0547075376000461
94 0.05450518985554
95 0.0543041362270473
96 0.0541043610637185
97 0.0539058521332936
98 0.0537085959793621
99 0.053512578776198
};
\addlegendentry{\large SrTiO$_3$ (b)}
\end{axis}

\end{tikzpicture}

%% file: Figure/Result/Figure_7/Fig_7_Ambiguity_Atomic_Decomposition_new.tex
\pgfplotsset{every tick label/.append style={font=\myfont}}
\begin{tikzpicture}

\begin{groupplot}[group style={group size=9 by 3,vertical sep = 1.5em,horizontal sep=1.5em},width = 0.25*\textwidth, height=(0.25)*\textwidth]
\nextgroupplot[
title = \myfont GT (a),
ticks=none,
hide x axis,
ticks=none,
ylabel = \myfont Slice 1,
ylabel style={rotate=0},
tick align=outside,
tick pos=left,
x grid style={white!69.0196078431373!black},
xmin=-0.5, xmax=39.5,
xtick style={color=black},
y dir=reverse,
y grid style={white!69.0196078431373!black},
ymin=-0.5, ymax=39.5,
ytick style={color=black}
]
\addplot graphics [includegraphics cmd=\pgfimage,xmin=-0.5, xmax=39.5, ymin=39.5, ymax=-0.5] {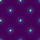};

\nextgroupplot[
title = \myfont L (a),
hide x axis,
hide y axis,
tick align=outside,
tick pos=left,
x grid style={white!69.0196078431373!black},
xmin=-0.5, xmax=39.5,
xtick style={color=black},
y dir=reverse,
y grid style={white!69.0196078431373!black},
ymin=-0.5, ymax=39.5,
ytick style={color=black}
]
\addplot graphics [includegraphics cmd=\pgfimage,xmin=-0.5, xmax=39.5, ymin=39.5, ymax=-0.5] {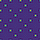};

\nextgroupplot[
title = \myfont S (a),
hide x axis,
hide y axis,
point meta max=0.410591125488281,
tick align=outside,
tick pos=left,
x grid style={white!69.0196078431373!black},
xmin=-0.5, xmax=39.5,
xtick style={color=black},
y dir=reverse,
y grid style={white!69.0196078431373!black},
ymin=-0.5, ymax=39.5,
ytick style={color=black}
]
\addplot graphics [includegraphics cmd=\pgfimage,xmin=-0.5, xmax=39.5, ymin=39.5, ymax=-0.5] {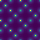};

\nextgroupplot[
title = \myfont GT (b),
colormap/viridis,
hide x axis,
hide y axis,
point meta max=0.427196651697159,
tick align=outside,
tick pos=left,
x grid style={white!69.0196078431373!black},
xmin=-0.5, xmax=39.5,
xtick style={color=black},
y dir=reverse,
y grid style={white!69.0196078431373!black},
ymin=-0.5, ymax=39.5,
ytick style={color=black}
]
\addplot graphics [includegraphics cmd=\pgfimage,xmin=-0.5, xmax=39.5, ymin=39.5, ymax=-0.5] {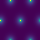};

\nextgroupplot[
title = \myfont L (b),
colormap/viridis,
hide x axis,
hide y axis,
point meta max=0.111907802522182,
tick align=outside,
tick pos=left,
x grid style={white!69.0196078431373!black},
xmin=-0.5, xmax=39.5,
xtick style={color=black},
y dir=reverse,
y grid style={white!69.0196078431373!black},
ymin=-0.5, ymax=39.5,
ytick style={color=black}
]
\addplot graphics [includegraphics cmd=\pgfimage,xmin=-0.5, xmax=39.5, ymin=39.5, ymax=-0.5] {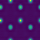};

\nextgroupplot[
title = \myfont S (b),
colormap/viridis,
hide x axis,
hide y axis,
point meta max=0.472353041172028,
tick align=outside,
tick pos=left,
x grid style={white!69.0196078431373!black},
xmin=-0.5, xmax=39.5,
xtick style={color=black},
y dir=reverse,
y grid style={white!69.0196078431373!black},
ymin=-0.5, ymax=39.5,
ytick style={color=black}
]
\addplot graphics [includegraphics cmd=\pgfimage,xmin=-0.5, xmax=39.5, ymin=39.5, ymax=-0.5] {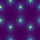};

\nextgroupplot[
title = \myfont GT (c),
colorbar style={xtick={0.25,0.5},xticklabels={0.25,0.50}},
colormap/viridis,
hide x axis,
hide y axis,
point meta max=0.672440230846405,
tick align=outside,
tick pos=left,
x grid style={white!69.0196078431373!black},
xmin=-0.5, xmax=39.5,
xtick style={color=black},
y dir=reverse,
y grid style={white!69.0196078431373!black},
ymin=-0.5, ymax=39.5,
ytick style={color=black}
]
\addplot graphics [includegraphics cmd=\pgfimage,xmin=-0.5, xmax=39.5, ymin=39.5, ymax=-0.5] {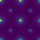};

\nextgroupplot[
title = \myfont L (c),
colorbar style={xtick={0,0.05},xticklabels={0.00,0.05}},
colormap/viridis,
hide x axis,
hide y axis,
point meta max=0.0805451571941376,
tick align=outside,
tick pos=left,
x grid style={white!69.0196078431373!black},
xmin=-0.5, xmax=39.5,
xtick style={color=black},
y dir=reverse,
y grid style={white!69.0196078431373!black},
ymin=-0.5, ymax=39.5,
ytick style={color=black}
]
\addplot graphics [includegraphics cmd=\pgfimage,xmin=-0.5, xmax=39.5, ymin=39.5, ymax=-0.5] {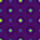};

\nextgroupplot[
title = \myfont S (c),
colormap/viridis,
hide x axis,
hide y axis,
point meta max=0.802953839302063,
tick align=outside,
tick pos=left,
x grid style={white!69.0196078431373!black},
xmin=-0.5, xmax=39.5,
xtick style={color=black},
y dir=reverse,
y grid style={white!69.0196078431373!black},
ymin=-0.5, ymax=39.5,
ytick style={color=black}
]
\addplot graphics [includegraphics cmd=\pgfimage,xmin=-0.5, xmax=39.5, ymin=39.5, ymax=-0.5] {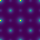};

\nextgroupplot[
ticks=none,
ylabel = \myfont Slice 2,
ylabel style={rotate=0},
hide x axis, 
point meta max=0.504094004631042,
tick align=outside,
tick pos=left,
x grid style={white!69.0196078431373!black},
xmin=-0.5, xmax=39.5,
xtick style={color=black},
y dir=reverse,
y grid style={white!69.0196078431373!black},
ymin=-0.5, ymax=39.5,
ytick style={color=black}
]
\addplot graphics [includegraphics cmd=\pgfimage,xmin=-0.5, xmax=39.5, ymin=39.5, ymax=-0.5] {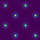};

\nextgroupplot[
hide y axis,
ticks=none,
hide x axis, 
tick align=outside,
tick pos=left,
x grid style={white!69.0196078431373!black},
xmin=-0.5, xmax=39.5,
xtick style={color=black},
y dir=reverse,
y grid style={white!69.0196078431373!black},
ymin=-0.5, ymax=39.5,
ytick style={color=black}
]
\addplot graphics [includegraphics cmd=\pgfimage,xmin=-0.5, xmax=39.5, ymin=39.5, ymax=-0.5] {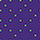};

\nextgroupplot[
colormap/viridis,
hide x axis,
hide y axis,
tick align=outside,
tick pos=left,
x grid style={white!69.0196078431373!black},
xmin=-0.5, xmax=39.5,
xtick style={color=black},
y dir=reverse,
y grid style={white!69.0196078431373!black},
ymin=-0.5, ymax=39.5,
ytick style={color=black}
]
\addplot graphics [includegraphics cmd=\pgfimage,xmin=-0.5, xmax=39.5, ymin=39.5, ymax=-0.5] {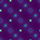};

\nextgroupplot[
colormap/viridis,
hide x axis,
hide y axis,
tick align=outside,
tick pos=left,
x grid style={white!69.0196078431373!black},
xmin=-0.5, xmax=39.5,
xtick style={color=black},
y dir=reverse,
y grid style={white!69.0196078431373!black},
ymin=-0.5, ymax=39.5,
ytick style={color=black}
]
\addplot graphics [includegraphics cmd=\pgfimage,xmin=-0.5, xmax=39.5, ymin=39.5, ymax=-0.5] {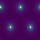};

\nextgroupplot[
colormap/viridis,
hide x axis,
hide y axis,
tick align=outside,
tick pos=left,
x grid style={white!69.0196078431373!black},
xmin=-0.5, xmax=39.5,
xtick style={color=black},
y dir=reverse,
y grid style={white!69.0196078431373!black},
ymin=-0.5, ymax=39.5,
ytick style={color=black}
]
\addplot graphics [includegraphics cmd=\pgfimage,xmin=-0.5, xmax=39.5, ymin=39.5, ymax=-0.5] {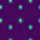};

\nextgroupplot[
colormap/viridis,
hide x axis,
hide y axis,
tick align=outside,
tick pos=left,
x grid style={white!69.0196078431373!black},
xmin=-0.5, xmax=39.5,
xtick style={color=black},
y dir=reverse,
y grid style={white!69.0196078431373!black},
ymin=-0.5, ymax=39.5,
ytick style={color=black}
]
\addplot graphics [includegraphics cmd=\pgfimage,xmin=-0.5, xmax=39.5, ymin=39.5, ymax=-0.5] {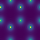};

\nextgroupplot[
colormap/viridis,
hide x axis,
hide y axis,
tick align=outside,
tick pos=left,
x grid style={white!69.0196078431373!black},
xmin=-0.5, xmax=39.5,
xtick style={color=black},
y dir=reverse,
y grid style={white!69.0196078431373!black},
ymin=-0.5, ymax=39.5,
ytick style={color=black}
]
\addplot graphics [includegraphics cmd=\pgfimage,xmin=-0.5, xmax=39.5, ymin=39.5, ymax=-0.5] {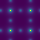};

\nextgroupplot[
colormap/viridis,
hide x axis,
hide y axis,
tick align=outside,
tick pos=left,
x grid style={white!69.0196078431373!black},
xmin=-0.5, xmax=39.5,
xtick style={color=black},
y dir=reverse,
y grid style={white!69.0196078431373!black},
ymin=-0.5, ymax=39.5,
ytick style={color=black}
]
\addplot graphics [includegraphics cmd=\pgfimage,xmin=-0.5, xmax=39.5, ymin=39.5, ymax=-0.5] {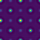};

\nextgroupplot[
hide y axis,
hide x axis,
tick align=outside,
tick pos=left,
x grid style={white!69.0196078431373!black},
xmin=-0.5, xmax=39.5,
xtick style={color=black},
y dir=reverse,
y grid style={white!69.0196078431373!black},
ymin=-0.5, ymax=39.5,
ytick style={color=black}
]
\addplot graphics [includegraphics cmd=\pgfimage,xmin=-0.5, xmax=39.5, ymin=39.5, ymax=-0.5] {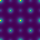};

\nextgroupplot[
colorbar horizontal,
colorbar style={xtick={0.2,0.8},xticklabels={0,0.5}},
colormap/viridis,
ticks=none,
ylabel = \myfont Slice 3,
ylabel style={rotate=0},
hide x axis,  
tick align=outside,
tick pos=left,
x grid style={white!69.0196078431373!black},
xmin=-0.5, xmax=39.5,
xtick style={color=black},
y dir=reverse,
y grid style={white!69.0196078431373!black},
ymin=-0.5, ymax=39.5,
ytick style={color=black}
]
\addplot graphics [includegraphics cmd=\pgfimage,xmin=-0.5, xmax=39.5, ymin=39.5, ymax=-0.5] {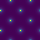};

\nextgroupplot[
colorbar horizontal,
colorbar style={xtick={0.2,0.8},xticklabels={0,0.03}},
colormap/viridis,
hide x axis,
hide y axis,
tick align=outside,
tick pos=left,
x grid style={white!69.0196078431373!black},
xmin=-0.5, xmax=39.5,
xtick style={color=black},
y dir=reverse,
y grid style={white!69.0196078431373!black},
ymin=-0.5, ymax=39.5,
ytick style={color=black}
]
\addplot graphics [includegraphics cmd=\pgfimage,xmin=-0.5, xmax=39.5, ymin=39.5, ymax=-0.5] {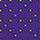};

\nextgroupplot[
colorbar horizontal,
colorbar style={xtick={0.3,0.9},xticklabels={-0.04,0.4}},
colormap/viridis,
hide x axis,
hide y axis,
tick align=outside,
tick pos=left,
x grid style={white!69.0196078431373!black},
xmin=-0.5, xmax=39.5,
xtick style={color=black},
y dir=reverse,
y grid style={white!69.0196078431373!black},
ymin=-0.5, ymax=39.5,
ytick style={color=black}
]
\addplot graphics [includegraphics cmd=\pgfimage,xmin=-0.5, xmax=39.5, ymin=39.5, ymax=-0.5] {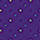};

\nextgroupplot[
colorbar horizontal,
colormap/viridis,
hide x axis,
hide y axis,
colorbar style={xtick={0.2,0.8},xticklabels={0,0.8}},
tick align=outside,
tick pos=left,
x grid style={white!69.0196078431373!black},
xmin=-0.5, xmax=39.5,
xtick style={color=black},
y dir=reverse,
y grid style={white!69.0196078431373!black},
ymin=-0.5, ymax=39.5,
ytick style={color=black}
]
\addplot graphics [includegraphics cmd=\pgfimage,xmin=-0.5, xmax=39.5, ymin=39.5, ymax=-0.5] {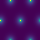};

\nextgroupplot[
colorbar horizontal,
colormap/viridis,
colorbar style={xtick={0.2,0.8},xticklabels={0,0.14}},
hide x axis,
hide y axis,
tick align=outside,
tick pos=left,
x grid style={white!69.0196078431373!black},
xmin=-0.5, xmax=39.5,
xtick style={color=black},
y dir=reverse,
y grid style={white!69.0196078431373!black},
ymin=-0.5, ymax=39.5,
ytick style={color=black}
]
\addplot graphics [includegraphics cmd=\pgfimage,xmin=-0.5, xmax=39.5, ymin=39.5, ymax=-0.5] {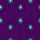};

\nextgroupplot[
colorbar horizontal,
colormap/viridis,
hide x axis,
hide y axis,
colorbar style={xtick={0.3,0.9},xticklabels={-0.05,0.5}},
tick align=outside,
tick pos=left,
x grid style={white!69.0196078431373!black},
xmin=-0.5, xmax=39.5,
xtick style={color=black},
y dir=reverse,
y grid style={white!69.0196078431373!black},
ymin=-0.5, ymax=39.5,
ytick style={color=black}
]
\addplot graphics [includegraphics cmd=\pgfimage,xmin=-0.5, xmax=39.5, ymin=39.5, ymax=-0.5] {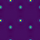};

\nextgroupplot[
colorbar horizontal,
colorbar style={xtick={0.25,0.8},xticklabels={0,0.7}},
colormap/viridis,
hide x axis,
hide y axis,
tick align=outside,
tick pos=left,
x grid style={white!69.0196078431373!black},
xmin=-0.5, xmax=39.5,
xtick style={color=black},
y dir=reverse,
y grid style={white!69.0196078431373!black},
ymin=-0.5, ymax=39.5,
ytick style={color=black}
]
\addplot graphics [includegraphics cmd=\pgfimage,xmin=-0.5, xmax=39.5, ymin=39.5, ymax=-0.5] {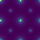};

\nextgroupplot[
colorbar horizontal,
colormap/viridis,
hide x axis,
hide y axis,
colorbar style={xtick={0.2,0.8},xticklabels={0,0.13}},
tick align=outside,
tick pos=left,
x grid style={white!69.0196078431373!black},
xmin=-0.5, xmax=39.5,
xtick style={color=black},
y dir=reverse,
y grid style={white!69.0196078431373!black},
ymin=-0.5, ymax=39.5,
ytick style={color=black}
]
\addplot graphics [includegraphics cmd=\pgfimage,xmin=-0.5, xmax=39.5, ymin=39.5, ymax=-0.5] {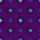};

\nextgroupplot[
colorbar horizontal,
colormap/viridis,
hide x axis,
hide y axis,
colorbar style={xtick={0.3,0.9},xticklabels={-0.04,0.8}},
tick align=outside,
tick pos=left,
x grid style={white!69.0196078431373!black},
xmin=-0.5, xmax=39.5,
xtick style={color=black},
y dir=reverse,
y grid style={white!69.0196078431373!black},
ymin=-0.5, ymax=39.5,
ytick style={color=black}
]
\addplot graphics [includegraphics cmd=\pgfimage,xmin=-0.5, xmax=39.5, ymin=39.5, ymax=-0.5] {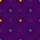};
\end{groupplot}

\end{tikzpicture}

%% file: Figure/Result/Figure_8/recon_error_w_woprobe_200.tex
\pgfplotsset{every tick label/.append style={font=\Large}}
\begin{tikzpicture}
\begin{groupplot}[group style={group size=1 by 2,vertical sep=7em}]
\nextgroupplot[
width=5in,
height=3in,
legend columns = 3,
title = \Large (a),
legend cell align={left},
legend style={at={(0.22,1.2)},anchor=south west,fill opacity=0.8, draw opacity=1, text opacity=1, draw=white!80!black},
tick align=outside,
tick pos=left,
x grid style={white!69.0196078431373!black},
xlabel={\Large Distance (nm)},
xmajorgrids,
xmin=-0.5, xmax=10.5,
xtick style={color=black},
y grid style={white!69.0196078431373!black},
ylabel={\Large Error},
ymajorgrids,
ymin=0, ymax=0.1,
ytick={0,0.05,0.1},
yticklabels={0,0.05,0.1},
ytick style={color=black}
]

\addplot [thick, black, dashed, mark=*, mark size=3, mark options={solid}]
table {%
0 0.0461472049355507
0.5 0.04306544115146
1 0.0280814611663421
1.5 0.022482726102074
2 0.0191821474581957
2.5 0.016516910555462
3 0.0155117788041631
3.5 0.0153865165387591
4 0.0153617952018976
4.5 0.015342591330409
5 0.015331226401031
5.5 0.0153321757291754
6 0.0153478089099129
6.5 0.0153966788202524
7 0.0154742185647289
7.5 0.0155264834562937
8 0.015519669589897
8.5 0.0154639997829994
9 0.0153929485629002
9.5 0.0153584911798437
10 0.0153389529635509
};
\addlegendentry{\Large GaAs}
\addplot [thick, red, dashed, mark=*, mark size=3, mark options={solid}]
table {%
0 0.0746101960539818
0.5 0.0756898472706477
1 0.0777792297303677
1.5 0.0418318373461564
2 0.0369240716099739
2.5 0.0313079288850228
3 0.0288414750248194
3.5 0.0286544666936
4 0.0291356947273016
4.5 0.0295995070288579
5 0.0297390166670084
5.5 0.0290179327130318
6 0.028073043252031
6.5 0.0274902172386646
7 0.027299901470542
7.5 0.0272964456429084
8 0.0272413833687703
8.5 0.0272806690384944
9 0.0273158966253201
9.5 0.0273448961476485
10 0.0279959073911111
};
\addlegendentry{\LARGE MoS$_2$}
\addplot [thick, blue, dashed, mark=*, mark size=3, mark options={solid}]
table {%
0 0.0580764052768548
0.5 0.0485976499815782
1 0.0421619427700837
1.5 0.0409612841904163
2 0.0396125142772992
2.5 0.0373420181373755
3 0.0355714398125807
3.5 0.0342381546894709
4 0.0333067905157804
4.5 0.0329723538209995
5 0.0328859618554513
5.5 0.0326801507423321
6 0.0324357636272907
6.5 0.0323646708081166
7 0.0324178946514924
7.5 0.0324997883290052
8 0.0325224337478479
8.5 0.0324799194931984
9 0.0324289624889692
9.5 0.0323778974513213
10 0.0323146333297094
};
\addlegendentry{\Large SrTiO$_3$}
\nextgroupplot[
width=5in,
height=3in,
title = \Large (b),
legend cell align={left},
legend style={fill opacity=0.8, draw opacity=1, text opacity=1, draw=white!80!black},
tick align=outside,
tick pos=left,
x grid style={white!69.0196078431373!black},
xlabel={\Large Distance (nm)},
xmajorgrids,
xmin=-0.5, xmax=10.5,
xtick style={color=black},
y grid style={white!69.0196078431373!black},
ylabel={\Large Error},
yticklabels={,,}
ymajorgrids,
ymin=0, ymax=0.1,
ytick={0,0.05,0.1},
yticklabels={0,0.05,0.1},
ytick style={color=black}
]
\addplot [thick, black, dashed, mark=*, mark size=3, mark options={solid}]
table {%
0 0.046637679139773
0.5 0.0466435129443804
1 0.0466615321735541
1.5 0.0466736579934756
2 0.0466810303429762
2.5 0.0466863674422105
3 0.0466965883970261
3.5 0.0467046710352103
4 0.0467069956163565
4.5 0.0467107171813647
5 0.0467122321327527
5.5 0.0467109071711699
6 0.0467084012925625
6.5 0.0467074376841386
7 0.0467077828943729
7.5 0.0467096716165543
8 0.0467118496696154
8.5 0.0467148311436176
9 0.0467179777721564
9.5 0.0467187849183877
10 0.0467167670528094
};
\addplot [thick, red, dashed, mark=*, mark size=3, mark options={solid}]
table {%
0 0.0701238264640172
0.5 0.0702928975224495
1 0.0704784480233987
1.5 0.07062936698397
2 0.0707246027886868
2.5 0.0707473841806253
3 0.0708391852676868
3.5 0.0709021451572577
4 0.0709359757602215
4.5 0.0709324243168036
5 0.0709442906081676
5.5 0.0709690389533838
6 0.0709968109925588
6.5 0.0710069860021273
7 0.0710298679769039
7.5 0.0710460046927134
8 0.0710479654371738
8.5 0.0710468379159768
9 0.0710651415089766
9.5 0.0710908186932405
10 0.0710663944482803
};
\addplot [thick, blue, dashed, mark=*, mark size=3, mark options={solid}]
table {%
0 0.0632386095821857
0.5 0.0632934719324112
1 0.0634095557034016
1.5 0.0634774727125963
2 0.0635185701151689
2.5 0.063550906876723
3 0.0636020600795746
3.5 0.0636161851386229
4 0.0636312228937944
4.5 0.0636611295243104
5 0.0636725537478924
5.5 0.0636654794216156
6 0.0636541818579038
6.5 0.0636521813770135
7 0.0636494271457195
7.5 0.0636535249650478
8 0.0636534914374352
8.5 0.0636594953636328
9 0.0636577519277732
9.5 0.0636563847462336
10 0.0636432208120823
};
\end{groupplot}

\end{tikzpicture}

%% file: Figure/Result/Figure_Both_Algo_Dist2nm/Both_Algo_Dist_2nm.tex
\pgfplotsset{every tick label/.append style={font=\myfont}}
\begin{tikzpicture}

\begin{groupplot}[group style={group size=9 by 3,vertical sep = 1.5em,horizontal sep=1.5em},width = 0.25*\textwidth, height=(0.25)*\textwidth]
\nextgroupplot[
colorbar style={ylabel={}},
colormap/viridis,
hide x axis,
ticks=none,
ylabel = \myfont Slice 1,
title = \myfont GT (a),
point meta max=0.458228230476379,
point meta min=-0.00135803618468344,
tick align=outside,
tick pos=left,
x grid style={white!69.0196078431373!black},
xmin=-0.5, xmax=39.5,
xtick style={color=black},
y dir=reverse,
y grid style={white!69.0196078431373!black},
ymin=-0.5, ymax=39.5,
ytick style={color=black}
]
\addplot graphics [includegraphics cmd=\pgfimage,xmin=-0.5, xmax=39.5, ymin=39.5, ymax=-0.5] {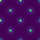};

\nextgroupplot[
colormap/viridis,
hide x axis,
hide y axis,
title = \myfont L (a),
point meta max=0.000374788360204548,
point meta min=-7.55657092668116e-05,
tick align=outside,
tick pos=left,
x grid style={white!69.0196078431373!black},
xmin=-0.5, xmax=39.5,
xtick style={color=black},
y dir=reverse,
y grid style={white!69.0196078431373!black},
ymin=-0.5, ymax=39.5,
ytick style={color=black}
]
\addplot graphics [includegraphics cmd=\pgfimage,xmin=-0.5, xmax=39.5, ymin=39.5, ymax=-0.5] {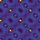};

\nextgroupplot[
colormap/viridis,
hide x axis,
hide y axis,
title = \myfont S (a),
colorbar style={xtick={0.3,0.8},xticklabels={-0.03,0.5}},
tick align=outside,
tick pos=left,
x grid style={white!69.0196078431373!black},
xmin=-0.5, xmax=39.5,
xtick style={color=black},
y dir=reverse,
y grid style={white!69.0196078431373!black},
ymin=-0.5, ymax=39.5,
ytick style={color=black}
]
\addplot graphics [includegraphics cmd=\pgfimage,xmin=-0.5, xmax=39.5, ymin=39.5, ymax=-0.5] {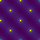};

\nextgroupplot[
colormap/viridis,
title = \myfont GT (b),
hide x axis,
hide y axis,
point meta max=0.427196651697159,
point meta min=0.00109258946031332,
tick align=outside,
tick pos=left,
x grid style={white!69.0196078431373!black},
xmin=-0.5, xmax=39.5,
xtick style={color=black},
y dir=reverse,
y grid style={white!69.0196078431373!black},
ymin=-0.5, ymax=39.5,
ytick style={color=black}
]
\addplot graphics [includegraphics cmd=\pgfimage,xmin=-0.5, xmax=39.5, ymin=39.5, ymax=-0.5] {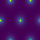};

\nextgroupplot[
colormap/viridis,
title = \myfont L (b),
hide x axis,
hide y axis,
point meta max=0.00241465889848769,
point meta min=-0.000440321920905262,
tick align=outside,
tick pos=left,
x grid style={white!69.0196078431373!black},
xmin=-0.5, xmax=39.5,
xtick style={color=black},
y dir=reverse,
y grid style={white!69.0196078431373!black},
ymin=-0.5, ymax=39.5,
ytick style={color=black}
]
\addplot graphics [includegraphics cmd=\pgfimage,xmin=-0.5, xmax=39.5, ymin=39.5, ymax=-0.5] {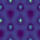};

\nextgroupplot[
colormap/viridis,
hide x axis,
hide y axis,
title = \myfont S (b),
point meta max=0.357105314731598,
point meta min=-0.0235141590237617,
tick align=outside,
tick pos=left,
x grid style={white!69.0196078431373!black},
xmin=-0.5, xmax=39.5,
xtick style={color=black},
y dir=reverse,
y grid style={white!69.0196078431373!black},
ymin=-0.5, ymax=39.5,
ytick style={color=black}
]
\addplot graphics [includegraphics cmd=\pgfimage,xmin=-0.5, xmax=39.5, ymin=39.5, ymax=-0.5] {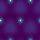};

\nextgroupplot[
colormap/viridis,
hide x axis,
hide y axis,
title = \myfont GT (c),
point meta max=0.672440230846405,
point meta min=0.0092761218547821,
tick align=outside,
tick pos=left,
x grid style={white!69.0196078431373!black},
xmin=-0.5, xmax=39.5,
xtick style={color=black},
y dir=reverse,
y grid style={white!69.0196078431373!black},
ymin=-0.5, ymax=39.5,
ytick style={color=black}
]
\addplot graphics [includegraphics cmd=\pgfimage,xmin=-0.5, xmax=39.5, ymin=39.5, ymax=-0.5] {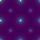};

\nextgroupplot[
colormap/viridis,
title = \myfont L (c),
hide x axis,
hide y axis,
point meta max=0.00245136790908873,
point meta min=-0.000281726388493553,
tick align=outside,
tick pos=left,
x grid style={white!69.0196078431373!black},
xmin=-0.5, xmax=39.5,
xtick style={color=black},
y dir=reverse,
y grid style={white!69.0196078431373!black},
ymin=-0.5, ymax=39.5,
ytick style={color=black}
]
\addplot graphics [includegraphics cmd=\pgfimage,xmin=-0.5, xmax=39.5, ymin=39.5, ymax=-0.5] {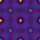};

\nextgroupplot[
colormap/viridis,
title = \myfont S (c),
hide x axis,
hide y axis,
point meta max=0.585296630859375,
point meta min=-0.0195708815008402,
tick align=outside,
tick pos=left,
x grid style={white!69.0196078431373!black},
xmin=-0.5, xmax=39.5,
xtick style={color=black},
y dir=reverse,
y grid style={white!69.0196078431373!black},
ymin=-0.5, ymax=39.5,
ytick style={color=black}
]
\addplot graphics [includegraphics cmd=\pgfimage,xmin=-0.5, xmax=39.5, ymin=39.5, ymax=-0.5] {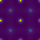};

\nextgroupplot[
colormap/viridis,
hide x axis,
ticks=none,
ylabel = \myfont Slice 2,
point meta max=0.504094004631042,
point meta min=-0.00156226276885718,
tick align=outside,
tick pos=left,
x grid style={white!69.0196078431373!black},
xmin=-0.5, xmax=39.5,
xtick style={color=black},
y dir=reverse,
y grid style={white!69.0196078431373!black},
ymin=-0.5, ymax=39.5,
ytick style={color=black}
]
\addplot graphics [includegraphics cmd=\pgfimage,xmin=-0.5, xmax=39.5, ymin=39.5, ymax=-0.5] {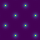};

\nextgroupplot[
colormap/viridis,
hide x axis,
hide y axis,
point meta max=0.000748052552808076,
point meta min=-0.00016763737949077,
tick align=outside,
tick pos=left,
x grid style={white!69.0196078431373!black},
xmin=-0.5, xmax=39.5,
xtick style={color=black},
y dir=reverse,
y grid style={white!69.0196078431373!black},
ymin=-0.5, ymax=39.5,
ytick style={color=black}
]
\addplot graphics [includegraphics cmd=\pgfimage,xmin=-0.5, xmax=39.5, ymin=39.5, ymax=-0.5] {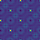};

\nextgroupplot[
colormap/viridis,
hide x axis,
hide y axis,
point meta max=0.469688415527344,
point meta min=-0.026364017277956,
tick align=outside,
tick pos=left,
x grid style={white!69.0196078431373!black},
xmin=-0.5, xmax=39.5,
xtick style={color=black},
y dir=reverse,
y grid style={white!69.0196078431373!black},
ymin=-0.5, ymax=39.5,
ytick style={color=black}
]
\addplot graphics [includegraphics cmd=\pgfimage,xmin=-0.5, xmax=39.5, ymin=39.5, ymax=-0.5] {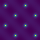};

\nextgroupplot[
colormap/viridis,
hide x axis,
hide y axis,
point meta max=0.830543518066406,
point meta min=0.00518810469657183,
tick align=outside,
tick pos=left,
x grid style={white!69.0196078431373!black},
xmin=-0.5, xmax=39.5,
xtick style={color=black},
y dir=reverse,
y grid style={white!69.0196078431373!black},
ymin=-0.5, ymax=39.5,
ytick style={color=black}
]
\addplot graphics [includegraphics cmd=\pgfimage,xmin=-0.5, xmax=39.5, ymin=39.5, ymax=-0.5] {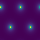};

\nextgroupplot[
colormap/viridis,
hide x axis,
hide y axis,
point meta max=0.00649806996807456,
point meta min=-0.000836835475638509,
tick align=outside,
tick pos=left,
x grid style={white!69.0196078431373!black},
xmin=-0.5, xmax=39.5,
xtick style={color=black},
y dir=reverse,
y grid style={white!69.0196078431373!black},
ymin=-0.5, ymax=39.5,
ytick style={color=black}
]
\addplot graphics [includegraphics cmd=\pgfimage,xmin=-0.5, xmax=39.5, ymin=39.5, ymax=-0.5] {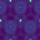};

\nextgroupplot[
colormap/viridis,
hide x axis,
hide y axis,
point meta max=0.726743519306183,
point meta min=-0.0382585227489471,
tick align=outside,
tick pos=left,
x grid style={white!69.0196078431373!black},
xmin=-0.5, xmax=39.5,
xtick style={color=black},
y dir=reverse,
y grid style={white!69.0196078431373!black},
ymin=-0.5, ymax=39.5,
ytick style={color=black}
]
\addplot graphics [includegraphics cmd=\pgfimage,xmin=-0.5, xmax=39.5, ymin=39.5, ymax=-0.5] {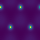};

\nextgroupplot[
colormap/viridis,
hide x axis,
hide y axis,
point meta max=0.495945811271667,
point meta min=0.00134838337544352,
tick align=outside,
tick pos=left,
x grid style={white!69.0196078431373!black},
xmin=-0.5, xmax=39.5,
xtick style={color=black},
y dir=reverse,
y grid style={white!69.0196078431373!black},
ymin=-0.5, ymax=39.5,
ytick style={color=black}
]
\addplot graphics [includegraphics cmd=\pgfimage,xmin=-0.5, xmax=39.5, ymin=39.5, ymax=-0.5] {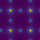};

\nextgroupplot[
colormap/viridis,
hide x axis,
hide y axis,
point meta max=0.00440111663192511,
point meta min=-0.000480574526591226,
tick align=outside,
tick pos=left,
x grid style={white!69.0196078431373!black},
xmin=-0.5, xmax=39.5,
xtick style={color=black},
y dir=reverse,
y grid style={white!69.0196078431373!black},
ymin=-0.5, ymax=39.5,
ytick style={color=black}
]
\addplot graphics [includegraphics cmd=\pgfimage,xmin=-0.5, xmax=39.5, ymin=39.5, ymax=-0.5] {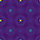};

\nextgroupplot[
colormap/viridis,
hide x axis,
hide y axis,
point meta max=0.429643630981445,
point meta min=-0.0306514501571655,
tick align=outside,
tick pos=left,
x grid style={white!69.0196078431373!black},
xmin=-0.5, xmax=39.5,
xtick style={color=black},
y dir=reverse,
y grid style={white!69.0196078431373!black},
ymin=-0.5, ymax=39.5,
ytick style={color=black}
]
\addplot graphics [includegraphics cmd=\pgfimage,xmin=-0.5, xmax=39.5, ymin=39.5, ymax=-0.5] {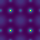};

\nextgroupplot[
colorbar horizontal,
colormap/viridis,
hide x axis,
ticks=none,
ylabel = \myfont Slice 3,
colorbar style={xtick={0.2,0.8},xticklabels={0,0.5}},
tick align=outside,
tick pos=left,
x grid style={white!69.0196078431373!black},
xmin=-0.5, xmax=39.5,
xtick style={color=black},
y dir=reverse,
y grid style={white!69.0196078431373!black},
ymin=-0.5, ymax=39.5,
ytick style={color=black}
]
\addplot graphics [includegraphics cmd=\pgfimage,xmin=-0.5, xmax=39.5, ymin=39.5, ymax=-0.5] {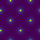};

\nextgroupplot[
colorbar horizontal,
colormap/viridis,
hide x axis,
hide y axis,
colorbar style={xtick={0.1,0.7},xticklabels={0,0.002}},
tick align=outside,
tick pos=left,
x grid style={white!69.0196078431373!black},
xmin=-0.5, xmax=39.5,
xtick style={color=black},
y dir=reverse,
y grid style={white!69.0196078431373!black},
ymin=-0.5, ymax=39.5,
ytick style={color=black}
]
\addplot graphics [includegraphics cmd=\pgfimage,xmin=-0.5, xmax=39.5, ymin=39.5, ymax=-0.5] {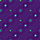};

\nextgroupplot[
colorbar horizontal,
colormap/viridis,
hide x axis,
hide y axis,
colorbar style={xtick={0.2,0.8},xticklabels={-0.03,0.5}},
tick align=outside,
tick pos=left,
x grid style={white!69.0196078431373!black},
xmin=-0.5, xmax=39.5,
xtick style={color=black},
y dir=reverse,
y grid style={white!69.0196078431373!black},
ymin=-0.5, ymax=39.5,
ytick style={color=black}
]
\addplot graphics [includegraphics cmd=\pgfimage,xmin=-0.5, xmax=39.5, ymin=39.5, ymax=-0.5] {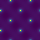};

\nextgroupplot[
colorbar horizontal,
colormap/viridis,
hide x axis,
hide y axis,
colorbar style={xtick={0.2,0.8},xticklabels={0,0.8}},
tick align=outside,
tick pos=left,
x grid style={white!69.0196078431373!black},
xmin=-0.5, xmax=39.5,
xtick style={color=black},
y dir=reverse,
y grid style={white!69.0196078431373!black},
ymin=-0.5, ymax=39.5,
ytick style={color=black}
]
\addplot graphics [includegraphics cmd=\pgfimage,xmin=-0.5, xmax=39.5, ymin=39.5, ymax=-0.5] {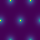};

\nextgroupplot[
colorbar horizontal,
colormap/viridis,
hide x axis,
hide y axis,
colorbar style={xtick={0.2,0.8},xticklabels={0,0.03}, xlabel = {} },
tick align=outside,
tick pos=left,
x grid style={white!69.0196078431373!black},
xmin=-0.5, xmax=39.5,
xtick style={color=black},
y dir=reverse,
y grid style={white!69.0196078431373!black},
ymin=-0.5, ymax=39.5,
ytick style={color=black}
]
\addplot graphics [includegraphics cmd=\pgfimage,xmin=-0.5, xmax=39.5, ymin=39.5, ymax=-0.5] {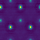};

\nextgroupplot[
colorbar horizontal,
colormap/viridis,
hide x axis,
hide y axis,
colorbar style={xtick={0.2,0.8},xticklabels={-0.04,0.73}},
tick align=outside,
tick pos=left,
x grid style={white!69.0196078431373!black},
xmin=-0.5, xmax=39.5,
xtick style={color=black},
y dir=reverse,
y grid style={white!69.0196078431373!black},
ymin=-0.5, ymax=39.5,
ytick style={color=black}
]
\addplot graphics [includegraphics cmd=\pgfimage,xmin=-0.5, xmax=39.5, ymin=39.5, ymax=-0.5] {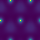};

\nextgroupplot[
colorbar horizontal,
colormap/viridis,
colorbar style={xtick={0.2,0.8},xticklabels={0,0.7}},
hide x axis,
hide y axis,
tick align=outside,
tick pos=left,
x grid style={white!69.0196078431373!black},
xmin=-0.5, xmax=39.5,
xtick style={color=black},
y dir=reverse,
y grid style={white!69.0196078431373!black},
ymin=-0.5, ymax=39.5,
ytick style={color=black}
]
\addplot graphics [includegraphics cmd=\pgfimage,xmin=-0.5, xmax=39.5, ymin=39.5, ymax=-0.5] {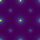};

\nextgroupplot[
colorbar horizontal,
colormap/viridis,
hide x axis,
hide y axis,
colorbar style={xtick={0.2,0.8},xticklabels={0,0.01}},
tick align=outside,
tick pos=left,
x grid style={white!69.0196078431373!black},
xmin=-0.5, xmax=39.5,
xtick style={color=black},
y dir=reverse,
y grid style={white!69.0196078431373!black},
ymin=-0.5, ymax=39.5,
ytick style={color=black}
]
\addplot graphics [includegraphics cmd=\pgfimage,xmin=-0.5, xmax=39.5, ymin=39.5, ymax=-0.5] {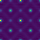};

\nextgroupplot[
colorbar horizontal,
colormap/viridis,
hide x axis,
hide y axis,
colorbar style={xtick={0.2,0.8},xticklabels={-0.03,0.6}},
tick align=outside,
tick pos=left,
x grid style={white!69.0196078431373!black},
xmin=-0.5, xmax=39.5,
xtick style={color=black},
y dir=reverse,
y grid style={white!69.0196078431373!black},
ymin=-0.5, ymax=39.5,
ytick style={color=black}
]
\addplot graphics [includegraphics cmd=\pgfimage,xmin=-0.5, xmax=39.5, ymin=39.5, ymax=-0.5] {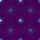};
\end{groupplot}

\end{tikzpicture}

%% file: Figure/Result/Figure_10/Fig_10_Increase_Distance_Sparse_Dec_1nm_2nm_new.tex
\pgfplotsset{every tick label/.append style={font=\myfont}}
\begin{tikzpicture}

\begin{groupplot}[group style={group size=9 by 9,vertical sep = 1em,horizontal sep=1em},width = 0.25*\textwidth, height=(0.25)*\textwidth]
\nextgroupplot[
colormap/viridis,
title = \myfont GT (a),
ticks=none,
hide x axis,
ticks=none,
ylabel = \myfont Slice 1,
ylabel style={rotate=0},
point meta max=0.458228230476379,
point meta min=-0.00135803618468344,
tick align=outside,
tick pos=left,
x grid style={white!69.0196078431373!black},
xmin=-0.5, xmax=39.5,
xtick style={color=black},
y dir=reverse,
y grid style={white!69.0196078431373!black},
ymin=-0.5, ymax=39.5,
ytick style={color=black}
]
\addplot graphics [includegraphics cmd=\pgfimage,xmin=-0.5, xmax=39.5, ymin=39.5, ymax=-0.5] {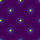};

\nextgroupplot[
title = \myfont 1 nm (a),
colormap/viridis,
hide x axis,
hide y axis,
point meta max=0.431624948978424,
point meta min=-0.0481205657124519,
tick align=outside,
tick pos=left,
x grid style={white!69.0196078431373!black},
xmin=-0.5, xmax=39.5,
xtick style={color=black},
y dir=reverse,
y grid style={white!69.0196078431373!black},
ymin=-0.5, ymax=39.5,
ytick style={color=black}
]
\addplot graphics [includegraphics cmd=\pgfimage,xmin=-0.5, xmax=39.5, ymin=39.5, ymax=-0.5] {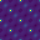};

\nextgroupplot[
title = \myfont 2 nm (a),
colormap/viridis,
hide x axis,
hide y axis,
point meta max=0.416686326265335,
point meta min=-0.0301901753991842,
tick align=outside,
tick pos=left,
x grid style={white!69.0196078431373!black},
xmin=-0.5, xmax=39.5,
xtick style={color=black},
y dir=reverse,
y grid style={white!69.0196078431373!black},
ymin=-0.5, ymax=39.5,
ytick style={color=black}
]
\addplot graphics [includegraphics cmd=\pgfimage,xmin=-0.5, xmax=39.5, ymin=39.5, ymax=-0.5] {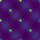};

\nextgroupplot[
title = \myfont GT (b),
colormap/viridis,
hide x axis,
hide y axis,
point meta max=0.427196651697159,
point meta min=0.00109258946031332,
tick align=outside,
tick pos=left,
x grid style={white!69.0196078431373!black},
xmin=-0.5, xmax=39.5,
xtick style={color=black},
y dir=reverse,
y grid style={white!69.0196078431373!black},
ymin=-0.5, ymax=39.5,
ytick style={color=black}
]
\addplot graphics [includegraphics cmd=\pgfimage,xmin=-0.5, xmax=39.5, ymin=39.5, ymax=-0.5] {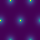};

\nextgroupplot[
title = \myfont 1 nm (b),
colormap/viridis,
hide x axis,
hide y axis,
point meta max=0.311410903930664,
point meta min=-0.0388218462467194,
tick align=outside,
tick pos=left,
x grid style={white!69.0196078431373!black},
xmin=-0.5, xmax=39.5,
xtick style={color=black},
y dir=reverse,
y grid style={white!69.0196078431373!black},
ymin=-0.5, ymax=39.5,
ytick style={color=black}
]
\addplot graphics [includegraphics cmd=\pgfimage,xmin=-0.5, xmax=39.5, ymin=39.5, ymax=-0.5] {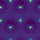};

\nextgroupplot[
title = \myfont 2 nm (b),
colorbar style={xtick={3.46944695195361e-18,0.25},xticklabels={0.00,0.25}},
colormap/viridis,
hide x axis,
hide y axis,
point meta max=0.324361383914948,
point meta min=-0.0273035801947117,
tick align=outside,
tick pos=left,
x grid style={white!69.0196078431373!black},
xmin=-0.5, xmax=39.5,
xtick style={color=black},
y dir=reverse,
y grid style={white!69.0196078431373!black},
ymin=-0.5, ymax=39.5,
ytick style={color=black}
]
\addplot graphics [includegraphics cmd=\pgfimage,xmin=-0.5, xmax=39.5, ymin=39.5, ymax=-0.5] {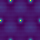};

\nextgroupplot[
title = \myfont GT (c),
colorbar style={xtick={0.25,0.5},xticklabels={0.25,0.50}},
colormap/viridis,
hide x axis,
hide y axis,
point meta max=0.672440230846405,
point meta min=0.0092761218547821,
tick align=outside,
tick pos=left,
x grid style={white!69.0196078431373!black},
xmin=-0.5, xmax=39.5,
xtick style={color=black},
y dir=reverse,
y grid style={white!69.0196078431373!black},
ymin=-0.5, ymax=39.5,
ytick style={color=black}
]
\addplot graphics [includegraphics cmd=\pgfimage,xmin=-0.5, xmax=39.5, ymin=39.5, ymax=-0.5] {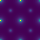};

\nextgroupplot[
title = \myfont 1 nm (c),
colormap/viridis,
hide x axis,
hide y axis,
point meta max=0.51936799287796,
point meta min=-0.0148673774674535,
tick align=outside,
tick pos=left,
x grid style={white!69.0196078431373!black},
xmin=-0.5, xmax=39.5,
xtick style={color=black},
y dir=reverse,
y grid style={white!69.0196078431373!black},
ymin=-0.5, ymax=39.5,
ytick style={color=black}
]
\addplot graphics [includegraphics cmd=\pgfimage,xmin=-0.5, xmax=39.5, ymin=39.5, ymax=-0.5] {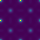};

\nextgroupplot[
title = \myfont 2 nm (c),
colormap/viridis,
hide x axis,
hide y axis,
point meta max=0.477798610925674,
point meta min=-0.0128755578771234,
tick align=outside,
tick pos=left,
x grid style={white!69.0196078431373!black},
xmin=-0.5, xmax=39.5,
xtick style={color=black},
y dir=reverse,
y grid style={white!69.0196078431373!black},
ymin=-0.5, ymax=39.5,
ytick style={color=black}
]
\addplot graphics [includegraphics cmd=\pgfimage,xmin=-0.5, xmax=39.5, ymin=39.5, ymax=-0.5] {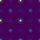};

\nextgroupplot[
colormap/viridis,
hide x axis,
ticks=none,
ylabel = \myfont Slice 2,
ylabel style={rotate=0},
point meta max=0.504094004631042,
point meta min=-0.00156226276885718,
tick align=outside,
tick pos=left,
x grid style={white!69.0196078431373!black},
xmin=-0.5, xmax=39.5,
xtick style={color=black},
y dir=reverse,
y grid style={white!69.0196078431373!black},
ymin=-0.5, ymax=39.5,
ytick style={color=black}
]
\addplot graphics [includegraphics cmd=\pgfimage,xmin=-0.5, xmax=39.5, ymin=39.5, ymax=-0.5] {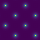};

\nextgroupplot[
colormap/viridis,
hide x axis,
hide y axis,
point meta max=0.295081824064255,
point meta min=-0.0387080200016499,
tick align=outside,
tick pos=left,
x grid style={white!69.0196078431373!black},
xmin=-0.5, xmax=39.5,
xtick style={color=black},
y dir=reverse,
y grid style={white!69.0196078431373!black},
ymin=-0.5, ymax=39.5,
ytick style={color=black}
]
\addplot graphics [includegraphics cmd=\pgfimage,xmin=-0.5, xmax=39.5, ymin=39.5, ymax=-0.5] {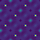};

\nextgroupplot[
colormap/viridis,
hide x axis,
hide y axis,
point meta max=0.416424006223679,
point meta min=-0.0346187986433506,
tick align=outside,
tick pos=left,
x grid style={white!69.0196078431373!black},
xmin=-0.5, xmax=39.5,
xtick style={color=black},
y dir=reverse,
y grid style={white!69.0196078431373!black},
ymin=-0.5, ymax=39.5,
ytick style={color=black}
]
\addplot graphics [includegraphics cmd=\pgfimage,xmin=-0.5, xmax=39.5, ymin=39.5, ymax=-0.5] {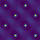};

\nextgroupplot[
colormap/viridis,
hide x axis,
hide y axis,
point meta max=0.830543518066406,
point meta min=0.00518810469657183,
tick align=outside,
tick pos=left,
x grid style={white!69.0196078431373!black},
xmin=-0.5, xmax=39.5,
xtick style={color=black},
y dir=reverse,
y grid style={white!69.0196078431373!black},
ymin=-0.5, ymax=39.5,
ytick style={color=black}
]
\addplot graphics [includegraphics cmd=\pgfimage,xmin=-0.5, xmax=39.5, ymin=39.5, ymax=-0.5] {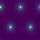};

\nextgroupplot[
colormap/viridis,
hide x axis,
hide y axis,
point meta max=0.303753346204758,
point meta min=-0.034093551337719,
tick align=outside,
tick pos=left,
x grid style={white!69.0196078431373!black},
xmin=-0.5, xmax=39.5,
xtick style={color=black},
y dir=reverse,
y grid style={white!69.0196078431373!black},
ymin=-0.5, ymax=39.5,
ytick style={color=black}
]
\addplot graphics [includegraphics cmd=\pgfimage,xmin=-0.5, xmax=39.5, ymin=39.5, ymax=-0.5] {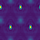};

\nextgroupplot[
colormap/viridis,
hide x axis,
hide y axis,
point meta max=0.649460911750793,
point meta min=-0.0401431247591972,
tick align=outside,
tick pos=left,
x grid style={white!69.0196078431373!black},
xmin=-0.5, xmax=39.5,
xtick style={color=black},
y dir=reverse,
y grid style={white!69.0196078431373!black},
ymin=-0.5, ymax=39.5,
ytick style={color=black}
]
\addplot graphics [includegraphics cmd=\pgfimage,xmin=-0.5, xmax=39.5, ymin=39.5, ymax=-0.5] {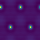};

\nextgroupplot[
colormap/viridis,
hide x axis,
hide y axis,
point meta max=0.495945811271667,
point meta min=0.00134838337544352,
tick align=outside,
tick pos=left,
x grid style={white!69.0196078431373!black},
xmin=-0.5, xmax=39.5,
xtick style={color=black},
y dir=reverse,
y grid style={white!69.0196078431373!black},
ymin=-0.5, ymax=39.5,
ytick style={color=black}
]
\addplot graphics [includegraphics cmd=\pgfimage,xmin=-0.5, xmax=39.5, ymin=39.5, ymax=-0.5] {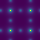};

\nextgroupplot[
colormap/viridis,
hide x axis,
hide y axis,
point meta max=0.364638924598694,
point meta min=-0.017701119184494,
tick align=outside,
tick pos=left,
x grid style={white!69.0196078431373!black},
xmin=-0.5, xmax=39.5,
xtick style={color=black},
y dir=reverse,
y grid style={white!69.0196078431373!black},
ymin=-0.5, ymax=39.5,
ytick style={color=black}
]
\addplot graphics [includegraphics cmd=\pgfimage,xmin=-0.5, xmax=39.5, ymin=39.5, ymax=-0.5] {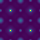};

\nextgroupplot[
colormap/viridis,
hide x axis,
hide y axis,
point meta max=0.398154318332672,
point meta min=-0.0222057495266199,
tick align=outside,
tick pos=left,
x grid style={white!69.0196078431373!black},
xmin=-0.5, xmax=39.5,
xtick style={color=black},
y dir=reverse,
y grid style={white!69.0196078431373!black},
ymin=-0.5, ymax=39.5,
ytick style={color=black}
]
\addplot graphics [includegraphics cmd=\pgfimage,xmin=-0.5, xmax=39.5, ymin=39.5, ymax=-0.5] {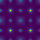};

\nextgroupplot[
colormap/viridis,
hide x axis,
ticks=none,
ylabel = \myfont Slice 3,
ylabel style={rotate=0},
point meta max=0.458228230476379,
point meta min=-0.00135803618468344,
tick align=outside,
tick pos=left,
x grid style={white!69.0196078431373!black},
xmin=-0.5, xmax=39.5,
xtick style={color=black},
y dir=reverse,
y grid style={white!69.0196078431373!black},
ymin=-0.5, ymax=39.5,
ytick style={color=black}
]
\addplot graphics [includegraphics cmd=\pgfimage,xmin=-0.5, xmax=39.5, ymin=39.5, ymax=-0.5] {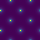};

\nextgroupplot[
colormap/viridis,
hide x axis,
hide y axis,
point meta max=0.23021487891674,
point meta min=-0.0193400308489799,
tick align=outside,
tick pos=left,
x grid style={white!69.0196078431373!black},
xmin=-0.5, xmax=39.5,
xtick style={color=black},
y dir=reverse,
y grid style={white!69.0196078431373!black},
ymin=-0.5, ymax=39.5,
ytick style={color=black}
]
\addplot graphics [includegraphics cmd=\pgfimage,xmin=-0.5, xmax=39.5, ymin=39.5, ymax=-0.5] {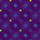};

\nextgroupplot[
colormap/viridis,
hide x axis,
hide y axis,
point meta max=0.3645398914814,
point meta min=-0.0295844972133636,
tick align=outside,
tick pos=left,
x grid style={white!69.0196078431373!black},
xmin=-0.5, xmax=39.5,
xtick style={color=black},
y dir=reverse,
y grid style={white!69.0196078431373!black},
ymin=-0.5, ymax=39.5,
ytick style={color=black}
]
\addplot graphics [includegraphics cmd=\pgfimage,xmin=-0.5, xmax=39.5, ymin=39.5, ymax=-0.5] {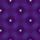};

\nextgroupplot[
colormap/viridis,
hide x axis,
hide y axis,
point meta max=0.427196651697159,
point meta min=0.00109258946031332,
tick align=outside,
tick pos=left,
x grid style={white!69.0196078431373!black},
xmin=-0.5, xmax=39.5,
xtick style={color=black},
y dir=reverse,
y grid style={white!69.0196078431373!black},
ymin=-0.5, ymax=39.5,
ytick style={color=black}
]
\addplot graphics [includegraphics cmd=\pgfimage,xmin=-0.5, xmax=39.5, ymin=39.5, ymax=-0.5] {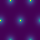};

\nextgroupplot[
colormap/viridis,
hide x axis,
hide y axis,
point meta max=0.29066064953804,
point meta min=-0.0292406994849443,
tick align=outside,
tick pos=left,
x grid style={white!69.0196078431373!black},
xmin=-0.5, xmax=39.5,
xtick style={color=black},
y dir=reverse,
y grid style={white!69.0196078431373!black},
ymin=-0.5, ymax=39.5,
ytick style={color=black}
]
\addplot graphics [includegraphics cmd=\pgfimage,xmin=-0.5, xmax=39.5, ymin=39.5, ymax=-0.5] {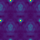};

\nextgroupplot[
colorbar style={xtick={3.46944695195361e-18,0.25},xticklabels={0.00,0.25}},
colormap/viridis,
hide x axis,
hide y axis,
point meta max=0.344182878732681,
point meta min=-0.0290266275405884,
tick align=outside,
tick pos=left,
x grid style={white!69.0196078431373!black},
xmin=-0.5, xmax=39.5,
xtick style={color=black},
y dir=reverse,
y grid style={white!69.0196078431373!black},
ymin=-0.5, ymax=39.5,
ytick style={color=black}
]
\addplot graphics [includegraphics cmd=\pgfimage,xmin=-0.5, xmax=39.5, ymin=39.5, ymax=-0.5] {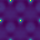};

\nextgroupplot[
colorbar style={xtick={0.25,0.5},xticklabels={0.25,0.50}},
colormap/viridis,
hide x axis,
hide y axis,
point meta max=0.672440230846405,
point meta min=0.0092761218547821,
tick align=outside,
tick pos=left,
x grid style={white!69.0196078431373!black},
xmin=-0.5, xmax=39.5,
xtick style={color=black},
y dir=reverse,
y grid style={white!69.0196078431373!black},
ymin=-0.5, ymax=39.5,
ytick style={color=black}
]
\addplot graphics [includegraphics cmd=\pgfimage,xmin=-0.5, xmax=39.5, ymin=39.5, ymax=-0.5] {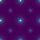};

\nextgroupplot[
colormap/viridis,
hide x axis,
hide y axis,
point meta max=0.434961974620819,
point meta min=-0.0200665649026632,
tick align=outside,
tick pos=left,
x grid style={white!69.0196078431373!black},
xmin=-0.5, xmax=39.5,
xtick style={color=black},
y dir=reverse,
y grid style={white!69.0196078431373!black},
ymin=-0.5, ymax=39.5,
ytick style={color=black}
]
\addplot graphics [includegraphics cmd=\pgfimage,xmin=-0.5, xmax=39.5, ymin=39.5, ymax=-0.5] {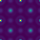};

\nextgroupplot[
colormap/viridis,
hide x axis,
hide y axis,
point meta max=0.433829784393311,
point meta min=-0.0264446213841438,
tick align=outside,
tick pos=left,
x grid style={white!69.0196078431373!black},
xmin=-0.5, xmax=39.5,
xtick style={color=black},
y dir=reverse,
y grid style={white!69.0196078431373!black},
ymin=-0.5, ymax=39.5,
ytick style={color=black}
]
\addplot graphics [includegraphics cmd=\pgfimage,xmin=-0.5, xmax=39.5, ymin=39.5, ymax=-0.5] {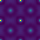};

\nextgroupplot[
colormap/viridis,
hide x axis,
ticks=none,
ylabel = \myfont Slice 4,
ylabel style={rotate=0},
point meta max=0.504094004631042,
point meta min=-0.00156226276885718,
tick align=outside,
tick pos=left,
x grid style={white!69.0196078431373!black},
xmin=-0.5, xmax=39.5,
xtick style={color=black},
y dir=reverse,
y grid style={white!69.0196078431373!black},
ymin=-0.5, ymax=39.5,
ytick style={color=black}
]
\addplot graphics [includegraphics cmd=\pgfimage,xmin=-0.5, xmax=39.5, ymin=39.5, ymax=-0.5] {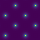};

\nextgroupplot[
colorbar style={xtick={-3.46944695195361e-18,0.25},xticklabels={0.00,0.25}},
colormap/viridis,
hide x axis,
hide y axis,
point meta max=0.258146047592163,
point meta min=-0.0247435346245766,
tick align=outside,
tick pos=left,
x grid style={white!69.0196078431373!black},
xmin=-0.5, xmax=39.5,
xtick style={color=black},
y dir=reverse,
y grid style={white!69.0196078431373!black},
ymin=-0.5, ymax=39.5,
ytick style={color=black}
]
\addplot graphics [includegraphics cmd=\pgfimage,xmin=-0.5, xmax=39.5, ymin=39.5, ymax=-0.5] {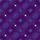};

\nextgroupplot[
colorbar style={xtick={3.46944695195361e-18,0.25},xticklabels={0.00,0.25}},
colormap/viridis,
hide x axis,
hide y axis,
point meta max=0.399349689483643,
point meta min=-0.0279363412410021,
tick align=outside,
tick pos=left,
x grid style={white!69.0196078431373!black},
xmin=-0.5, xmax=39.5,
xtick style={color=black},
y dir=reverse,
y grid style={white!69.0196078431373!black},
ymin=-0.5, ymax=39.5,
ytick style={color=black}
]
\addplot graphics [includegraphics cmd=\pgfimage,xmin=-0.5, xmax=39.5, ymin=39.5, ymax=-0.5] {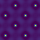};

\nextgroupplot[
colormap/viridis,
hide x axis,
hide y axis,
point meta max=0.427196651697159,
point meta min=0.00109258946031332,
tick align=outside,
tick pos=left,
x grid style={white!69.0196078431373!black},
xmin=-0.5, xmax=39.5,
xtick style={color=black},
y dir=reverse,
y grid style={white!69.0196078431373!black},
ymin=-0.5, ymax=39.5,
ytick style={color=black}
]
\addplot graphics [includegraphics cmd=\pgfimage,xmin=-0.5, xmax=39.5, ymin=39.5, ymax=-0.5] {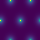};

\nextgroupplot[
colormap/viridis,
hide x axis,
hide y axis,
point meta max=0.267505288124084,
point meta min=-0.0254802536219358,
tick align=outside,
tick pos=left,
x grid style={white!69.0196078431373!black},
xmin=-0.5, xmax=39.5,
xtick style={color=black},
y dir=reverse,
y grid style={white!69.0196078431373!black},
ymin=-0.5, ymax=39.5,
ytick style={color=black}
]
\addplot graphics [includegraphics cmd=\pgfimage,xmin=-0.5, xmax=39.5, ymin=39.5, ymax=-0.5] {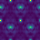};

\nextgroupplot[
colormap/viridis,
hide x axis,
hide y axis,
point meta max=0.341051310300827,
point meta min=-0.0219536554068327,
tick align=outside,
tick pos=left,
x grid style={white!69.0196078431373!black},
xmin=-0.5, xmax=39.5,
xtick style={color=black},
y dir=reverse,
y grid style={white!69.0196078431373!black},
ymin=-0.5, ymax=39.5,
ytick style={color=black}
]
\addplot graphics [includegraphics cmd=\pgfimage,xmin=-0.5, xmax=39.5, ymin=39.5, ymax=-0.5] {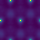};

\nextgroupplot[
colormap/viridis,
hide x axis,
hide y axis,
point meta max=0.495945811271667,
point meta min=0.00134838337544352,
tick align=outside,
tick pos=left,
x grid style={white!69.0196078431373!black},
xmin=-0.5, xmax=39.5,
xtick style={color=black},
y dir=reverse,
y grid style={white!69.0196078431373!black},
ymin=-0.5, ymax=39.5,
ytick style={color=black}
]
\addplot graphics [includegraphics cmd=\pgfimage,xmin=-0.5, xmax=39.5, ymin=39.5, ymax=-0.5] {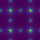};

\nextgroupplot[
colormap/viridis,
hide x axis,
hide y axis,
point meta max=0.371323347091675,
point meta min=-0.022379020228982,
tick align=outside,
tick pos=left,
x grid style={white!69.0196078431373!black},
xmin=-0.5, xmax=39.5,
xtick style={color=black},
y dir=reverse,
y grid style={white!69.0196078431373!black},
ymin=-0.5, ymax=39.5,
ytick style={color=black}
]
\addplot graphics [includegraphics cmd=\pgfimage,xmin=-0.5, xmax=39.5, ymin=39.5, ymax=-0.5] {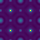};

\nextgroupplot[
colormap/viridis,
hide x axis,
hide y axis,
point meta max=0.369010508060455,
point meta min=-0.0211786832660437,
tick align=outside,
tick pos=left,
x grid style={white!69.0196078431373!black},
xmin=-0.5, xmax=39.5,
xtick style={color=black},
y dir=reverse,
y grid style={white!69.0196078431373!black},
ymin=-0.5, ymax=39.5,
ytick style={color=black}
]
\addplot graphics [includegraphics cmd=\pgfimage,xmin=-0.5, xmax=39.5, ymin=39.5, ymax=-0.5] {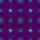};

\nextgroupplot[
colormap/viridis,
hide x axis,
ticks=none,
ylabel = \myfont Slice 5,
ylabel style={rotate=0},
point meta max=0.458228230476379,
point meta min=-0.00135803618468344,
tick align=outside,
tick pos=left,
x grid style={white!69.0196078431373!black},
xmin=-0.5, xmax=39.5,
xtick style={color=black},
y dir=reverse,
y grid style={white!69.0196078431373!black},
ymin=-0.5, ymax=39.5,
ytick style={color=black}
]
\addplot graphics [includegraphics cmd=\pgfimage,xmin=-0.5, xmax=39.5, ymin=39.5, ymax=-0.5] {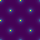};

\nextgroupplot[
colorbar style={xtick={6.93889390390723e-18,0.2},xticklabels={0.0,0.2}},
colormap/viridis,
hide x axis,
hide y axis,
point meta max=0.225702479481697,
point meta min=-0.0356566794216633,
tick align=outside,
tick pos=left,
x grid style={white!69.0196078431373!black},
xmin=-0.5, xmax=39.5,
xtick style={color=black},
y dir=reverse,
y grid style={white!69.0196078431373!black},
ymin=-0.5, ymax=39.5,
ytick style={color=black}
]
\addplot graphics [includegraphics cmd=\pgfimage,xmin=-0.5, xmax=39.5, ymin=39.5, ymax=-0.5] {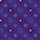};

\nextgroupplot[
colormap/viridis,
hide x axis,
hide y axis,
point meta max=0.36752849817276,
point meta min=-0.0280124898999929,
tick align=outside,
tick pos=left,
x grid style={white!69.0196078431373!black},
xmin=-0.5, xmax=39.5,
xtick style={color=black},
y dir=reverse,
y grid style={white!69.0196078431373!black},
ymin=-0.5, ymax=39.5,
ytick style={color=black}
]
\addplot graphics [includegraphics cmd=\pgfimage,xmin=-0.5, xmax=39.5, ymin=39.5, ymax=-0.5] {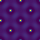};

\nextgroupplot[
colormap/viridis,
hide x axis,
hide y axis,
point meta max=0.830543518066406,
point meta min=0.00518810469657183,
tick align=outside,
tick pos=left,
x grid style={white!69.0196078431373!black},
xmin=-0.5, xmax=39.5,
xtick style={color=black},
y dir=reverse,
y grid style={white!69.0196078431373!black},
ymin=-0.5, ymax=39.5,
ytick style={color=black}
]
\addplot graphics [includegraphics cmd=\pgfimage,xmin=-0.5, xmax=39.5, ymin=39.5, ymax=-0.5] {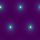};

\nextgroupplot[
colormap/viridis,
hide x axis,
hide y axis,
point meta max=0.418602406978607,
point meta min=-0.0307930801063776,
tick align=outside,
tick pos=left,
x grid style={white!69.0196078431373!black},
xmin=-0.5, xmax=39.5,
xtick style={color=black},
y dir=reverse,
y grid style={white!69.0196078431373!black},
ymin=-0.5, ymax=39.5,
ytick style={color=black}
]
\addplot graphics [includegraphics cmd=\pgfimage,xmin=-0.5, xmax=39.5, ymin=39.5, ymax=-0.5] {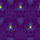};

\nextgroupplot[
colormap/viridis,
hide x axis,
hide y axis,
point meta max=0.596176981925964,
point meta min=-0.0441955663263798,
tick align=outside,
tick pos=left,
x grid style={white!69.0196078431373!black},
xmin=-0.5, xmax=39.5,
xtick style={color=black},
y dir=reverse,
y grid style={white!69.0196078431373!black},
ymin=-0.5, ymax=39.5,
ytick style={color=black}
]
\addplot graphics [includegraphics cmd=\pgfimage,xmin=-0.5, xmax=39.5, ymin=39.5, ymax=-0.5] {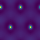};

\nextgroupplot[
colorbar style={xtick={0.25,0.5},xticklabels={0.25,0.50}},
colormap/viridis,
hide x axis,
hide y axis,
point meta max=0.672440230846405,
point meta min=0.0092761218547821,
tick align=outside,
tick pos=left,
x grid style={white!69.0196078431373!black},
xmin=-0.5, xmax=39.5,
xtick style={color=black},
y dir=reverse,
y grid style={white!69.0196078431373!black},
ymin=-0.5, ymax=39.5,
ytick style={color=black}
]
\addplot graphics [includegraphics cmd=\pgfimage,xmin=-0.5, xmax=39.5, ymin=39.5, ymax=-0.5] {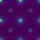};

\nextgroupplot[
colormap/viridis,
hide x axis,
hide y axis,
point meta max=0.42089056968689,
point meta min=-0.0224141012877226,
tick align=outside,
tick pos=left,
x grid style={white!69.0196078431373!black},
xmin=-0.5, xmax=39.5,
xtick style={color=black},
y dir=reverse,
y grid style={white!69.0196078431373!black},
ymin=-0.5, ymax=39.5,
ytick style={color=black}
]
\addplot graphics [includegraphics cmd=\pgfimage,xmin=-0.5, xmax=39.5, ymin=39.5, ymax=-0.5] {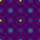};

\nextgroupplot[
colormap/viridis,
hide x axis,
hide y axis,
point meta max=0.464630037546158,
point meta min=-0.0299720410257578,
tick align=outside,
tick pos=left,
x grid style={white!69.0196078431373!black},
xmin=-0.5, xmax=39.5,
xtick style={color=black},
y dir=reverse,
y grid style={white!69.0196078431373!black},
ymin=-0.5, ymax=39.5,
ytick style={color=black}
]
\addplot graphics [includegraphics cmd=\pgfimage,xmin=-0.5, xmax=39.5, ymin=39.5, ymax=-0.5] {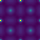};

\nextgroupplot[
colormap/viridis,
hide x axis,
ticks=none,
ylabel = \myfont Slice 6,
ylabel style={rotate=0},
point meta max=0.504094004631042,
point meta min=-0.00156226276885718,
tick align=outside,
tick pos=left,
x grid style={white!69.0196078431373!black},
xmin=-0.5, xmax=39.5,
xtick style={color=black},
y dir=reverse,
y grid style={white!69.0196078431373!black},
ymin=-0.5, ymax=39.5,
ytick style={color=black}
]
\addplot graphics [includegraphics cmd=\pgfimage,xmin=-0.5, xmax=39.5, ymin=39.5, ymax=-0.5] {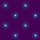};

\nextgroupplot[
colormap/viridis,
hide x axis,
hide y axis,
point meta max=0.252851128578186,
point meta min=-0.025577237829566,
tick align=outside,
tick pos=left,
x grid style={white!69.0196078431373!black},
xmin=-0.5, xmax=39.5,
xtick style={color=black},
y dir=reverse,
y grid style={white!69.0196078431373!black},
ymin=-0.5, ymax=39.5,
ytick style={color=black}
]
\addplot graphics [includegraphics cmd=\pgfimage,xmin=-0.5, xmax=39.5, ymin=39.5, ymax=-0.5] {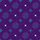};

\nextgroupplot[
colormap/viridis,
hide x axis,
hide y axis,
point meta max=0.400350749492645,
point meta min=-0.0274834614247084,
tick align=outside,
tick pos=left,
x grid style={white!69.0196078431373!black},
xmin=-0.5, xmax=39.5,
xtick style={color=black},
y dir=reverse,
y grid style={white!69.0196078431373!black},
ymin=-0.5, ymax=39.5,
ytick style={color=black}
]
\addplot graphics [includegraphics cmd=\pgfimage,xmin=-0.5, xmax=39.5, ymin=39.5, ymax=-0.5] {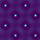};

\nextgroupplot[
colormap/viridis,
hide x axis,
hide y axis,
point meta max=0.427196651697159,
point meta min=0.00109258946031332,
tick align=outside,
tick pos=left,
x grid style={white!69.0196078431373!black},
xmin=-0.5, xmax=39.5,
xtick style={color=black},
y dir=reverse,
y grid style={white!69.0196078431373!black},
ymin=-0.5, ymax=39.5,
ytick style={color=black}
]
\addplot graphics [includegraphics cmd=\pgfimage,xmin=-0.5, xmax=39.5, ymin=39.5, ymax=-0.5] {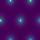};

\nextgroupplot[
colormap/viridis,
hide x axis,
hide y axis,
point meta max=0.229150623083115,
point meta min=-0.0435585752129555,
tick align=outside,
tick pos=left,
x grid style={white!69.0196078431373!black},
xmin=-0.5, xmax=39.5,
xtick style={color=black},
y dir=reverse,
y grid style={white!69.0196078431373!black},
ymin=-0.5, ymax=39.5,
ytick style={color=black}
]
\addplot graphics [includegraphics cmd=\pgfimage,xmin=-0.5, xmax=39.5, ymin=39.5, ymax=-0.5] {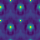};

\nextgroupplot[
colormap/viridis,
hide x axis,
hide y axis,
point meta max=0.279523700475693,
point meta min=-0.0411471612751484,
tick align=outside,
tick pos=left,
x grid style={white!69.0196078431373!black},
xmin=-0.5, xmax=39.5,
xtick style={color=black},
y dir=reverse,
y grid style={white!69.0196078431373!black},
ymin=-0.5, ymax=39.5,
ytick style={color=black}
]
\addplot graphics [includegraphics cmd=\pgfimage,xmin=-0.5, xmax=39.5, ymin=39.5, ymax=-0.5] {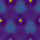};

\nextgroupplot[
colormap/viridis,
hide x axis,
hide y axis,
point meta max=0.495945811271667,
point meta min=0.00134838337544352,
tick align=outside,
tick pos=left,
x grid style={white!69.0196078431373!black},
xmin=-0.5, xmax=39.5,
xtick style={color=black},
y dir=reverse,
y grid style={white!69.0196078431373!black},
ymin=-0.5, ymax=39.5,
ytick style={color=black}
]
\addplot graphics [includegraphics cmd=\pgfimage,xmin=-0.5, xmax=39.5, ymin=39.5, ymax=-0.5] {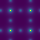};

\nextgroupplot[
colormap/viridis,
hide x axis,
hide y axis,
point meta max=0.360214173793793,
point meta min=-0.0224490575492382,
tick align=outside,
tick pos=left,
x grid style={white!69.0196078431373!black},
xmin=-0.5, xmax=39.5,
xtick style={color=black},
y dir=reverse,
y grid style={white!69.0196078431373!black},
ymin=-0.5, ymax=39.5,
ytick style={color=black}
]
\addplot graphics [includegraphics cmd=\pgfimage,xmin=-0.5, xmax=39.5, ymin=39.5, ymax=-0.5] {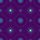};

\nextgroupplot[
colormap/viridis,
hide x axis,
hide y axis,
point meta max=0.372209191322327,
point meta min=-0.0278071705251932,
tick align=outside,
tick pos=left,
x grid style={white!69.0196078431373!black},
xmin=-0.5, xmax=39.5,
xtick style={color=black},
y dir=reverse,
y grid style={white!69.0196078431373!black},
ymin=-0.5, ymax=39.5,
ytick style={color=black}
]
\addplot graphics [includegraphics cmd=\pgfimage,xmin=-0.5, xmax=39.5, ymin=39.5, ymax=-0.5] {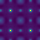};

\nextgroupplot[
colormap/viridis,
hide x axis,
ticks=none,
ylabel = \myfont Slice 7,
ylabel style={rotate=0},
point meta max=0.458228230476379,
point meta min=-0.00135803618468344,
tick align=outside,
tick pos=left,
x grid style={white!69.0196078431373!black},
xmin=-0.5, xmax=39.5,
xtick style={color=black},
y dir=reverse,
y grid style={white!69.0196078431373!black},
ymin=-0.5, ymax=39.5,
ytick style={color=black}
]
\addplot graphics [includegraphics cmd=\pgfimage,xmin=-0.5, xmax=39.5, ymin=39.5, ymax=-0.5] {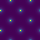};

\nextgroupplot[
colormap/viridis,
hide x axis,
hide y axis,
point meta max=0.234138682484627,
point meta min=-0.029674094170332,
tick align=outside,
tick pos=left,
x grid style={white!69.0196078431373!black},
xmin=-0.5, xmax=39.5,
xtick style={color=black},
y dir=reverse,
y grid style={white!69.0196078431373!black},
ymin=-0.5, ymax=39.5,
ytick style={color=black}
]
\addplot graphics [includegraphics cmd=\pgfimage,xmin=-0.5, xmax=39.5, ymin=39.5, ymax=-0.5] {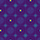};

\nextgroupplot[
colormap/viridis,
hide x axis,
hide y axis,
point meta max=0.371089100837708,
point meta min=-0.0250629615038633,
tick align=outside,
tick pos=left,
x grid style={white!69.0196078431373!black},
xmin=-0.5, xmax=39.5,
xtick style={color=black},
y dir=reverse,
y grid style={white!69.0196078431373!black},
ymin=-0.5, ymax=39.5,
ytick style={color=black}
]
\addplot graphics [includegraphics cmd=\pgfimage,xmin=-0.5, xmax=39.5, ymin=39.5, ymax=-0.5] {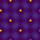};

\nextgroupplot[
colormap/viridis,
hide x axis,
hide y axis,
point meta max=0.830543518066406,
point meta min=0.00518810469657183,
tick align=outside,
tick pos=left,
x grid style={white!69.0196078431373!black},
xmin=-0.5, xmax=39.5,
xtick style={color=black},
y dir=reverse,
y grid style={white!69.0196078431373!black},
ymin=-0.5, ymax=39.5,
ytick style={color=black}
]
\addplot graphics [includegraphics cmd=\pgfimage,xmin=-0.5, xmax=39.5, ymin=39.5, ymax=-0.5] {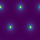};

\nextgroupplot[
colormap/viridis,
hide x axis,
hide y axis,
point meta max=0.389874160289764,
point meta min=-0.0644280612468719,
tick align=outside,
tick pos=left,
x grid style={white!69.0196078431373!black},
xmin=-0.5, xmax=39.5,
xtick style={color=black},
y dir=reverse,
y grid style={white!69.0196078431373!black},
ymin=-0.5, ymax=39.5,
ytick style={color=black}
]
\addplot graphics [includegraphics cmd=\pgfimage,xmin=-0.5, xmax=39.5, ymin=39.5, ymax=-0.5] {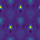};

\nextgroupplot[
colormap/viridis,
hide x axis,
hide y axis,
point meta max=0.598581612110138,
point meta min=-0.0330221056938171,
tick align=outside,
tick pos=left,
x grid style={white!69.0196078431373!black},
xmin=-0.5, xmax=39.5,
xtick style={color=black},
y dir=reverse,
y grid style={white!69.0196078431373!black},
ymin=-0.5, ymax=39.5,
ytick style={color=black}
]
\addplot graphics [includegraphics cmd=\pgfimage,xmin=-0.5, xmax=39.5, ymin=39.5, ymax=-0.5] {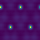};

\nextgroupplot[
colorbar style={xtick={0.25,0.5},xticklabels={0.25,0.50}},
colormap/viridis,
hide x axis,
hide y axis,
point meta max=0.672440230846405,
point meta min=0.0092761218547821,
tick align=outside,
tick pos=left,
x grid style={white!69.0196078431373!black},
xmin=-0.5, xmax=39.5,
xtick style={color=black},
y dir=reverse,
y grid style={white!69.0196078431373!black},
ymin=-0.5, ymax=39.5,
ytick style={color=black}
]
\addplot graphics [includegraphics cmd=\pgfimage,xmin=-0.5, xmax=39.5, ymin=39.5, ymax=-0.5] {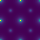};

\nextgroupplot[
colormap/viridis,
hide x axis,
hide y axis,
point meta max=0.418990463018417,
point meta min=-0.0206130649894476,
tick align=outside,
tick pos=left,
x grid style={white!69.0196078431373!black},
xmin=-0.5, xmax=39.5,
xtick style={color=black},
y dir=reverse,
y grid style={white!69.0196078431373!black},
ymin=-0.5, ymax=39.5,
ytick style={color=black}
]
\addplot graphics [includegraphics cmd=\pgfimage,xmin=-0.5, xmax=39.5, ymin=39.5, ymax=-0.5] {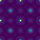};

\nextgroupplot[
colormap/viridis,
hide x axis,
hide y axis,
point meta max=0.483466923236847,
point meta min=-0.0278838519006968,
tick align=outside,
tick pos=left,
x grid style={white!69.0196078431373!black},
xmin=-0.5, xmax=39.5,
xtick style={color=black},
y dir=reverse,
y grid style={white!69.0196078431373!black},
ymin=-0.5, ymax=39.5,
ytick style={color=black}
]
\addplot graphics [includegraphics cmd=\pgfimage,xmin=-0.5, xmax=39.5, ymin=39.5, ymax=-0.5] {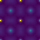};

\nextgroupplot[
colormap/viridis,
hide x axis,
ticks=none,
ylabel = \myfont Slice 8,
ylabel style={rotate=0},
point meta max=0.504094004631042,
point meta min=-0.00156226276885718,
tick align=outside,
tick pos=left,
x grid style={white!69.0196078431373!black},
xmin=-0.5, xmax=39.5,
xtick style={color=black},
y dir=reverse,
y grid style={white!69.0196078431373!black},
ymin=-0.5, ymax=39.5,
ytick style={color=black}
]
\addplot graphics [includegraphics cmd=\pgfimage,xmin=-0.5, xmax=39.5, ymin=39.5, ymax=-0.5] {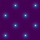};

\nextgroupplot[
colormap/viridis,
hide x axis,
hide y axis,
point meta max=0.288041472434998,
point meta min=-0.0237519573420286,
tick align=outside,
tick pos=left,
x grid style={white!69.0196078431373!black},
xmin=-0.5, xmax=39.5,
xtick style={color=black},
y dir=reverse,
y grid style={white!69.0196078431373!black},
ymin=-0.5, ymax=39.5,
ytick style={color=black}
]
\addplot graphics [includegraphics cmd=\pgfimage,xmin=-0.5, xmax=39.5, ymin=39.5, ymax=-0.5] {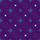};

\nextgroupplot[
colormap/viridis,
hide x axis,
hide y axis,
point meta max=0.416246086359024,
point meta min=-0.0227808114141226,
tick align=outside,
tick pos=left,
x grid style={white!69.0196078431373!black},
xmin=-0.5, xmax=39.5,
xtick style={color=black},
y dir=reverse,
y grid style={white!69.0196078431373!black},
ymin=-0.5, ymax=39.5,
ytick style={color=black}
]
\addplot graphics [includegraphics cmd=\pgfimage,xmin=-0.5, xmax=39.5, ymin=39.5, ymax=-0.5] {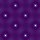};

\nextgroupplot[
colormap/viridis,
hide x axis,
hide y axis,
point meta max=0.427196651697159,
point meta min=0.00109258946031332,
tick align=outside,
tick pos=left,
x grid style={white!69.0196078431373!black},
xmin=-0.5, xmax=39.5,
xtick style={color=black},
y dir=reverse,
y grid style={white!69.0196078431373!black},
ymin=-0.5, ymax=39.5,
ytick style={color=black}
]
\addplot graphics [includegraphics cmd=\pgfimage,xmin=-0.5, xmax=39.5, ymin=39.5, ymax=-0.5] {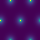};

\nextgroupplot[
colorbar style={xtick={-6.93889390390723e-18,0.25},xticklabels={0.00,0.25}},
colormap/viridis,
hide x axis,
hide y axis,
point meta max=0.299150288105011,
point meta min=-0.0428503043949604,
tick align=outside,
tick pos=left,
x grid style={white!69.0196078431373!black},
xmin=-0.5, xmax=39.5,
xtick style={color=black},
y dir=reverse,
y grid style={white!69.0196078431373!black},
ymin=-0.5, ymax=39.5,
ytick style={color=black}
]
\addplot graphics [includegraphics cmd=\pgfimage,xmin=-0.5, xmax=39.5, ymin=39.5, ymax=-0.5] {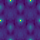};

\nextgroupplot[
colormap/viridis,
hide x axis,
hide y axis,
point meta max=0.356584817171097,
point meta min=-0.0263603758066893,
tick align=outside,
tick pos=left,
x grid style={white!69.0196078431373!black},
xmin=-0.5, xmax=39.5,
xtick style={color=black},
y dir=reverse,
y grid style={white!69.0196078431373!black},
ymin=-0.5, ymax=39.5,
ytick style={color=black}
]
\addplot graphics [includegraphics cmd=\pgfimage,xmin=-0.5, xmax=39.5, ymin=39.5, ymax=-0.5] {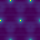};

\nextgroupplot[
colormap/viridis,
hide x axis,
hide y axis,
point meta max=0.495945811271667,
point meta min=0.00134838337544352,
tick align=outside,
tick pos=left,
x grid style={white!69.0196078431373!black},
xmin=-0.5, xmax=39.5,
xtick style={color=black},
y dir=reverse,
y grid style={white!69.0196078431373!black},
ymin=-0.5, ymax=39.5,
ytick style={color=black}
]
\addplot graphics [includegraphics cmd=\pgfimage,xmin=-0.5, xmax=39.5, ymin=39.5, ymax=-0.5] {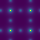};

\nextgroupplot[
colormap/viridis,
hide x axis,
hide y axis,
point meta max=0.353903502225876,
point meta min=-0.0197095908224583,
tick align=outside,
tick pos=left,
x grid style={white!69.0196078431373!black},
xmin=-0.5, xmax=39.5,
xtick style={color=black},
y dir=reverse,
y grid style={white!69.0196078431373!black},
ymin=-0.5, ymax=39.5,
ytick style={color=black}
]
\addplot graphics [includegraphics cmd=\pgfimage,xmin=-0.5, xmax=39.5, ymin=39.5, ymax=-0.5] {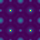};

\nextgroupplot[
colorbar style={xtick={-3.46944695195361e-18,0.25},xticklabels={0.00,0.25}},
colormap/viridis,
hide x axis,
hide y axis,
point meta max=0.371125370264053,
point meta min=-0.0255983434617519,
tick align=outside,
tick pos=left,
x grid style={white!69.0196078431373!black},
xmin=-0.5, xmax=39.5,
xtick style={color=black},
y dir=reverse,
y grid style={white!69.0196078431373!black},
ymin=-0.5, ymax=39.5,
ytick style={color=black}
]
\addplot graphics [includegraphics cmd=\pgfimage,xmin=-0.5, xmax=39.5, ymin=39.5, ymax=-0.5] {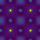};

\nextgroupplot[
colorbar horizontal,
colormap/viridis,
hide x axis,
ticks=none,
ylabel = \myfont Slice 9,
ylabel style={rotate=0},
colorbar style={xtick={0.2,0.8},xticklabels={0,0.5}},
tick align=outside,
tick pos=left,
x grid style={white!69.0196078431373!black},
xmin=-0.5, xmax=39.5,
xtick style={color=black},
y dir=reverse,
y grid style={white!69.0196078431373!black},
ymin=-0.5, ymax=39.5,
ytick style={color=black}
]
\addplot graphics [includegraphics cmd=\pgfimage,xmin=-0.5, xmax=39.5, ymin=39.5, ymax=-0.5] {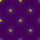};

\nextgroupplot[
colorbar horizontal,
colormap/viridis,
hide x axis,
hide y axis,
colorbar style={xtick={0.2,0.8},xticklabels={-0.05,0.4}},
tick align=outside,
tick pos=left,
x grid style={white!69.0196078431373!black},
xmin=-0.5, xmax=39.5,
xtick style={color=black},
y dir=reverse,
y grid style={white!69.0196078431373!black},
ymin=-0.5, ymax=39.5,
ytick style={color=black}
]
\addplot graphics [includegraphics cmd=\pgfimage,xmin=-0.5, xmax=39.5, ymin=39.5, ymax=-0.5] {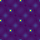};

\nextgroupplot[
colorbar horizontal,
colormap/viridis,
hide x axis,
hide y axis,
colorbar style={xtick={0.2,0.8},xticklabels={-0.03,0.4}},
tick align=outside,
tick pos=left,
x grid style={white!69.0196078431373!black},
xmin=-0.5, xmax=39.5,
xtick style={color=black},
y dir=reverse,
y grid style={white!69.0196078431373!black},
ymin=-0.5, ymax=39.5,
ytick style={color=black}
]
\addplot graphics [includegraphics cmd=\pgfimage,xmin=-0.5, xmax=39.5, ymin=39.5, ymax=-0.5] {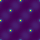};

\nextgroupplot[
colorbar horizontal,
colormap/viridis,
hide x axis,
hide y axis,
colorbar style={xtick={0.2,0.8},xticklabels={0,0.8}},
tick align=outside,
tick pos=left,
x grid style={white!69.0196078431373!black},
xmin=-0.5, xmax=39.5,
xtick style={color=black},
y dir=reverse,
y grid style={white!69.0196078431373!black},
ymin=-0.5, ymax=39.5,
ytick style={color=black}
]
\addplot graphics [includegraphics cmd=\pgfimage,xmin=-0.5, xmax=39.5, ymin=39.5, ymax=-0.5] {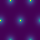};

\nextgroupplot[
colorbar horizontal,
colormap/viridis,
hide x axis,
hide y axis,
colorbar style={xtick={0.2,0.8},xticklabels={-0.06,0.4}},
point meta min=-0.0315597839653492,
tick align=outside,
tick pos=left,
x grid style={white!69.0196078431373!black},
xmin=-0.5, xmax=39.5,
xtick style={color=black},
y dir=reverse,
y grid style={white!69.0196078431373!black},
ymin=-0.5, ymax=39.5,
ytick style={color=black}
]
\addplot graphics [includegraphics cmd=\pgfimage,xmin=-0.5, xmax=39.5, ymin=39.5, ymax=-0.5] {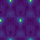};

\nextgroupplot[
colorbar horizontal,
colormap/viridis,
hide x axis,
hide y axis,
colorbar style={xtick={0.2,0.8},xticklabels={-0.04,0.6}},
tick align=outside,
tick pos=left,
x grid style={white!69.0196078431373!black},
xmin=-0.5, xmax=39.5,
xtick style={color=black},
y dir=reverse,
y grid style={white!69.0196078431373!black},
ymin=-0.5, ymax=39.5,
ytick style={color=black}
]
\addplot graphics [includegraphics cmd=\pgfimage,xmin=-0.5, xmax=39.5, ymin=39.5, ymax=-0.5] {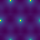};

\nextgroupplot[
colorbar horizontal,
colorbar style={xtick={0.2,0.8},xticklabels={0,0.7}},
colormap/viridis,
hide x axis,
hide y axis,
tick align=outside,
tick pos=left,
x grid style={white!69.0196078431373!black},
xmin=-0.5, xmax=39.5,
xtick style={color=black},
y dir=reverse,
y grid style={white!69.0196078431373!black},
ymin=-0.5, ymax=39.5,
ytick style={color=black}
]
\addplot graphics [includegraphics cmd=\pgfimage,xmin=-0.5, xmax=39.5, ymin=39.5, ymax=-0.5] {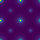};

\nextgroupplot[
colorbar horizontal,
colorbar style={xtick={0.2,0.8},xticklabels={-0.03,0.5}},
colormap/viridis,
hide x axis,
hide y axis,
tick align=outside,
tick pos=left,
x grid style={white!69.0196078431373!black},
xmin=-0.5, xmax=39.5,
xtick style={color=black},
y dir=reverse,
y grid style={white!69.0196078431373!black},
ymin=-0.5, ymax=39.5,
ytick style={color=black}
]
\addplot graphics [includegraphics cmd=\pgfimage,xmin=-0.5, xmax=39.5, ymin=39.5, ymax=-0.5] {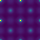};

\nextgroupplot[
colorbar horizontal,
colormap/viridis,
hide x axis,
hide y axis,
colorbar style={xtick={0.2,0.8},xticklabels={-0.04,0.6}},
tick align=outside,
tick pos=left,
x grid style={white!69.0196078431373!black},
xmin=-0.5, xmax=39.5,
xtick style={color=black},
y dir=reverse,
y grid style={white!69.0196078431373!black},
ymin=-0.5, ymax=39.5,
ytick style={color=black}
]
\addplot graphics [includegraphics cmd=\pgfimage,xmin=-0.5, xmax=39.5, ymin=39.5, ymax=-0.5] {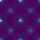};
\end{groupplot}

\end{tikzpicture}

%% file: Figure/Result/Figure_Low_Acc_Iter_5000_Slice3/Low_Acc_Iter_5000_Slice3.tex
\pgfplotsset{every tick label/.append style={font=\huge}}
\begin{tikzpicture}
\begin{groupplot}[group style={group size=6 by 3,vertical sep=1em, horizontal sep=1em},width = 0.25*\textwidth, height=(0.25)*\textwidth]
\nextgroupplot[
colorbar style={ylabel={}},
colormap/viridis,
title = \huge GT (a),
hide x axis,
ticks=none,
ylabel = \huge Slice 1,
point meta max=0.458228230476379,
point meta min=-0.00135803618468344,
tick align=outside,
tick pos=left,
x grid style={white!69.0196078431373!black},
xmin=-0.5, xmax=39.5,
xtick style={color=black},
y dir=reverse,
y grid style={white!69.0196078431373!black},
ymin=-0.5, ymax=39.5,
ytick style={color=black}
]
\addplot graphics [includegraphics cmd=\pgfimage,xmin=-0.5, xmax=39.5, ymin=39.5, ymax=-0.5] {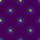};

\nextgroupplot[
colorbar style={ylabel={}},
colormap/viridis,
hide x axis,
hide y axis,
title = \huge S (a),
point meta max=0.60926216840744,
point meta min=-0.143314450979233,
tick align=outside,
tick pos=left,
x grid style={white!69.0196078431373!black},
xmin=-0.5, xmax=39.5,
xtick style={color=black},
y dir=reverse,
y grid style={white!69.0196078431373!black},
ymin=-0.5, ymax=39.5,
ytick style={color=black}
]
\addplot graphics [includegraphics cmd=\pgfimage,xmin=-0.5, xmax=39.5, ymin=39.5, ymax=-0.5] {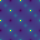};

\nextgroupplot[
colorbar style={ylabel={}},
colormap/viridis,
hide x axis,
hide y axis,
title = \huge GT (b),
point meta max=0.427196651697159,
point meta min=0.00109258946031332,
tick align=outside,
tick pos=left,
x grid style={white!69.0196078431373!black},
xmin=-0.5, xmax=39.5,
xtick style={color=black},
y dir=reverse,
y grid style={white!69.0196078431373!black},
ymin=-0.5, ymax=39.5,
ytick style={color=black}
]
\addplot graphics [includegraphics cmd=\pgfimage,xmin=-0.5, xmax=39.5, ymin=39.5, ymax=-0.5] {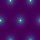};

\nextgroupplot[
colorbar style={ylabel={}},
colormap/viridis,
hide x axis,
hide y axis,
title = \huge S (b),
point meta max=0.485836058855057,
point meta min=-0.0376329161226749,
tick align=outside,
tick pos=left,
x grid style={white!69.0196078431373!black},
xmin=-0.5, xmax=39.5,
xtick style={color=black},
y dir=reverse,
y grid style={white!69.0196078431373!black},
ymin=-0.5, ymax=39.5,
ytick style={color=black}
]
\addplot graphics [includegraphics cmd=\pgfimage,xmin=-0.5, xmax=39.5, ymin=39.5, ymax=-0.5] {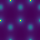};

\nextgroupplot[
colorbar style={ ylabel={}},
colormap/viridis,
hide x axis,
hide y axis,
title = \huge GT (c),
point meta max=0.672440230846405,
point meta min=0.0092761218547821,
tick align=outside,
tick pos=left,
x grid style={white!69.0196078431373!black},
xmin=-0.5, xmax=39.5,
xtick style={color=black},
y dir=reverse,
y grid style={white!69.0196078431373!black},
ymin=-0.5, ymax=39.5,
ytick style={color=black}
]
\addplot graphics [includegraphics cmd=\pgfimage,xmin=-0.5, xmax=39.5, ymin=39.5, ymax=-0.5] {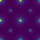};

\nextgroupplot[
colorbar style={ylabel={}},
colormap/viridis,
hide x axis,
hide y axis,
title = \huge S (c),
point meta max=0.771924614906311,
point meta min=-0.023497486487031,
tick align=outside,
tick pos=left,
x grid style={white!69.0196078431373!black},
xmin=-0.5, xmax=39.5,
xtick style={color=black},
y dir=reverse,
y grid style={white!69.0196078431373!black},
ymin=-0.5, ymax=39.5,
ytick style={color=black}
]
\addplot graphics [includegraphics cmd=\pgfimage,xmin=-0.5, xmax=39.5, ymin=39.5, ymax=-0.5] {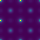};

\nextgroupplot[
colorbar style={ylabel={}},
colormap/viridis,
hide x axis,
ticks=none,
ylabel = \huge Slice 2,
point meta max=0.504094004631042,
point meta min=-0.00156226276885718,
tick align=outside,
tick pos=left,
x grid style={white!69.0196078431373!black},
xmin=-0.5, xmax=39.5,
xtick style={color=black},
y dir=reverse,
y grid style={white!69.0196078431373!black},
ymin=-0.5, ymax=39.5,
ytick style={color=black}
]
\addplot graphics [includegraphics cmd=\pgfimage,xmin=-0.5, xmax=39.5, ymin=39.5, ymax=-0.5] {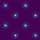};

\nextgroupplot[
colorbar style={ylabel={}},
colormap/viridis,
hide x axis,
hide y axis,
point meta max=0.393629133701324,
point meta min=-0.0271165408194065,
tick align=outside,
tick pos=left,
x grid style={white!69.0196078431373!black},
xmin=-0.5, xmax=39.5,
xtick style={color=black},
y dir=reverse,
y grid style={white!69.0196078431373!black},
ymin=-0.5, ymax=39.5,
ytick style={color=black}
]
\addplot graphics [includegraphics cmd=\pgfimage,xmin=-0.5, xmax=39.5, ymin=39.5, ymax=-0.5] {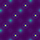};

\nextgroupplot[
colorbar style={ylabel={}},
colormap/viridis,
hide x axis,
hide y axis,
point meta max=0.830543518066406,
point meta min=0.00518810469657183,
tick align=outside,
tick pos=left,
x grid style={white!69.0196078431373!black},
xmin=-0.5, xmax=39.5,
xtick style={color=black},
y dir=reverse,
y grid style={white!69.0196078431373!black},
ymin=-0.5, ymax=39.5,
ytick style={color=black}
]
\addplot graphics [includegraphics cmd=\pgfimage,xmin=-0.5, xmax=39.5, ymin=39.5, ymax=-0.5] {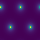};

\nextgroupplot[
colorbar style={ylabel={}},
colormap/viridis,
hide x axis,
hide y axis,
point meta max=0.652714729309082,
point meta min=-0.0397517271339893,
tick align=outside,
tick pos=left,
x grid style={white!69.0196078431373!black},
xmin=-0.5, xmax=39.5,
xtick style={color=black},
y dir=reverse,
y grid style={white!69.0196078431373!black},
ymin=-0.5, ymax=39.5,
ytick style={color=black}
]
\addplot graphics [includegraphics cmd=\pgfimage,xmin=-0.5, xmax=39.5, ymin=39.5, ymax=-0.5] {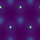};

\nextgroupplot[
colorbar style={ylabel={}},
colormap/viridis,
hide x axis,
hide y axis,
point meta max=0.495945811271667,
point meta min=0.00134838337544352,
tick align=outside,
tick pos=left,
x grid style={white!69.0196078431373!black},
xmin=-0.5, xmax=39.5,
xtick style={color=black},
y dir=reverse,
y grid style={white!69.0196078431373!black},
ymin=-0.5, ymax=39.5,
ytick style={color=black}
]
\addplot graphics [includegraphics cmd=\pgfimage,xmin=-0.5, xmax=39.5, ymin=39.5, ymax=-0.5] {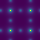};

\nextgroupplot[
colorbar style={ylabel={}},
colormap/viridis,
hide x axis,
hide y axis,
point meta max=0.509053647518158,
point meta min=-0.0366245917975903,
tick align=outside,
tick pos=left,
x grid style={white!69.0196078431373!black},
xmin=-0.5, xmax=39.5,
xtick style={color=black},
y dir=reverse,
y grid style={white!69.0196078431373!black},
ymin=-0.5, ymax=39.5,
ytick style={color=black}
]
\addplot graphics [includegraphics cmd=\pgfimage,xmin=-0.5, xmax=39.5, ymin=39.5, ymax=-0.5] {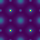};

\nextgroupplot[
colorbar horizontal,
colorbar style={ylabel={}},
colormap/viridis,
hide x axis,
ticks=none,
ylabel = \huge Slice 3,
colorbar style={xtick={0.2,0.8},xticklabels={0,0.5}},
tick align=outside,
tick pos=left,
x grid style={white!69.0196078431373!black},
xmin=-0.5, xmax=39.5,
xtick style={color=black},
y dir=reverse,
y grid style={white!69.0196078431373!black},
ymin=-0.5, ymax=39.5,
ytick style={color=black}
]
\addplot graphics [includegraphics cmd=\pgfimage,xmin=-0.5, xmax=39.5, ymin=39.5, ymax=-0.5] {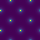};
 
\nextgroupplot[
colorbar horizontal,
colorbar style={ylabel={}},
colormap/viridis,
hide x axis,
hide y axis,
colorbar style={xtick={0.2,0.8},xticklabels={-0.14,0.6}},
tick align=outside,
tick pos=left,
x grid style={white!69.0196078431373!black},
xmin=-0.5, xmax=39.5,
xtick style={color=black},
y dir=reverse,
y grid style={white!69.0196078431373!black},
ymin=-0.5, ymax=39.5,
ytick style={color=black}
]
\addplot graphics [includegraphics cmd=\pgfimage,xmin=-0.5, xmax=39.5, ymin=39.5, ymax=-0.5] {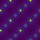};

\nextgroupplot[
colorbar horizontal,
colorbar style={ylabel={}},
colormap/viridis,
hide x axis,
hide y axis,
colorbar style={xtick={0.2,0.8},xticklabels={0,0.8}},
tick align=outside,
tick pos=left,
x grid style={white!69.0196078431373!black},
xmin=-0.5, xmax=39.5,
xtick style={color=black},
y dir=reverse,
y grid style={white!69.0196078431373!black},
ymin=-0.5, ymax=39.5,
ytick style={color=black}
]
\addplot graphics [includegraphics cmd=\pgfimage,xmin=-0.5, xmax=39.5, ymin=39.5, ymax=-0.5] {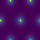};

\nextgroupplot[
colorbar horizontal,
colorbar style={ylabel={}},
colormap/viridis,
hide x axis,
hide y axis,
colorbar style={xtick={0.2,0.8},xticklabels={-0.04,0.6}},
tick align=outside,
tick pos=left,
x grid style={white!69.0196078431373!black},
xmin=-0.5, xmax=39.5,
xtick style={color=black},
y dir=reverse,
y grid style={white!69.0196078431373!black},
ymin=-0.5, ymax=39.5,
ytick style={color=black}
]
\addplot graphics [includegraphics cmd=\pgfimage,xmin=-0.5, xmax=39.5, ymin=39.5, ymax=-0.5] {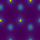};

\nextgroupplot[
colorbar horizontal,
colorbar style={ ylabel={}},
colormap/viridis,
hide x axis,
hide y axis,
colorbar style={xtick={0.2,0.8},xticklabels={0,0.7}},
tick align=outside,
tick pos=left,
x grid style={white!69.0196078431373!black},
xmin=-0.5, xmax=39.5,
xtick style={color=black},
y dir=reverse,
y grid style={white!69.0196078431373!black},
ymin=-0.5, ymax=39.5,
ytick style={color=black}
]
\addplot graphics [includegraphics cmd=\pgfimage,xmin=-0.5, xmax=39.5, ymin=39.5, ymax=-0.5] {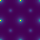};

\nextgroupplot[
colorbar horizontal,
colorbar style={ylabel={}},
colormap/viridis,
hide x axis,
hide y axis,
colorbar style={xtick={0.2,0.8},xticklabels={-0.03,0.8}},
tick align=outside,
tick pos=left,
x grid style={white!69.0196078431373!black},
xmin=-0.5, xmax=39.5,
xtick style={color=black},
y dir=reverse,
y grid style={white!69.0196078431373!black},
ymin=-0.5, ymax=39.5,
ytick style={color=black}
]
\addplot graphics [includegraphics cmd=\pgfimage,xmin=-0.5, xmax=39.5, ymin=39.5, ymax=-0.5] {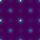};
\end{groupplot}

\end{tikzpicture}

%% file: Figure/Result/Figure_Low_Acc_Iter_10000_Slice6/Low_Acc_Iter_10000_Slice6.tex
\pgfplotsset{every tick label/.append style={font=\huge}}
\begin{tikzpicture}

\begin{groupplot}[group style={group size=6 by 6,vertical sep=1em,horizontal sep=1em},width = 0.25*\textwidth, height=(0.25)*\textwidth]
\nextgroupplot[
colorbar style={ylabel={}},
colormap/viridis,
title =\huge GT (a),
hide x axis,
ticks=none,
ylabel = \huge Slice 1,
point meta max=0.458228230476379,
point meta min=-0.00135803618468344,
tick align=outside,
tick pos=left,
x grid style={white!69.0196078431373!black},
xmin=-0.5, xmax=39.5,
xtick style={color=black},
y dir=reverse,
y grid style={white!69.0196078431373!black},
ymin=-0.5, ymax=39.5,
ytick style={color=black}
]
\addplot graphics [includegraphics cmd=\pgfimage,xmin=-0.5, xmax=39.5, ymin=39.5, ymax=-0.5] {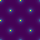};

\nextgroupplot[
colormap/viridis,
hide x axis,
hide y axis,
title = \huge S (a),
point meta max=0.583492994308472,
point meta min=-0.104496091604233,
tick align=outside,
tick pos=left,
x grid style={white!69.0196078431373!black},
xmin=-0.5, xmax=39.5,
xtick style={color=black},
y dir=reverse,
y grid style={white!69.0196078431373!black},
ymin=-0.5, ymax=39.5,
ytick style={color=black}
]
\addplot graphics [includegraphics cmd=\pgfimage,xmin=-0.5, xmax=39.5, ymin=39.5, ymax=-0.5] {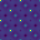};

\nextgroupplot[
colormap/viridis,
hide x axis,
hide y axis,
title =\huge  GT (b),
point meta max=0.427196651697159,
point meta min=0.00109258946031332,
tick align=outside,
tick pos=left,
x grid style={white!69.0196078431373!black},
xmin=-0.5, xmax=39.5,
xtick style={color=black},
y dir=reverse,
y grid style={white!69.0196078431373!black},
ymin=-0.5, ymax=39.5,
ytick style={color=black}
]
\addplot graphics [includegraphics cmd=\pgfimage,xmin=-0.5, xmax=39.5, ymin=39.5, ymax=-0.5] {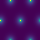};

\nextgroupplot[
colormap/viridis,
hide x axis,
hide y axis,
title = \huge  S (b),
point meta max=0.558730006217957,
point meta min=-0.045606717467308,
tick align=outside,
tick pos=left,
x grid style={white!69.0196078431373!black},
xmin=-0.5, xmax=39.5,
xtick style={color=black},
y dir=reverse,
y grid style={white!69.0196078431373!black},
ymin=-0.5, ymax=39.5,
ytick style={color=black}
]
\addplot graphics [includegraphics cmd=\pgfimage,xmin=-0.5, xmax=39.5, ymin=39.5, ymax=-0.5] {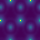};

\nextgroupplot[
colormap/viridis,
hide x axis,
hide y axis,
title = \huge GT (c),
point meta max=0.672440230846405,
point meta min=0.0092761218547821,
tick align=outside,
tick pos=left,
x grid style={white!69.0196078431373!black},
xmin=-0.5, xmax=39.5,
xtick style={color=black},
y dir=reverse,
y grid style={white!69.0196078431373!black},
ymin=-0.5, ymax=39.5,
ytick style={color=black}
]
\addplot graphics [includegraphics cmd=\pgfimage,xmin=-0.5, xmax=39.5, ymin=39.5, ymax=-0.5] {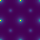};

\nextgroupplot[
colormap/viridis,
hide x axis,
hide y axis,
title = \huge  S (c),
point meta max=0.853756248950958,
point meta min=-0.0368711799383163,
tick align=outside,
tick pos=left,
x grid style={white!69.0196078431373!black},
xmin=-0.5, xmax=39.5,
xtick style={color=black},
y dir=reverse,
y grid style={white!69.0196078431373!black},
ymin=-0.5, ymax=39.5,
ytick style={color=black}
]
\addplot graphics [includegraphics cmd=\pgfimage,xmin=-0.5, xmax=39.5, ymin=39.5, ymax=-0.5] {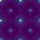};

\nextgroupplot[
colormap/viridis,
hide x axis,
ticks=none,
ylabel = \huge Slice 2,
point meta max=0.504094004631042,
point meta min=-0.00156226276885718,
tick align=outside,
tick pos=left,
x grid style={white!69.0196078431373!black},
xmin=-0.5, xmax=39.5,
xtick style={color=black},
y dir=reverse,
y grid style={white!69.0196078431373!black},
ymin=-0.5, ymax=39.5,
ytick style={color=black}
]
\addplot graphics [includegraphics cmd=\pgfimage,xmin=-0.5, xmax=39.5, ymin=39.5, ymax=-0.5] {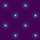};

\nextgroupplot[
colormap/viridis,
hide x axis,
hide y axis,
point meta max=0.256572723388672,
point meta min=-0.0714732110500336,
tick align=outside,
tick pos=left,
x grid style={white!69.0196078431373!black},
xmin=-0.5, xmax=39.5,
xtick style={color=black},
y dir=reverse,
y grid style={white!69.0196078431373!black},
ymin=-0.5, ymax=39.5,
ytick style={color=black}
]
\addplot graphics [includegraphics cmd=\pgfimage,xmin=-0.5, xmax=39.5, ymin=39.5, ymax=-0.5] {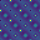};

\nextgroupplot[
colormap/viridis,
hide x axis,
hide y axis,
point meta max=0.830543518066406,
point meta min=0.00518810469657183,
tick align=outside,
tick pos=left,
x grid style={white!69.0196078431373!black},
xmin=-0.5, xmax=39.5,
xtick style={color=black},
y dir=reverse,
y grid style={white!69.0196078431373!black},
ymin=-0.5, ymax=39.5,
ytick style={color=black}
]
\addplot graphics [includegraphics cmd=\pgfimage,xmin=-0.5, xmax=39.5, ymin=39.5, ymax=-0.5] {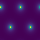};

\nextgroupplot[
colormap/viridis,
hide x axis,
hide y axis,
point meta max=0.55182957649231,
point meta min=-0.0353236049413681,
tick align=outside,
tick pos=left,
x grid style={white!69.0196078431373!black},
xmin=-0.5, xmax=39.5,
xtick style={color=black},
y dir=reverse,
y grid style={white!69.0196078431373!black},
ymin=-0.5, ymax=39.5,
ytick style={color=black}
]
\addplot graphics [includegraphics cmd=\pgfimage,xmin=-0.5, xmax=39.5, ymin=39.5, ymax=-0.5] {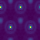};

\nextgroupplot[
colormap/viridis,
hide x axis,
hide y axis,
point meta max=0.495945811271667,
point meta min=0.00134838337544352,
tick align=outside,
tick pos=left,
x grid style={white!69.0196078431373!black},
xmin=-0.5, xmax=39.5,
xtick style={color=black},
y dir=reverse,
y grid style={white!69.0196078431373!black},
ymin=-0.5, ymax=39.5,
ytick style={color=black}
]
\addplot graphics [includegraphics cmd=\pgfimage,xmin=-0.5, xmax=39.5, ymin=39.5, ymax=-0.5] {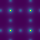};

\nextgroupplot[
colormap/viridis,
hide x axis,
hide y axis,
point meta max=0.471088111400604,
point meta min=-0.0369863361120224,
tick align=outside,
tick pos=left,
x grid style={white!69.0196078431373!black},
xmin=-0.5, xmax=39.5,
xtick style={color=black},
y dir=reverse,
y grid style={white!69.0196078431373!black},
ymin=-0.5, ymax=39.5,
ytick style={color=black}
]
\addplot graphics [includegraphics cmd=\pgfimage,xmin=-0.5, xmax=39.5, ymin=39.5, ymax=-0.5] {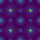};

\nextgroupplot[
colormap/viridis,
hide x axis,
ticks=none,
ylabel = \huge Slice 3,
point meta max=0.458228230476379,
point meta min=-0.00135803618468344,
tick align=outside,
tick pos=left,
x grid style={white!69.0196078431373!black},
xmin=-0.5, xmax=39.5,
xtick style={color=black},
y dir=reverse,
y grid style={white!69.0196078431373!black},
ymin=-0.5, ymax=39.5,
ytick style={color=black}
]
\addplot graphics [includegraphics cmd=\pgfimage,xmin=-0.5, xmax=39.5, ymin=39.5, ymax=-0.5] {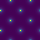};

\nextgroupplot[
colormap/viridis,
hide x axis,
hide y axis,
point meta max=0.241374492645264,
point meta min=-0.0289389751851559,
tick align=outside,
tick pos=left,
x grid style={white!69.0196078431373!black},
xmin=-0.5, xmax=39.5,
xtick style={color=black},
y dir=reverse,
y grid style={white!69.0196078431373!black},
ymin=-0.5, ymax=39.5,
ytick style={color=black}
]
\addplot graphics [includegraphics cmd=\pgfimage,xmin=-0.5, xmax=39.5, ymin=39.5, ymax=-0.5] {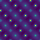};

\nextgroupplot[
colormap/viridis,
hide x axis,
hide y axis,
point meta max=0.427196651697159,
point meta min=0.00109258946031332,
tick align=outside,
tick pos=left,
x grid style={white!69.0196078431373!black},
xmin=-0.5, xmax=39.5,
xtick style={color=black},
y dir=reverse,
y grid style={white!69.0196078431373!black},
ymin=-0.5, ymax=39.5,
ytick style={color=black}
]
\addplot graphics [includegraphics cmd=\pgfimage,xmin=-0.5, xmax=39.5, ymin=39.5, ymax=-0.5] {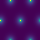};

\nextgroupplot[
colormap/viridis,
hide x axis,
hide y axis,
point meta max=0.371159315109253,
point meta min=-0.0266385860741138,
tick align=outside,
tick pos=left,
x grid style={white!69.0196078431373!black},
xmin=-0.5, xmax=39.5,
xtick style={color=black},
y dir=reverse,
y grid style={white!69.0196078431373!black},
ymin=-0.5, ymax=39.5,
ytick style={color=black}
]
\addplot graphics [includegraphics cmd=\pgfimage,xmin=-0.5, xmax=39.5, ymin=39.5, ymax=-0.5] {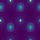};

\nextgroupplot[
colormap/viridis,
hide x axis,
hide y axis,
point meta max=0.672440230846405,
point meta min=0.0092761218547821,
tick align=outside,
tick pos=left,
x grid style={white!69.0196078431373!black},
xmin=-0.5, xmax=39.5,
xtick style={color=black},
y dir=reverse,
y grid style={white!69.0196078431373!black},
ymin=-0.5, ymax=39.5,
ytick style={color=black}
]
\addplot graphics [includegraphics cmd=\pgfimage,xmin=-0.5, xmax=39.5, ymin=39.5, ymax=-0.5] {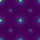};

\nextgroupplot[
colormap/viridis,
hide x axis,
hide y axis,
point meta max=0.541455268859863,
point meta min=-0.0270484779030085,
tick align=outside,
tick pos=left,
x grid style={white!69.0196078431373!black},
xmin=-0.5, xmax=39.5,
xtick style={color=black},
y dir=reverse,
y grid style={white!69.0196078431373!black},
ymin=-0.5, ymax=39.5,
ytick style={color=black}
]
\addplot graphics [includegraphics cmd=\pgfimage,xmin=-0.5, xmax=39.5, ymin=39.5, ymax=-0.5] {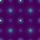};

\nextgroupplot[
colormap/viridis,
hide x axis,
ticks=none,
ylabel = \huge Slice 4,
point meta max=0.504094004631042,
point meta min=-0.00156226276885718,
tick align=outside,
tick pos=left,
x grid style={white!69.0196078431373!black},
xmin=-0.5, xmax=39.5,
xtick style={color=black},
y dir=reverse,
y grid style={white!69.0196078431373!black},
ymin=-0.5, ymax=39.5,
ytick style={color=black}
]
\addplot graphics [includegraphics cmd=\pgfimage,xmin=-0.5, xmax=39.5, ymin=39.5, ymax=-0.5] {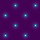};

\nextgroupplot[
colormap/viridis,
hide x axis,
hide y axis,
point meta max=0.207408413290977,
point meta min=-0.0214910488575697,
tick align=outside,
tick pos=left,
x grid style={white!69.0196078431373!black},
xmin=-0.5, xmax=39.5,
xtick style={color=black},
y dir=reverse,
y grid style={white!69.0196078431373!black},
ymin=-0.5, ymax=39.5,
ytick style={color=black}
]
\addplot graphics [includegraphics cmd=\pgfimage,xmin=-0.5, xmax=39.5, ymin=39.5, ymax=-0.5] {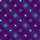};

\nextgroupplot[
colormap/viridis,
hide x axis,
hide y axis,
point meta max=0.427196651697159,
point meta min=0.00109258946031332,
tick align=outside,
tick pos=left,
x grid style={white!69.0196078431373!black},
xmin=-0.5, xmax=39.5,
xtick style={color=black},
y dir=reverse,
y grid style={white!69.0196078431373!black},
ymin=-0.5, ymax=39.5,
ytick style={color=black}
]
\addplot graphics [includegraphics cmd=\pgfimage,xmin=-0.5, xmax=39.5, ymin=39.5, ymax=-0.5] {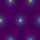};

\nextgroupplot[
colormap/viridis,
hide x axis,
hide y axis,
point meta max=0.521803855895996,
point meta min=-0.0488835498690605,
tick align=outside,
tick pos=left,
x grid style={white!69.0196078431373!black},
xmin=-0.5, xmax=39.5,
xtick style={color=black},
y dir=reverse,
y grid style={white!69.0196078431373!black},
ymin=-0.5, ymax=39.5,
ytick style={color=black}
]
\addplot graphics [includegraphics cmd=\pgfimage,xmin=-0.5, xmax=39.5, ymin=39.5, ymax=-0.5] {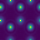};

\nextgroupplot[
colormap/viridis,
hide x axis,
hide y axis,
point meta max=0.495945811271667,
point meta min=0.00134838337544352,
tick align=outside,
tick pos=left,
x grid style={white!69.0196078431373!black},
xmin=-0.5, xmax=39.5,
xtick style={color=black},
y dir=reverse,
y grid style={white!69.0196078431373!black},
ymin=-0.5, ymax=39.5,
ytick style={color=black}
]
\addplot graphics [includegraphics cmd=\pgfimage,xmin=-0.5, xmax=39.5, ymin=39.5, ymax=-0.5] {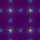};

\nextgroupplot[
colormap/viridis,
hide x axis,
hide y axis,
point meta max=0.494835764169693,
point meta min=-0.025128586217761,
tick align=outside,
tick pos=left,
x grid style={white!69.0196078431373!black},
xmin=-0.5, xmax=39.5,
xtick style={color=black},
y dir=reverse,
y grid style={white!69.0196078431373!black},
ymin=-0.5, ymax=39.5,
ytick style={color=black}
]
\addplot graphics [includegraphics cmd=\pgfimage,xmin=-0.5, xmax=39.5, ymin=39.5, ymax=-0.5] {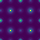};

\nextgroupplot[
colormap/viridis,
hide x axis,
ticks=none,
ylabel =\huge Slice 5,
point meta max=0.458228230476379,
point meta min=-0.00135803618468344,
tick align=outside,
tick pos=left,
x grid style={white!69.0196078431373!black},
xmin=-0.5, xmax=39.5,
xtick style={color=black},
y dir=reverse,
y grid style={white!69.0196078431373!black},
ymin=-0.5, ymax=39.5,
ytick style={color=black}
]
\addplot graphics [includegraphics cmd=\pgfimage,xmin=-0.5, xmax=39.5, ymin=39.5, ymax=-0.5] {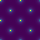};

\nextgroupplot[
colormap/viridis,
hide x axis,
hide y axis,
point meta max=0.222940593957901,
point meta min=-0.0176398847252131,
tick align=outside,
tick pos=left,
x grid style={white!69.0196078431373!black},
xmin=-0.5, xmax=39.5,
xtick style={color=black},
y dir=reverse,
y grid style={white!69.0196078431373!black},
ymin=-0.5, ymax=39.5,
ytick style={color=black}
]
\addplot graphics [includegraphics cmd=\pgfimage,xmin=-0.5, xmax=39.5, ymin=39.5, ymax=-0.5] {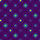};

\nextgroupplot[
colormap/viridis,
hide x axis,
hide y axis,
point meta max=0.830543518066406,
point meta min=0.00518810469657183,
tick align=outside,
tick pos=left,
x grid style={white!69.0196078431373!black},
xmin=-0.5, xmax=39.5,
xtick style={color=black},
y dir=reverse,
y grid style={white!69.0196078431373!black},
ymin=-0.5, ymax=39.5,
ytick style={color=black}
]
\addplot graphics [includegraphics cmd=\pgfimage,xmin=-0.5, xmax=39.5, ymin=39.5, ymax=-0.5] {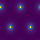};

\nextgroupplot[
colormap/viridis,
hide x axis,
hide y axis,
point meta max=0.581143021583557,
point meta min=-0.028544319793582,
tick align=outside,
tick pos=left,
x grid style={white!69.0196078431373!black},
xmin=-0.5, xmax=39.5,
xtick style={color=black},
y dir=reverse,
y grid style={white!69.0196078431373!black},
ymin=-0.5, ymax=39.5,
ytick style={color=black}
]
\addplot graphics [includegraphics cmd=\pgfimage,xmin=-0.5, xmax=39.5, ymin=39.5, ymax=-0.5] {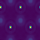};

\nextgroupplot[
colormap/viridis,
hide x axis,
hide y axis,
point meta max=0.672440230846405,
point meta min=0.0092761218547821,
tick align=outside,
tick pos=left,
x grid style={white!69.0196078431373!black},
xmin=-0.5, xmax=39.5,
xtick style={color=black},
y dir=reverse,
y grid style={white!69.0196078431373!black},
ymin=-0.5, ymax=39.5,
ytick style={color=black}
]
\addplot graphics [includegraphics cmd=\pgfimage,xmin=-0.5, xmax=39.5, ymin=39.5, ymax=-0.5] {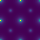};

\nextgroupplot[
colormap/viridis,
hide x axis,
hide y axis,
point meta max=0.508975386619568,
point meta min=-0.0257929675281048,
tick align=outside,
tick pos=left,
x grid style={white!69.0196078431373!black},
xmin=-0.5, xmax=39.5,
xtick style={color=black},
y dir=reverse,
y grid style={white!69.0196078431373!black},
ymin=-0.5, ymax=39.5,
ytick style={color=black}
]
\addplot graphics [includegraphics cmd=\pgfimage,xmin=-0.5, xmax=39.5, ymin=39.5, ymax=-0.5] {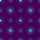};

\nextgroupplot[
colorbar horizontal,
colormap/viridis,
hide x axis,
ticks=none,
ylabel = \huge Slice 6,
colorbar style={xtick={0.2,0.8},xticklabels={0,0.5}},
tick align=outside,
tick pos=left,
x grid style={white!69.0196078431373!black},
xmin=-0.5, xmax=39.5,
xtick style={color=black},
y dir=reverse,
y grid style={white!69.0196078431373!black},
ymin=-0.5, ymax=39.5,
ytick style={color=black}
]
\addplot graphics [includegraphics cmd=\pgfimage,xmin=-0.5, xmax=39.5, ymin=39.5, ymax=-0.5] {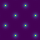};

\nextgroupplot[
colorbar horizontal,
colormap/viridis,
hide x axis,
hide y axis,
colorbar style={xtick={0.2,0.8},xticklabels={-0.1,0.6}},
tick align=outside,
tick pos=left,
x grid style={white!69.0196078431373!black},
xmin=-0.5, xmax=39.5,
xtick style={color=black},
y dir=reverse,
y grid style={white!69.0196078431373!black},
ymin=-0.5, ymax=39.5,
ytick style={color=black}
]
\addplot graphics [includegraphics cmd=\pgfimage,xmin=-0.5, xmax=39.5, ymin=39.5, ymax=-0.5] {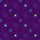};

\nextgroupplot[
colorbar horizontal,
colormap/viridis,
hide x axis,
hide y axis,
colorbar style={xtick={0.2,0.8},xticklabels={0,0.8}},
tick align=outside,
tick pos=left,
x grid style={white!69.0196078431373!black},
xmin=-0.5, xmax=39.5,
xtick style={color=black},
y dir=reverse,
y grid style={white!69.0196078431373!black},
ymin=-0.5, ymax=39.5,
ytick style={color=black}
]
\addplot graphics [includegraphics cmd=\pgfimage,xmin=-0.5, xmax=39.5, ymin=39.5, ymax=-0.5] {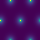};

\nextgroupplot[
colorbar horizontal,
colormap/viridis,
hide x axis,
hide y axis,
colorbar style={xtick={0.2,0.8},xticklabels={-0.05,0.6}},
tick align=outside,
tick pos=left,
x grid style={white!69.0196078431373!black},
xmin=-0.5, xmax=39.5,
xtick style={color=black},
y dir=reverse,
y grid style={white!69.0196078431373!black},
ymin=-0.5, ymax=39.5,
ytick style={color=black}
]
\addplot graphics [includegraphics cmd=\pgfimage,xmin=-0.5, xmax=39.5, ymin=39.5, ymax=-0.5] {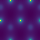};

\nextgroupplot[
colorbar horizontal,
colormap/viridis,
hide x axis,
hide y axis,
colorbar style={xtick={0.2,0.8},xticklabels={0,0.7}},
tick align=outside,
tick pos=left,
x grid style={white!69.0196078431373!black},
xmin=-0.5, xmax=39.5,
xtick style={color=black},
y dir=reverse,
y grid style={white!69.0196078431373!black},
ymin=-0.5, ymax=39.5,
ytick style={color=black}
]
\addplot graphics [includegraphics cmd=\pgfimage,xmin=-0.5, xmax=39.5, ymin=39.5, ymax=-0.5] {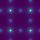};

\nextgroupplot[
colorbar horizontal,
colormap/viridis,
hide x axis,
hide y axis,
colorbar style={xtick={0.2,0.8},xticklabels={-0.04,0.8}},
tick align=outside,
tick pos=left,
x grid style={white!69.0196078431373!black},
xmin=-0.5, xmax=39.5,
xtick style={color=black},
y dir=reverse,
y grid style={white!69.0196078431373!black},
ymin=-0.5, ymax=39.5,
ytick style={color=black}
]
\addplot graphics [includegraphics cmd=\pgfimage,xmin=-0.5, xmax=39.5, ymin=39.5, ymax=-0.5] {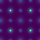};
\end{groupplot}

\end{tikzpicture}

%% file: Figure/Result/Figure_13/Fig_13_Probe_Est_OTF_Low_Acc_Sparse_3_Slice.tex
\pgfplotsset{every tick label/.append style={font=\huge}}
\begin{tikzpicture}

\begin{groupplot}[group style={group size=6 by 2,vertical sep = 1em,horizontal sep=1em},width = 0.25*\textwidth, height=(0.25)*\textwidth]
\nextgroupplot[
hide x axis,
ticks=none,
ylabel = \huge Amplitude,
tick align=outside,
tick pos=left,
title={\huge GT (a)},
x grid style={white!69.0196078431373!black},
xmin=-0.5, xmax=39.5,
xtick style={color=black},
y dir=reverse,
y grid style={white!69.0196078431373!black},
ymin=-0.5, ymax=39.5,
ytick style={color=black}
]
\addplot graphics [includegraphics cmd=\pgfimage,xmin=-0.5, xmax=39.5, ymin=39.5, ymax=-0.5] {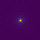};

\nextgroupplot[
title={\huge Est. (a)},
tick align=outside,
tick pos=left,
x grid style={white!69.0196078431373!black},
xmin=-0.5, xmax=39.5,
xtick style={color=black},
y dir=reverse,
y grid style={white!69.0196078431373!black},
ymin=-0.5, ymax=39.5,
ytick style={color=black}
]
\addplot graphics [includegraphics cmd=\pgfimage,xmin=-0.5, xmax=39.5, ymin=39.5, ymax=-0.5] {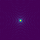};

\nextgroupplot[
hide x axis,
hide y axis,
tick align=outside,
tick pos=left,
title={\huge GT (b)},
x grid style={white!69.0196078431373!black},
xmin=-0.5, xmax=39.5,
xtick style={color=black},
y dir=reverse,
y grid style={white!69.0196078431373!black},
ymin=-0.5, ymax=39.5,
ytick style={color=black}
]
\addplot graphics [includegraphics cmd=\pgfimage,xmin=-0.5, xmax=39.5, ymin=39.5, ymax=-0.5] {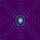};

\nextgroupplot[
hide x axis,
hide y axis,
tick align=outside,
tick pos=left,
title={\huge Est. (b)},
x grid style={white!69.0196078431373!black},
xmin=-0.5, xmax=39.5,
xtick style={color=black},
y dir=reverse,
y grid style={white!69.0196078431373!black},
ymin=-0.5, ymax=39.5,
ytick style={color=black}
]
\addplot graphics [includegraphics cmd=\pgfimage,xmin=-0.5, xmax=39.5, ymin=39.5, ymax=-0.5] {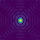};

\nextgroupplot[
hide x axis,
hide y axis,
tick align=outside,
tick pos=left,
title={\huge GT (c)},
x grid style={white!69.0196078431373!black},
xmin=-0.5, xmax=39.5,
xtick style={color=black},
y dir=reverse,
y grid style={white!69.0196078431373!black},
ymin=-0.5, ymax=39.5,
ytick style={color=black}
]
\addplot graphics [includegraphics cmd=\pgfimage,xmin=-0.5, xmax=39.5, ymin=39.5, ymax=-0.5] {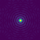};

\nextgroupplot[
colorbar,
colorbar style={ylabel={\huge Norm. Amp.}},
colormap/viridis,
hide x axis,
hide y axis,
point meta max=1,
point meta min=0.000171732172020711,
hide x axis,
hide y axis,
tick align=outside,
tick pos=left,
title={\huge Est. (c)},
x grid style={white!69.0196078431373!black},
xmin=-0.5, xmax=39.5,
xtick style={color=black},
y dir=reverse,
y grid style={white!69.0196078431373!black},
ymin=-0.5, ymax=39.5,
ytick style={color=black}
]
\addplot graphics [includegraphics cmd=\pgfimage,xmin=-0.5, xmax=39.5, ymin=39.5, ymax=-0.5] {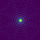};

\nextgroupplot[
hide x axis,
ticks=none,
ylabel = \huge Phase,
tick align=outside,
tick pos=left,
x grid style={white!69.0196078431373!black},
xmin=-0.5, xmax=39.5,
xtick style={color=black},
y dir=reverse,
y grid style={white!69.0196078431373!black},
ymin=-0.5, ymax=39.5,
ytick style={color=black}
]
\addplot graphics [includegraphics cmd=\pgfimage,xmin=-0.5, xmax=39.5, ymin=39.5, ymax=-0.5] {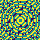};

\nextgroupplot[
hide x axis,
hide y axis,
tick align=outside,
tick pos=left,
x grid style={white!69.0196078431373!black},
xmin=-0.5, xmax=39.5,
xtick style={color=black},
y dir=reverse,
y grid style={white!69.0196078431373!black},
ymin=-0.5, ymax=39.5,
ytick style={color=black}
]
\addplot graphics [includegraphics cmd=\pgfimage,xmin=-0.5, xmax=39.5, ymin=39.5, ymax=-0.5] {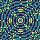};

\nextgroupplot[
hide x axis,
hide y axis,
tick align=outside,
tick pos=left,
x grid style={white!69.0196078431373!black},
xmin=-0.5, xmax=39.5,
xtick style={color=black},
y dir=reverse,
y grid style={white!69.0196078431373!black},
ymin=-0.5, ymax=39.5,
ytick style={color=black}
]
\addplot graphics [includegraphics cmd=\pgfimage,xmin=-0.5, xmax=39.5, ymin=39.5, ymax=-0.5] {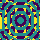};

\nextgroupplot[
hide x axis,
hide y axis,
tick align=outside,
tick pos=left,
x grid style={white!69.0196078431373!black},
xmin=-0.5, xmax=39.5,
xtick style={color=black},
y dir=reverse,
y grid style={white!69.0196078431373!black},
ymin=-0.5, ymax=39.5,
ytick style={color=black}
]
\addplot graphics [includegraphics cmd=\pgfimage,xmin=-0.5, xmax=39.5, ymin=39.5, ymax=-0.5] {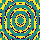};

\nextgroupplot[
hide x axis,
hide y axis,
tick align=outside,
tick pos=left,
x grid style={white!69.0196078431373!black},
xmin=-0.5, xmax=39.5,
xtick style={color=black},
y dir=reverse,
y grid style={white!69.0196078431373!black},
ymin=-0.5, ymax=39.5,
ytick style={color=black}
]
\addplot graphics [includegraphics cmd=\pgfimage,xmin=-0.5, xmax=39.5, ymin=39.5, ymax=-0.5] {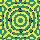};

\nextgroupplot[
colorbar,
colorbar style={ylabel={\huge Radian}},
colormap/viridis,
hide x axis,
hide y axis,
point meta max=3.14120984077454,
point meta min=-3.14141488075256,
tick align=outside,
tick pos=left,
x grid style={white!69.0196078431373!black},
xmin=-0.5, xmax=39.5,
xtick style={color=black},
y dir=reverse,
y grid style={white!69.0196078431373!black},
ymin=-0.5, ymax=39.5,
ytick style={color=black}
]
\addplot graphics [includegraphics cmd=\pgfimage,xmin=-0.5, xmax=39.5, ymin=39.5, ymax=-0.5] {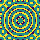};
\end{groupplot}

\end{tikzpicture}

%% file: Figure/Result/Experimental/Slice_5_MoS2/Slice_5_MoS2.tex
\pgfplotsset{every tick label/.append style={font=\LARGE}}
\begin{tikzpicture}

\begin{groupplot}[group style={group size=5 by 2,vertical sep = 1em,horizontal sep=1em},width = 0.25*\textwidth, height=(0.25)*\textwidth]
\nextgroupplot[
ylabel style={rotate=-90},
ylabel = \LARGE (a),
ticks=none,
hide x axis,
tick align=outside,
tick pos=left,
title={\LARGE Slice 1},
x grid style={white!69.0196078431373!black},
xmin=-0.5, xmax=97.5,
xtick style={color=black},
y dir=reverse,
y grid style={white!69.0196078431373!black},
ymin=-0.5, ymax=97.5,
ytick style={color=black}
]
\addplot graphics [includegraphics cmd=\pgfimage,xmin=-0.5, xmax=97.5, ymin=97.5, ymax=-0.5] {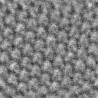};

\nextgroupplot[
hide x axis,
hide y axis,
tick align=outside,
tick pos=left,
title={\LARGE Slice 2},
x grid style={white!69.0196078431373!black},
xmin=-0.5, xmax=97.5,
xtick style={color=black},
y dir=reverse,
y grid style={white!69.0196078431373!black},
ymin=-0.5, ymax=97.5,
ytick style={color=black}
]
\addplot graphics [includegraphics cmd=\pgfimage,xmin=-0.5, xmax=97.5, ymin=97.5, ymax=-0.5] {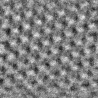};

\nextgroupplot[
hide x axis,
hide y axis,
tick align=outside,
tick pos=left,
title={\LARGE Slice 3},
x grid style={white!69.0196078431373!black},
xmin=-0.5, xmax=97.5,
xtick style={color=black},
y dir=reverse,
y grid style={white!69.0196078431373!black},
ymin=-0.5, ymax=97.5,
ytick style={color=black}
]
\addplot graphics [includegraphics cmd=\pgfimage,xmin=-0.5, xmax=97.5, ymin=97.5, ymax=-0.5] {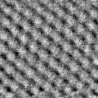};

\nextgroupplot[
hide x axis,
hide y axis,
tick align=outside,
tick pos=left,
title={\LARGE Slice 4},
x grid style={white!69.0196078431373!black},
xmin=-0.5, xmax=97.5,
xtick style={color=black},
y dir=reverse,
y grid style={white!69.0196078431373!black},
ymin=-0.5, ymax=97.5,
ytick style={color=black}
]
\addplot graphics [includegraphics cmd=\pgfimage,xmin=-0.5, xmax=97.5, ymin=97.5, ymax=-0.5] {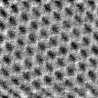};

\nextgroupplot[
title={\LARGE Slice 5},
hide x axis,
hide y axis, 
tick align=outside,
tick pos=left, 
x grid style={white!69.0196078431373!black},
xmin=-0.5, xmax=97.5,
xtick style={color=black},
y dir=reverse,
y grid style={white!69.0196078431373!black},
ymin=-0.5, ymax=97.5,
ytick style={color=black}
]
\addplot graphics [includegraphics cmd=\pgfimage,xmin=-0.5, xmax=97.5, ymin=97.5, ymax=-0.5] {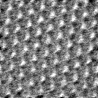};

\nextgroupplot[
ylabel style={rotate=-90},
ylabel = \LARGE (b),
ticks=none,
hide x axis,
tick align=outside,
tick pos=left,
x grid style={white!69.0196078431373!black},
xmin=-0.5, xmax=97.5,
xtick style={color=black},
y dir=reverse,
y grid style={white!69.0196078431373!black},
ymin=-0.5, ymax=97.5,
ytick style={color=black}
]
\addplot graphics [includegraphics cmd=\pgfimage,xmin=-0.5, xmax=97.5, ymin=97.5, ymax=-0.5] {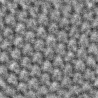};

\nextgroupplot[
colorbar horizontal,
colorbar style={xtick={-0.1, -0.05, 0, 0.05, 0.1},xticklabels={-0.1, -0.05, 0, 0.05, 0.1},width=13.5cm,at={(-1.2,-0.2)},xlabel={}},
colormap/blackwhite,
point meta max=0.1,
point meta min=-0.1,
hide x axis,
hide y axis,
tick align=outside,
tick pos=left,
x grid style={white!69.0196078431373!black},
xmin=-0.5, xmax=97.5,
xtick style={color=black},
y dir=reverse,
y grid style={white!69.0196078431373!black},
ymin=-0.5, ymax=97.5,
ytick style={color=black}
]
\addplot graphics [includegraphics cmd=\pgfimage,xmin=-0.5, xmax=97.5, ymin=97.5, ymax=-0.5] {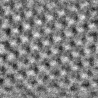};

\nextgroupplot[
hide x axis,
hide y axis,
tick align=outside,
tick pos=left,
x grid style={white!69.0196078431373!black},
xmin=-0.5, xmax=97.5,
xtick style={color=black},
y dir=reverse,
y grid style={white!69.0196078431373!black},
ymin=-0.5, ymax=97.5,
ytick style={color=black}
]
\addplot graphics [includegraphics cmd=\pgfimage,xmin=-0.5, xmax=97.5, ymin=97.5, ymax=-0.5] {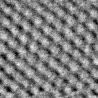};

\nextgroupplot[
hide x axis,
hide y axis,
tick align=outside,
tick pos=left,
x grid style={white!69.0196078431373!black},
xmin=-0.5, xmax=97.5,
xtick style={color=black},
y dir=reverse,
y grid style={white!69.0196078431373!black},
ymin=-0.5, ymax=97.5,
ytick style={color=black}
]
\addplot graphics [includegraphics cmd=\pgfimage,xmin=-0.5, xmax=97.5, ymin=97.5, ymax=-0.5] {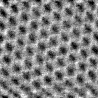};

\nextgroupplot[
hide x axis,
hide y axis,
tick align=outside,
tick pos=left,
x grid style={white!69.0196078431373!black},
xmin=-0.5, xmax=97.5,
xtick style={color=black},
y dir=reverse,
y grid style={white!69.0196078431373!black},
ymin=-0.5, ymax=97.5,
ytick style={color=black}
]
\addplot graphics [includegraphics cmd=\pgfimage,xmin=-0.5, xmax=97.5, ymin=97.5, ymax=-0.5] {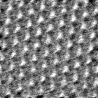};
\end{groupplot}
 
\end{tikzpicture}

%% file: Figure/Result/Experimental/Projection_3D_2D/Proj3Dto2D.tex
\pgfplotsset{every tick label/.append style={font=\normalsize}}
\begin{tikzpicture}
\begin{groupplot}[group style={group size=2 by 1,horizontal sep=6em},width = 0.25*\textwidth, height=(0.25)*\textwidth]
\nextgroupplot[
colorbar,
colorbar style={ylabel={ }},
colormap/blackwhite,
hide x axis,
hide y axis,
point meta max=0.21577313542366,
point meta min=-0.208285182714462,
tick align=outside,
tick pos=left,
title={\normalsize (a)},
x grid style={white!69.0196078431373!black},
xmin=-0.5, xmax=97.5,
xtick style={color=black},
y dir=reverse,
y grid style={white!69.0196078431373!black},
ymin=-0.5, ymax=97.5,
ytick style={color=black}
]
\addplot graphics [includegraphics cmd=\pgfimage,xmin=-0.5, xmax=97.5, ymin=97.5, ymax=-0.5] {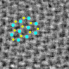};

\nextgroupplot[
colorbar,
colorbar style={ylabel={}},
colormap/blackwhite,
hide x axis,
hide y axis,
point meta max=0.227930992841721,
point meta min=-0.218955889344215,
tick align=outside,
tick pos=left,
title={\normalsize (b)},
x grid style={white!69.0196078431373!black},
xmin=-0.5, xmax=97.5,
xtick style={color=black},
y dir=reverse,
y grid style={white!69.0196078431373!black},
ymin=-0.5, ymax=97.5,
ytick style={color=black}
]
\addplot graphics [includegraphics cmd=\pgfimage,xmin=-0.5, xmax=97.5, ymin=97.5, ymax=-0.5] {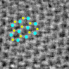};
\end{groupplot}

\end{tikzpicture}

%% file: Figure/Result/Experimental/Probe_Sparse/Probe_Recon.tex
\pgfplotsset{every tick label/.append style={font=\normalsize}}
\begin{tikzpicture}
\begin{groupplot}[group style={group size=2 by 1,horizontal sep=6em},width = 0.25*\textwidth, height=(0.25)*\textwidth]
\nextgroupplot[
colorbar,
colorbar style={ylabel={\normalsize Norm. Amp.}},
colormap/blackwhite,
hide x axis,
hide y axis,
point meta max=1,
point meta min=9.96048710744433e-05,
tick align=outside,
tick pos=left,
title={\normalsize (a)},
x grid style={white!69.0196078431373!black},
xmin=-0.5, xmax=97.5,
xtick style={color=black},
y dir=reverse,
y grid style={white!69.0196078431373!black},
ymin=-0.5, ymax=97.5,
ytick style={color=black}
]
\addplot graphics [includegraphics cmd=\pgfimage,xmin=-0.5, xmax=97.5, ymin=97.5, ymax=-0.5] {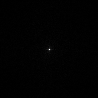};
 
\nextgroupplot[
colorbar,
colorbar style={ylabel={\normalsize Radian}},
colormap/blackwhite,
hide x axis,
hide y axis,
point meta max=3.14149269218696,
point meta min=-3.14142961791758,
tick align=outside,
tick pos=left,
title={\normalsize (b)},
x grid style={white!69.0196078431373!black},
xmin=-0.5, xmax=97.5,
xtick style={color=black},
y dir=reverse,
y grid style={white!69.0196078431373!black},
ymin=-0.5, ymax=97.5,
ytick style={color=black}
]
\addplot graphics [includegraphics cmd=\pgfimage,xmin=-0.5, xmax=97.5, ymin=97.5, ymax=-0.5] {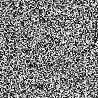};
\end{groupplot}

\end{tikzpicture}